\documentclass[%
reprint,
nofootinbib,
amsmath,
amssymb,
aps,
superscriptaddress, 
showkeys,
showpacs
]{revtex4-1} 

\usepackage{graphicx}
\usepackage{dcolumn}
\usepackage{bm}

\mathchardef\mhyphen="2D		
\usepackage{hyperref}		
\usepackage{booktabs}		
\usepackage{multirow}		

\usepackage{array}
\newcommand\MyBox[2]{
  \fbox{\lower0.75cm
    \vbox to 1.7cm{\vfil
      \hbox to 1.7cm{\hfil\parbox{1.4cm}{#1\\#2}\hfil}
      \vfil}
  }
}

\AtBeginDocument{
\heavyrulewidth=.08em
\lightrulewidth=.05em
\cmidrulewidth=.03em
\belowrulesep=.65ex
\belowbottomsep=0pt
\aboverulesep=.4ex
\abovetopsep=0pt
\cmidrulesep=\doublerulesep
\cmidrulekern=.5em
\defaultaddspace=.5em
}
\begin{document}


\title{Deep learning for gravitational-wave data analysis: \\A resampling white-box approach}

\author{Manuel D. Morales}
\email{Spokesperson: manueld.morales@academicos.udg.mx}
\affiliation{Departamento de F\'isica, Centro Universitario de Ciencias Exactas e Ingenier\'ias, Universidad de Guadalajara, Av. Revoluci\'on 1500, Guadalajara, Jalisco, 44430, M\'exico}

\author{Javier M. Antelis}
\email{mauricio.antelis@tec.mx}
\affiliation{Tecnol\'ogico de Monterrey, Escuela de Ingenier\'ia y Ciencias, Av. Gral. Ram\'on Corona 2514, Zapopan, Jalisco C.P. 45138, M\'exico}

\author{Claudia Moreno}
\email{claudia.moreno@cucei.udg.mx}
\affiliation{Departamento de F\'isica, Centro Universitario de Ciencias Exactas e Ingenier\'ias, Universidad de Guadalajara, Av. Revoluci\'on 1500, Guadalajara, Jalisco, 44430, M\'exico}

\author{Alexander I. Nesterov}
\email{nesterov@cencar.udg.mx}
\affiliation{Departamento de F\'isica, Centro Universitario de Ciencias Exactas e Ingenier\'ias, Universidad de Guadalajara, Av. Revoluci\'on 1500, Guadalajara, Jalisco, 44430, M\'exico}

\date{\today}

\begin{abstract}
In this work, we apply Convolutional Neural Networks (CNNs) to detect gravitational wave (GW) signals of compact binary coalescences, using single-interferometer data from LIGO detectors. As novel contribution, we adopted a resampling white-box approach to advance towards a statistical understanding of uncertainties intrinsic to CNNs in GW data analysis. Resampling is performed by repeated $k$-fold cross-validation experiments, and for a white-box approach, behavior of CNNs is mathematically described in detail. Through a Morlet wavelet transform, strain time series are converted to time-frequency images, which in turn are reduced before generating input datasets. Moreover, to reproduce more realistic experimental conditions, we worked only with data of non-Gaussian noise and hardware injections, removing freedom to set signal-to-noise ratio (SNR) values in GW templates by hand. After hyperparameter adjustments, we found that resampling smooths stochasticity of mini-batch stochastic gradient descend by reducing mean accuracy perturbations in a factor of $3.6$. CNNs were quite precise to detect noise but not sensitive enough to recall GW signals, meaning that CNNs are better for noise reduction than generation of GW triggers. However, applying a post-analysis, we found that for GW signals of SNR $\geq 21.80$ with H1 data and SNR $\geq 26.80$ with L1 data, CNNs could remain as tentative alternatives for detecting GW signals. Besides, with receiving operating characteristic curves we found that CNNs show much better performances than those of Naive Bayes and Support Vector Machines models and, with a significance level of $5\%$, we estimated that predictions of CNNs are significant different from those of a random classifier. Finally, we elucidated that performance of CNNs is highly class dependent because of the distribution of probabilistic scores outputted by the softmax layer.
\end{abstract}



\maketitle

\section{Introduction}

Beginning with the discovery of LIGO and Virgo collaborations~\cite{bA16}, observation of gravitational wave (GW) events emitted by compact binary coalescences (CBCs) have made GW astronomy, to some extent, a routine practice. Until now, eleven GW events (\texttt{GWTC-1-confident}) has been observed during O1 and O2 scientific runs~\cite{bA19a}, within which ten correspond to GW from binary black hole (BBH) sources and, notably, one from a binary neutron star (BNS) system~\cite{bA17a} with a reported electromagnetic counterpart~\cite{bA17b} that usher in a era of multimessenger astronomy. More recently, more than a dozen GW events has been detected during O3 run, including the first observation of a BBH with asymmetric masses~\cite{rA20a}, and a CBC of a black hole and other compact object that could be either a low-mass black hole or a heavy neutron star, without detected electromagnetic counterpart~\cite{rA20b}. O3 runs have been recorded with the network of LIGO Livingston (L1) and Hanford (H1) twin observatories and Virgo (V1) observatory.

Sensitivity of GW detectors have been remarkably increased these last years. Here GW data analysis had a crucial role quantifying and stirring as much as possible effects of non-Gaussian noise, mainly short-time transients called ``glitches''~\cite{bA13} for which there is still no fully explanations about their physical causes. Monitoring signal-to-noise ratio (SNR) of all detections is an essential procedure here, being indeed the heart of the standard algorithm used for detection and characterization of GWs in LIGO and Virgo, namely Matched Filter (MF)~\cite{bA12,sB13}. This technique is not original from GW data analysis, but rather has been broadly used in electrical engineering, particularly signal processing, from decades~\cite{gT60}. In general, MF is based on the asumption that signals to be search are embedded in additive Gaussian noise. And in particular, for GW data analysis, the very starting point are raw strain data (scalar time series) outputted by each detector of a network of detectors, having beforehand a wide bank of theoretical GW templates. Then, MF performs a one-by-one searching with all possible combinations, correlating whitened strain data with each template. In particular, each correlation actually define a SNR value and, consequently, the goal is selecting those matchs that maximize SNR, mediated by several statistical and coincidente tests between detectors of the network~\cite{jAcM17}. A standard pipeline that currently includes MF and is used in LIGO and Virgo analyses is PyCBC~\cite{sU16,aNtCdDsR18}.

Searchings performed by LIGO and Virgo for CBC signals have generally resorted to a bank theoretical templates with inspiral, merger, and ringdown stages, in order to cover a limited space of mass and spin values. Here MF has shown be a powerfull tool having a crucial role in all confirmed detections of GWs emitted by CBCs. However, there are significant reasons for exploring alternative detection strategies. To begin, noise outputted by interferometric detectors is non-Gaussian, which conflicts with assumption of Gaussian noise in MF techniques. Generally, whitening techniques~\cite{eCgCmG04} are applied to raw strain data before detection and characterization algorithms but glitches remain, raising the need of applying consistence statistical tests~\cite{bA05} and coincidence procedures as cross-correlation~\cite{bA19b,jAcM19} for a network of detectors, which rasing the problem of how to systematically deal with single detector data.

Elsewhere, one would want to perform more general GW searchings. Natural next steps is to include precessing spins~\cite{iHsPaBaB16}, orbital eccentricity~\cite{eH17}, and neutron star tidal deformability~\cite{jT20}, among others aspects. The more general a GW searching the much wider the parameter space and MF becomes computationally prohibitive.

Besides, there are sources other than CBCs that could produce nondeterministic GWs~\cite{bA19c} which, in principle, cannot been anticipated through theoretical waveforms and, therefore, make impossible any analysis with MF. Phenomena as core-collapse supernovae (CCSNe)~\cite{clFkcbN11}, pulsar glitches~\cite{nAglC01}, and neutron stars collapsing into black holes~\cite{lBiHlR07}, among others, fall into this category. Here it should be stressed that LIGO and Virgo pipelines currently have a standard data analysis algorithm for all-sky searching of GW transients without prior theoretical templates, namely coherent WaveBurst (cWB)~\cite{sKiYaMgM08}. Under the hood, this algorithm includes standard cleaning and whitening procedures, Wilson-Daubechies-Meyer wavelet transform, clustering of time-frequency samples based on power excesses, and reconstruction of waveforms by maximum likelihood estimation. cWB can also analize data from a network of detectors~\cite{sK16}. However, cWB rests upon the same assumption as MF, againly, that noise background is Gaussian, which is not the case given the outputs of interferometric detectors as it was mentioned, even after applying whitenings.

Actually, if we are not interested in replacing standard algorithms as MF and/or cWB, still it is pertinent to have alternative algorithms just for independent verifications and/or increasing the confidence level of GW detections.

In this context, Machine Learning (ML)~\cite{Bishop-Book} and its successor, Deep Learning (DL)~\cite{Goodfellow-Book}, emerge as promising alternatives or, at least, complementary tools to MF and/or cWB. These techniques assume nothing regarding background noise. In addition, they have shown be remarkably useful to analyze enormous mounts of data through sequential or online learning processes, which could be a significant improvement for more general GW searchings in real time. Even more, if we do not have deterministic templates to predict GW signals as those emmited by CCSNe or anticipate noise artifacts as non-Gaussian glitches, unsupervised ML and DL algorithms (which do not need prior labeled training samples) could be interesting for future explorations.

It is well known that, in recent years, ML techniques have had a remarkable boom, with many successful applications in both industry and academia: from recommender systems~\cite{Ricci-Book}, fraud detection~\cite{xZ18}, genomics~\cite{mLaDbAbF16}, and astronomy~\cite{cFcJ20}, among others. Overlapping between ML and physics is particularly interesting, because works as a cross-interaction. For instance, we have applications of ML to particular subfields of physics (e.g. cosmology~\cite{DESC16} and chemical physics~\cite{lZ18}), conceptual developments in ML algorithms inspired by theoretical physical insights~\cite{tCmWbKmW19}, and AI enhacements with quantum computing~\cite{Schuld-Book}. Carleo et al.~\cite{gC19} provide an illustrative review about ML and physical sciences.

In GW data analysis, implementation of ML algorithms have no more than a few years. Biswas et al.~\cite{rB19} contributed with a pioneering work, where Artificial Neural Networks (ANNs), Support Vector Machines, and Random Forest algorithms were used to detect glitches in data from H1 and L1 detectors, recorded during S4 and S6 runs. Competitive performance results, back then, were obtained. Later, works focused on several problems were published, with better results. For instance: ANN for detection of GWs associated with short gamma-ray bursts~\cite{kK15}, Dictionary Learning for denoising~\cite{aTFaMjFjI16}, Difference Boosting Neural Network and Hierarchical Clustering for detection of glitches~\cite{nM17}, ML and citizen science for glitches classification\cite{mZ17}, and background reduction for CCSNe searching with single-detector data~\cite{mC19}, among others.

Applications of DL, in particular Convolutional Neural Network (CNN) algorithms, are even more recent in GW data analysis. Gabbard et al.\cite{hGmWfHcM18}, George and Huerta~\cite{dGeH18a}, provided first works, in which CNNs were implemented to detect simulated GW signals from BBHs embedded in Gaussian noise. They claimed that performance of CNNs was similar and much better than MF, respectively. Later, George and Huerta extended their work by embedding simulated GWs and analytical glitches in real non-Gaussian noise data of LIGO detectors, with similar results~\cite{dGeH18b}. Thereafter, CNN algorithms has been applied for several instrumental and physical problems, showing more improvements. For instance: detection of glitches~\cite{mReC18}, trigger generation for locating coalescente time of GWs emitted by BBHs~\cite{tGnKiHbS19}, detection of GWs from BBHs and BNS~\cite{pK20}, detection of GWs emitted by CCSNe and using both phenomenological~\cite{pA18} and numerical~\cite{aIeCfMjP19} waveforms, and detection of continuous GWs from isolated neutron stars~\cite{aM19}, among others.

From a practical point of view, previous works of CNN algorithms applied to GW detection and characterization reach competitive performance results according to standard metrics as accuracy, loss functions, false alarm rates, etc., showing feasibility of DL in GW data analysis in first place. However, from a formal statistical point of view we warn that, beyond just applications, we are in need of deeper explorations that seriously take into account the inevitable uncertainty involved in DL algorithms before put them as real alternatives to standard pipelines in LIGO and Virgo.

For instance, Gebhard et al.~\cite{tGnKiHbS19} critically pointed out, a CNN algorithm does not input waveform templates as isolated samples, but also distribution of these templates considering the whole training and testing datasets. Indeed, in all mentioned previous works of DL applied to GW data analysis, distributions of samples were set class-balanced totally by hand. It is a common practice in ML and DL to draw on artificial balanced datasets to made CNN algorithms easily tractable with respect to hyperparameter tuning, choice of performance metrics, and cost missclasification. Nonetheless, when uncertainty is taken into account, real frequency of occurrence of samples cannot be ignored, because they define how reliable is our decision criteria for classifying, when an algorithm output an score for a single input sample. This kind details are very known in ML community~\cite{Japkowicz-Book}, and need to be seriously explored in GW data analysis beyond just hands-on approaches. Deeper multidisciplinary researches are necessary to advance in this field.

For this research, our general goal is to draw on CNN architectures to make a standard GW detection. In particular, beginning from a training set containing single-interferometer strain data, and then transforming it into time-frequency images with a Morlet wavelet transform, the aim is distinguish those samples which are only non-Gaussian noise from those samples that contain a GW embedded in non-Gaussian noise --in ML language, this is just a binary classification. Moreover, as a novel contribution, here we will take into consideration two ingredients to advance towards a statistically informed understanding of involved uncertainties, namely: resampling, and a white-box approach.

In this work we are not interested in reaching higher performance metric values than those reported in previous works, neither testing new CNN architectures, neither using latest real and/or simulated data; but rather facing the question about how to clever deal with uncertainties of CNNs which is an unescapable requesite if we claim that DL techniques are real alternatives to current pipelines. Moreover, as a secondary result that is useful for reproductibility purposes, we want to show that CNN algorithms, even considering repeated experiments, can be easily run in a single local CPU to reach good performance results, without resorting to expensive high-performance computing --result that we achieved mainly by transforming raw strain data to low resolution images in the time-frequency representation.

In particular, our resampling white-box approach has several subtleties, as follows.

Full problem of dealing with arbitrarily imbalanced datasets, for now, is beyond of this research. Nonetheless, even though working with balanced datasets, a good starting point is to include stochasticity by resampling, which in turn we define as repeated experiments of a global $k$-fold cross-validation (CV) routine. This stochasticity is different to that usually introduced in each learning epoch by taking a mini-batch of the whole training set for updating the model parameters (e.g. by a stochastic gradient descent algorithm) and, therefore, minimizing the cross-entropy. For this research, we will consider the above two sources of stochasticity.

Stochastic resampling helps to aliviate artificiality introduced by a balanced dataset, because the initial splitting into k folds is totally random and each whole $k$-fold CV experiment is not reproducible in a deterministic fashion. Besides, this approach put an experimental setup in which uncertainties are even more evident and need to be seriously treated beyond of just reporting metrics of single values. Indeed, in most common situations with really big datasets (i.e. millions, billions, or more samples), stastistical tools for decreasing system resources, and data changing over time, among others; CNN algorithms are generally set such that their predictions are not deterministic, leading to distributions of performance metrics instead of single value metrics --and demanding formal probabilistic analyses. Given these distributions, with their inherent uncertainties, a statistical paired-sample test is necessary to formally conclude how close or far is our CNNs to a totally random classifier, and that we performed in this work.

Our white-box approach works as a complementary tool to understand how uncertainties influence performance of CNN algorithms. First aspect here is to explicitely (mathematically) describe how each layer of a CNN works and why to choose these layers. This is actually the most basic explanatory procedure considering that, from a fundamental point of view, we still do not have analytical theories to explain low generalization errors in DL and to ensure, beforehand, good performances of a certain CNN architecture --in practice, we only have a lot of previously implemented CNN architectures for other problems, that need to be test as a just essay-error process for a new problem, actually. Then, this explicit description assist us to understand why, given output probabilistic scores, classification is class dependent and threshold-dependent given the distribution of these scores.

With regard to data, we decided to use recordings from S6 LIGO run, separately from H1 and L1 detectors, considering GW signals of CBCs only generated by hardware injections. This approach reproduces more realistic conditions and further strengthens performace results of CNNs, because it removes beforehand the instrumental freedom of generating distribution of numerical relativity templates with respect to SNR values totally by hand. Take in mind that, in real experimental conditions, SNR values of GW events, and even, their frequency of occurrence, are given as (unmodifiable) facts, because these directly depend on the nature of the astrophysical sources and the GW detectors.

Hereinafter, this paper will organize as follows. In Section~\ref{sec:methods} we describe technical details of the methodology and its implementation, including datasets, pre-processing, CNN architectures, model training, validation, and selection of performance metrics. Later, in Section ~\ref{sec:results} we present the results and discussion, including learning monitoring, hyperparameter adjustments, performance metrics, and statistical tests. Finally, in Section~\ref{sec:conclusions}, we include conclusions. In Appendices~\ref{sec:App_NB} and~\ref{sec:App_SVM} we present brief descriptions of Naive Bayes and Support Vector Machines models, respectively, which are classic ML algorithms that we compared with our CNN algorithms as it is shown in Section~\ref{sec:results}.

\section{Methods and Materials}
\label{sec:methods}

\subsection{Problem statement}
\label{sec:probstat}
As starting point of our problem, we have a $i \mhyphen th$ slice of raw strain data recorded at one of the LIGO detectors. In mathematical notation, this slice of data is expressed by a column vector of times series in $N$ dimensions,
\begin{eqnarray}
 s_{\text{raw}}{}^i (t) &=& \left[s^i(t_0), s^i(t_1), ..., s^i(t_{N-1})\right]^T~, \label{eq:raw_strain}
\end{eqnarray}
where the sampling time is $t_s = t_j - t_{j-1}$ with $j=1,2,...,N_{\text{slice}}-1$, the sampling frequency is $f_s = 1 / t_s$, and of course, the time length of the slice is $T_{\text{slice}}=N_{\text{slice}}/f_s$ (in seconds) with $N_{\text{slice}}$ points of data. The next step is to model theoretically the above slice of data, therefore we introduce the following expression:
\begin{eqnarray}
 s_{\text{raw}}{}^i (t) =
 \begin{cases}
  n^i{}(t)  &\text{if there is not a GW}~, \\
  n^i{}(t) + h^i(t) &\text{if there is a GW}~,
 \end{cases} \label{eq:GWdecision}
\end{eqnarray}
where $n^i(t)$ is the non-Gaussian noise from the detector and $h^i(t)$ the strain from a GW whose duration, arrival time and waveform are unknown. Notice that $h^i(t)$ actually is the response of the detector in its output when a GW is detected in its input, thus its amplitude not only depends on the intensity of the upcoming GW, but also on the location of the source in the sky, the polarization of the GW, and the location of the detector on the Earth's surface. If we work with data of one detector, Eq.~\ref{eq:GWdecision} is not afected but, if we consider data from two or more detectors together (i.e. data from a network of detectors), the GW strain $h^i(t)$ will need be multiplied by a specific scale factor depending on the detector, and noise $n^i{}(t)$ will be different.

For this research, the problem that we address is to decide if a segment of strain data contains only noise, or contains noise plus a unknown GW signal. Then, in practice, we will implement CNN algorithms that performs a binary classification, inputting strain samples of time lengh $T_{\text{win}} < T_{\text{slice}}$ and deciding, for each sample, if it does not contain a GW signal (i.e. the sample $\in$ class 1), or contains a GW signal (i.e. the sample $\in$ class 2).

\subsection{Dataset description}
\label{sec:datadescrip}

In order to build our strain samples, we use real data provided by LIGO detectors, which are freely available on the LIGO-Virgo Gravitational Wave Open Science Center (GWOSC), \url{https://www.gw-openscience.org}. We decided to use data of sixth science run (S6), recorded from July 07, 2009 to October 20, 2010. During this run, detectors achieved a sensitivity given by a power spectral denstity around $10^{-22}$, with uncertainties up to $\pm 15\%$ in amplitude~\cite{jA12}. Each downloaded strain data slice has a time length of $T_{\text{slice}}=4096$s and a sampling frequency of $f_s=4096 $Hz.

S6 contains hardware injections already added to the noise strain data. These injections are generated by physically displacing the legs of the detectors simulating the effects of GWs, and are used for experimental tests and calibration of the detectors. For this research, we solely work with hardware injections of GWs emitted by CBCs. For each injection, we known the coalescence (or merger) time $t_c$ in GPS, masses $m_1$ and $m_2$ in solar mass units $M_\odot$, the distance $D$ to the source in Mpc, the expected and the recovered signal-to-noise ratio, namely $SNR_{exp}$, $SNR_{rec}$, respectively. All this information is provided by LIGO on the aforementioned website.

\begin{figure*}[htp]
\begin{center}
  \includegraphics[width=8.2cm]{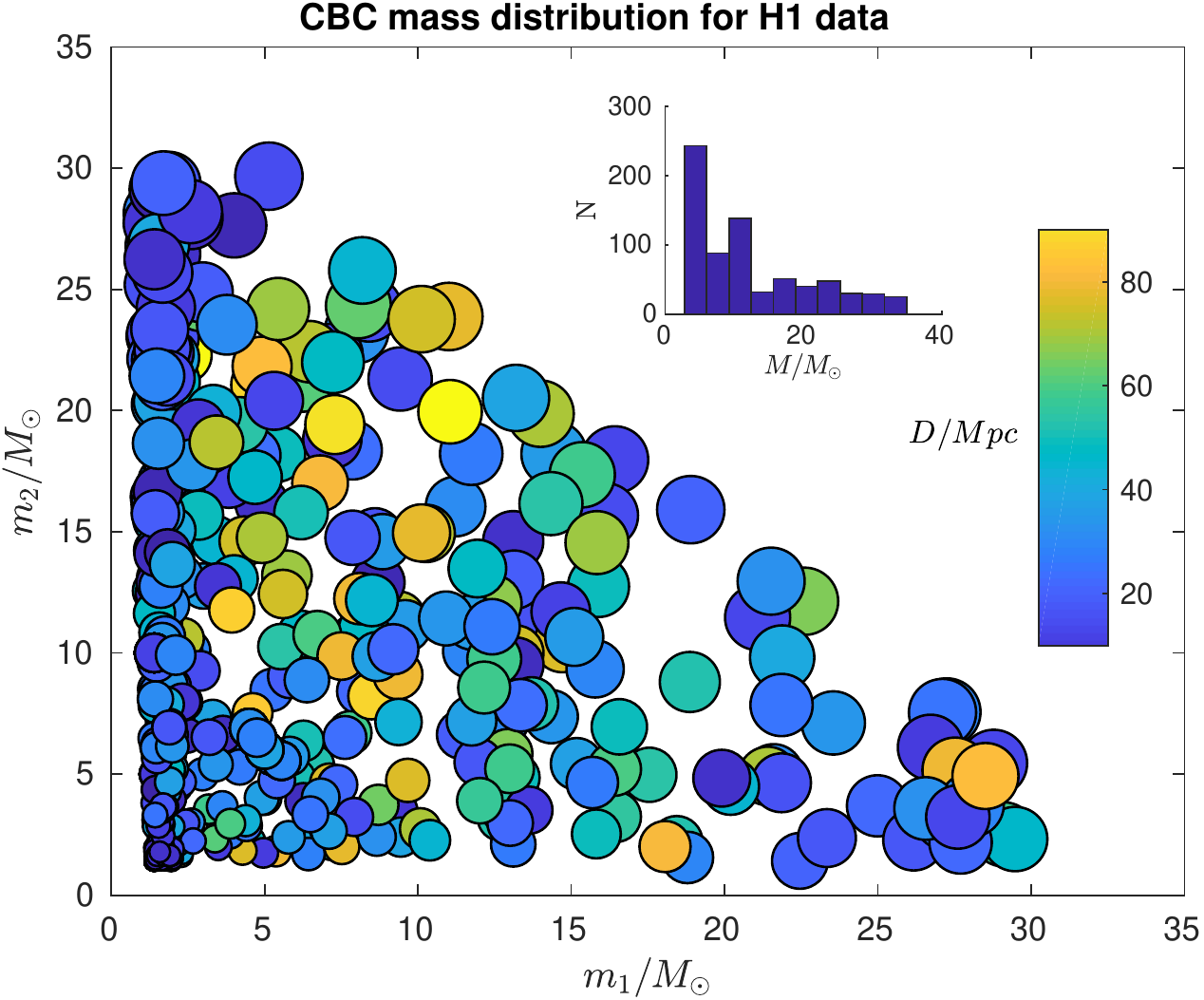}
  ~~~~
  \includegraphics[width=8.2cm]{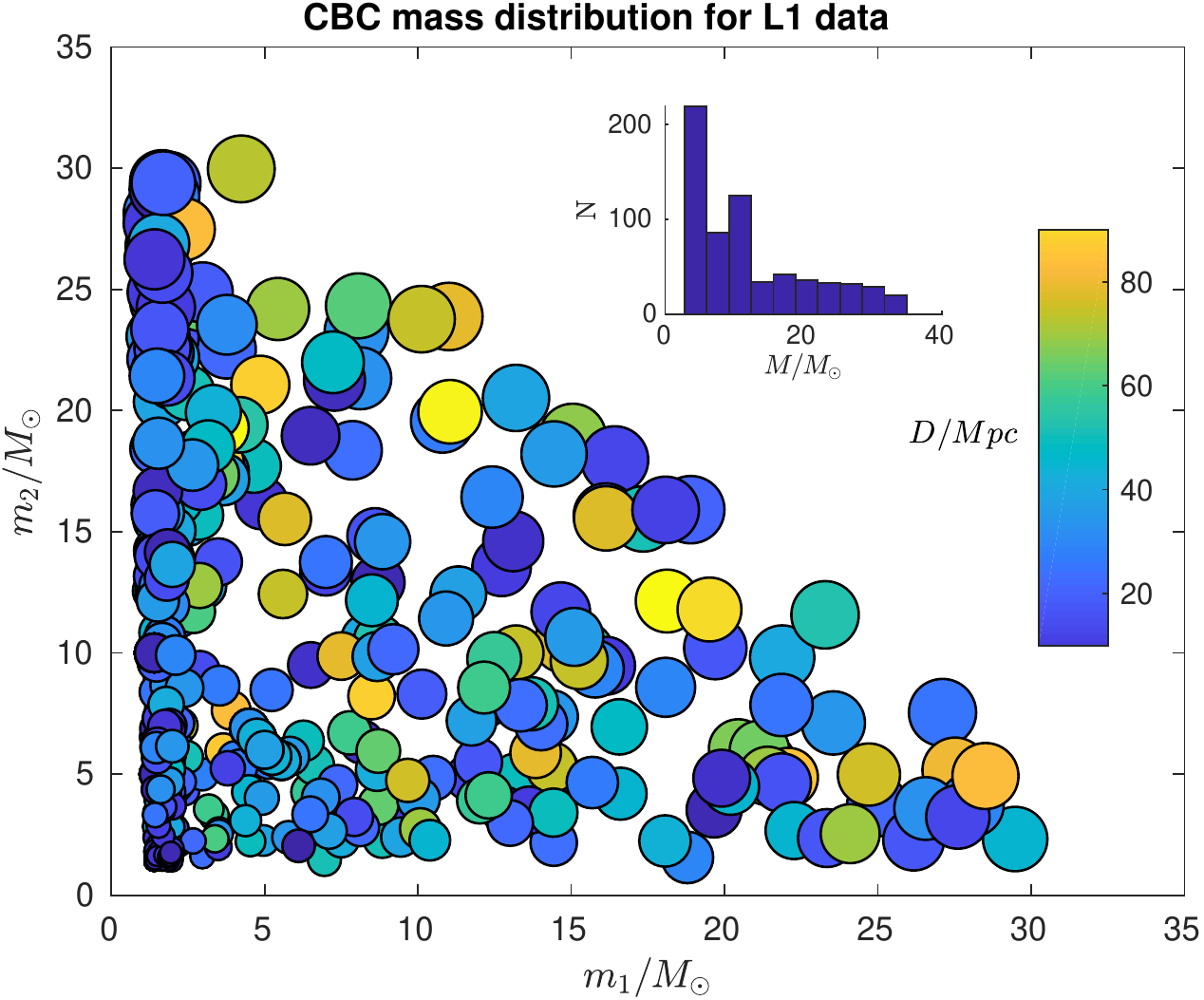}
   \caption{\label{fig:CBCmass_distribution} Distributions of CBC GW injections present in the S6 data from LIGO detectors H1 (left panel) and L1 (right panel). In particular, masses of and distances from the sources are shown. Except those GW events lying just adjacent to the vertical axis and some outliers, we can observe that the general trend is that as total mass $M=m_1+m_2$ increases, GW events tend to be less frequent and more distant. Data obtained from the LIGO-Virgo GWOSC, \url{https://www.gw-openscience.org}.}
\end{center}
\end{figure*}

For a detailed exploration, consider Fig.~\ref{fig:CBCmass_distribution}. This shows distributions of CBC injections in S6 data, from H1 and L1 detectors, giving information about masses of and distances from the sources. These injections simulate stellar BBHs located at the scale of nearby galaxies, covering a distance range from $10$Mpc to $100$Mpc. For data from both detectors there is a high occurrency of sources with total mass $M=m_1+m_2$ from $1M_\odot$ to $12M_\odot$ aprox. These GW events are included in the first three bins of the histograms, being mainly at distances about $10$Mpc to $30$Mpc according to the scatter plots --excepting some outliers lying in the subregions $m_1>m_2$ with distances $10$Mpc $\leqslant M \leqslant 12$Mpc. In addition, for total masses $M > 12M_\odot$, most of BBHs are more distant yet their frequency of occurrence clearly decreases --excepting those events appearing adjacent to the vertical axis, for $m_1$ taking its smaller values independent of $m_2$, describing BBHs at closer distances with high frequency of occurrence.

For this research, we discard data blocks with missing strain data, i.e. containing NaN (not a number) entries; and blind injections, i.e. those with no information except $t_c$. We downloaded $501$ segments of data from H1 and $402$ segments of data from L1, in which science data contained in time series $s_{\text{raw}}{}(t)$ are totally available.

\subsection{Data pre-processing}
\label{sec:datapreprocess}

With LIGO raw strain data segments at hand, the next methodological step is data pre-processing. This has three stages, namely data cleaning, construction of strain samples, and application of wavelet transform.

\subsubsection{Data cleaning}

\begin{figure*}[htp]
\begin{center}
  \includegraphics[width=8.0cm]{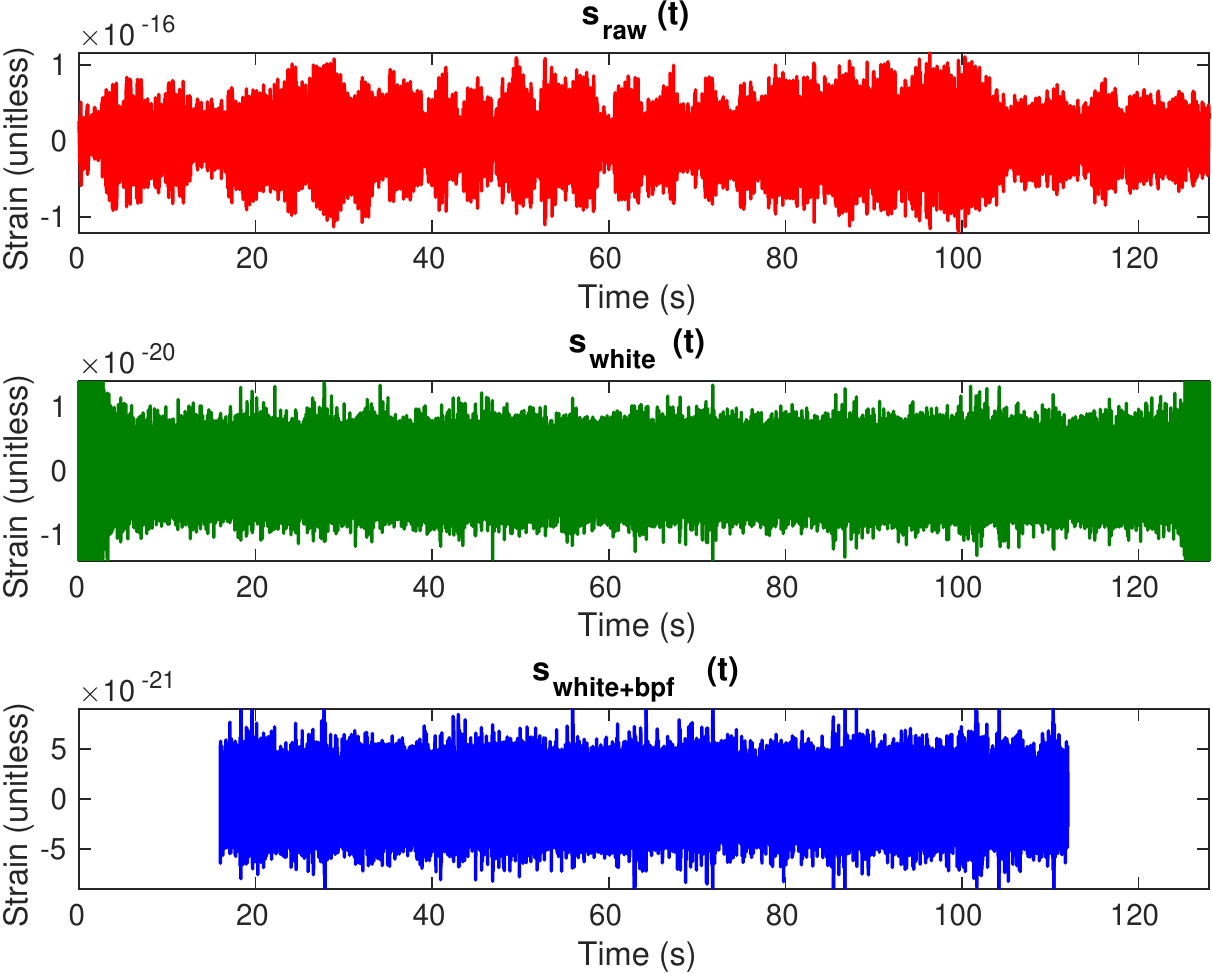}
  ~~~~
  \includegraphics[width=8.0cm]{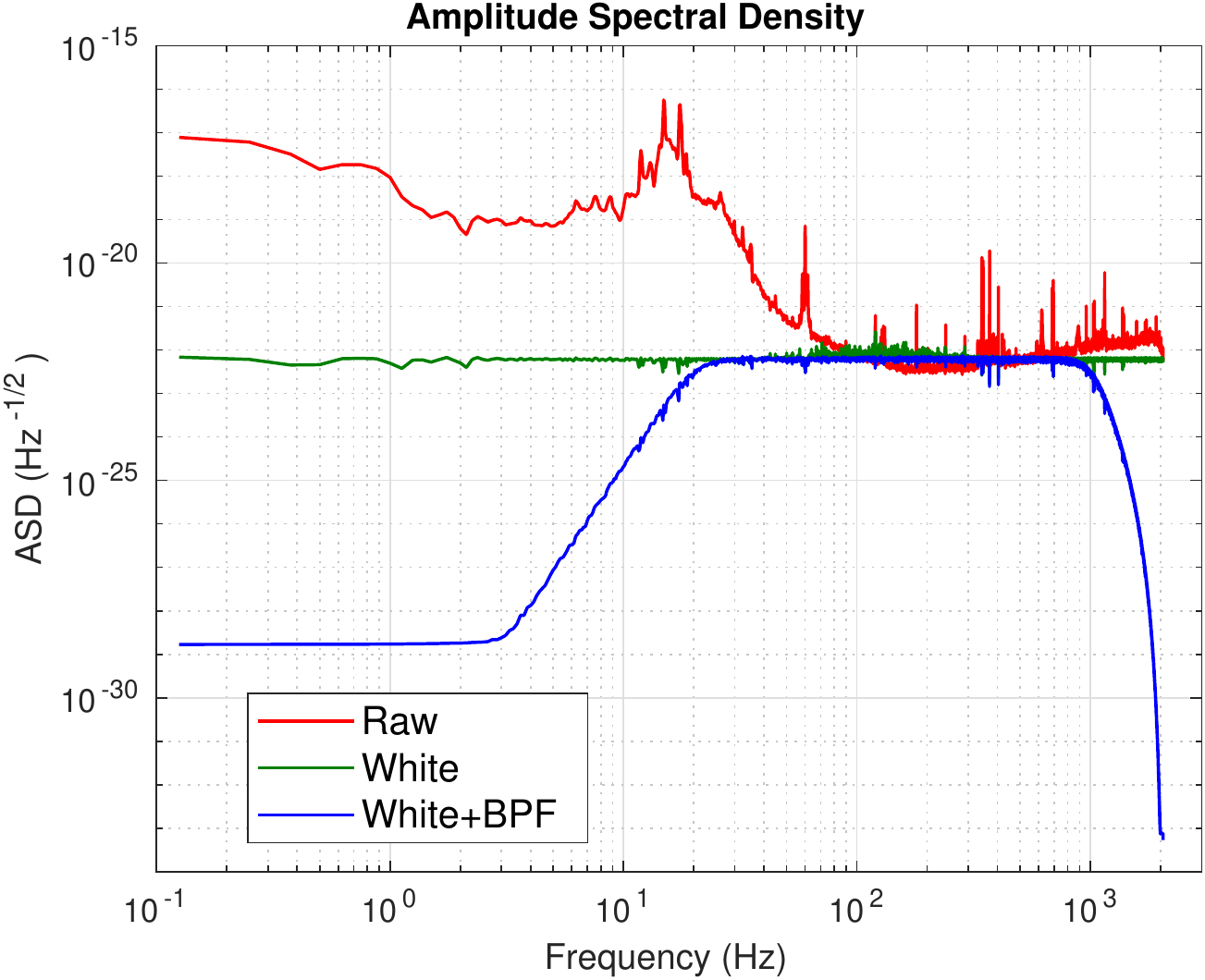}
  \caption{\label{fig:data_cleaning} Illustration of the raw strain data cleaning. The left panel shows three plots as time series segments. First, a segment of raw strain data of $128$s after we applied the blackman window. Second, the resulting strain data after the whitening, in which it is noticeable the spurious effects at the edges. Finally, the cleaned segment of data after applied the band-pass filtering and edges removal to the previous whitened segment of data. In the right panel we plot the same data, but now in the amplitude spectral density (ASD) vs. frequency space.}
\end{center}
\end{figure*}

Data cleaning or data conditioning is a standard stage to reduce the noise and flattening the power spectral density (PSD). It consists in three steps. Firstly, for each $i \mhyphen th$ slice of raw strain data that we introduced in Eq.~\ref{eq:raw_strain}, segments around coalescence times $t_c$ are extracted via blackman window of time length $128$s. Secondly, a whitening is applied to each $l \mhyphen th$ of these segments as:
\begin{equation}
 s_{\text{white}}{}^l(t) = \sum_{k=1}^{N}
 \frac{\tilde{s}_{\text{raw}}{}^l(f_k)}{\sqrt{S_{\text{raw}}{}^l(f_k)}}e^{i2\pi t k / N}~,
\end{equation}
where $\tilde{s}{}^l_{\text{raw}}(f_k)$ is the $N \mhyphen$point discrete Fourier Transform of the $128$s segment of raw strain data $s_{\text{raw}}{}^l(t)$, $S_{\text{raw}}{}^l(f_k)$ the $N \mhyphen$point two-sided PSD of the raw data, $N$ the number of points of data in the $128$s raw strain segment, and $i$ the imaginary unit. In theory, PSD is defined as the Fourier transform of the raw data autocorrelation. Then, we implemented the Welch's estimate~\cite{pW67}, where we compute the PSD between $0$Hz and $2048$Hz at a resolution of $1/128$Hz applying Hanning-windowed epochs of time length $128$s with overlap of $64$s. At the end, goal of whitening process is to approximate strain data to a Gaussian stochastic process defined by the following autocorrelation:
\begin{equation}
             R(\tau) = \langle s_{\text{white}}\left(t+\tau\right) s_{\text{white}}\left(\tau\right) \rangle
                     = \sigma^2\delta[\tau]~,
\end{equation}
with $\sigma$ denoting the variance and $\delta[\tau]$ the discrete time unit impulse function\footnote{Explicitly, the discrete time unit impulse function is defined by $\delta[\tau]=1$ for $\tau = 0$, and $\delta[\tau]=0$ for $\tau \neq 0$.}. Finally, a Butterworth band-pass filter from $20$Hz to $1000$Hz is applied to the already whitened segment. This filtering removes extreme frequencies that are out of our region of interest and discards $16$s on the edges of the segment to avoid roll-off effects, resulting in a new segment $s_{\text{white}}+s_{\text{bpf}}=s_{\text{clean}}$ of time length $T_{\text{clean}}=96$s. Fig.~\ref{fig:data_cleaning} shows the effects of applying the whitening and the band-pass filtering to the raw strain data, both in the time domain and the amplitude spectral density (ASD) wich is computed as $\sqrt{PSD}$. In particular, from the time domain plots, it can be seen that after the cleaning, amplitude of the strain data is reduced $5$ orders of magnitude, from $10^{-16}$ to $10^{-21}$ --effect of data corruption at the edges of data slice, after whitening, is clearly shown in the middle plot. Also notice that before applying data cleaning, ASD shows the known noise profile that describes the sensitivity of LIGO detectors, which is the sum of all contributions of noise sources, namely seismic activity, thermal fluctuations, and variable laser intensity, among others.

\subsubsection{Strain samples}
\label{subsec:strainsamples}

\begin{figure}[htp]
\begin{center}
  \includegraphics[width=8.2cm]{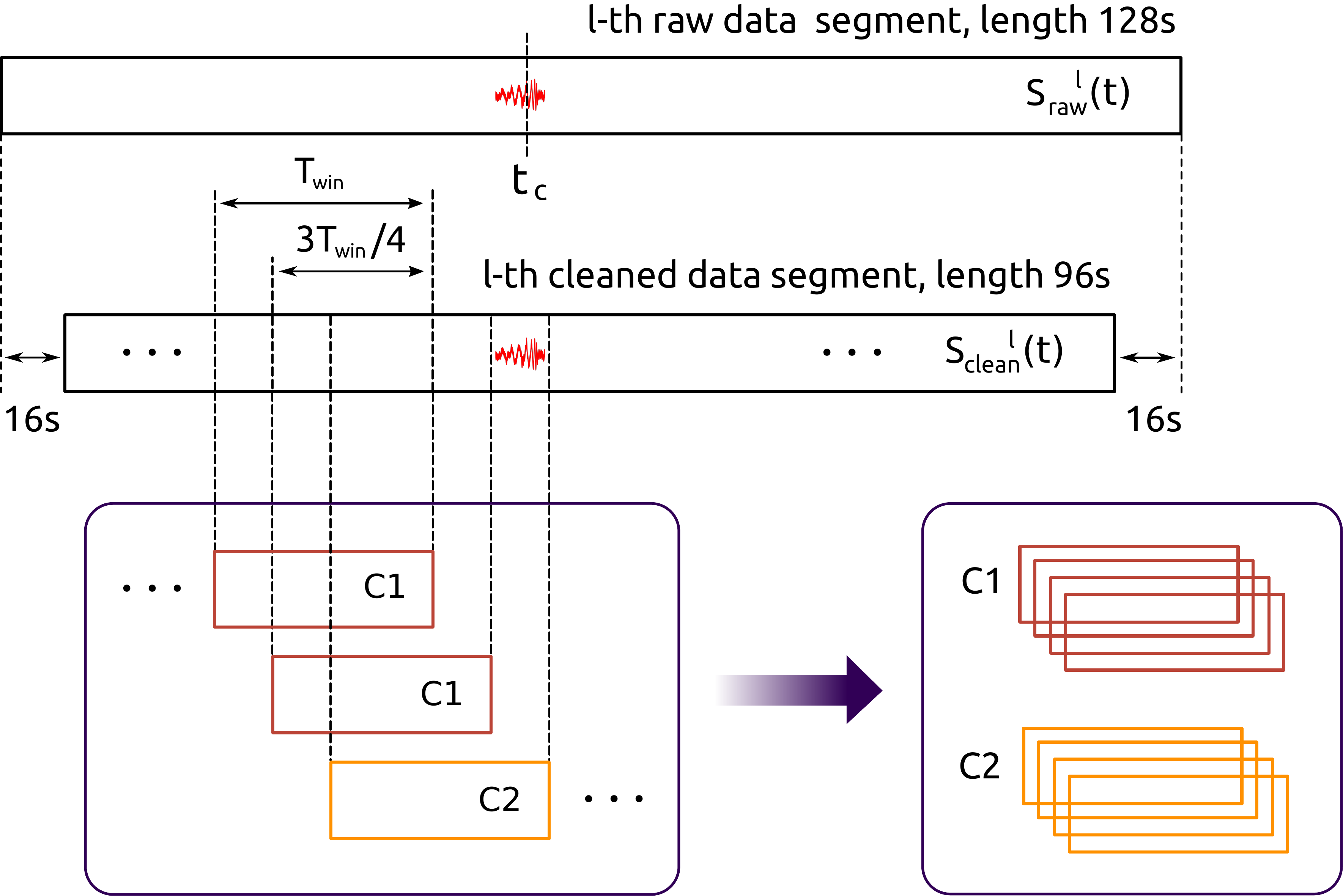}
  \caption{\label{fig:strain_samples} Strain samples generation by slidding windows. A CBC GW injection is located at coalescence time $t_c$ in $l \mhyphen th$ raw strain data segment of $128$s. After a cleaning, the segment is splitted in overlapped windows, individually identifying if a window has or does not have the GW injection. Finally, among all these windows, a set of eight strain samples is selected, four with noise alone (C1) and four with noise plus the mentioned CBC GW injection (C2). This set of eight samples will be part of the input dataset of our CNNs. This procedure is applied to all raw strain data segments.}
\end{center}
\end{figure}

The next stage of the pre-processing is the building of strain samples that will be inputted by our CNNs. This is schematically depicted in Fig.~\ref{fig:strain_samples}. This procedure has two steps. First, a $l \mhyphen th$ cleaned strain data segment, denoted as $s_{\text{clean}}{}^l(t)$, is splited in overlapped windows of duration $T_{\text{win}}$, identifying if it has or does not have an injection --time length $T_{\text{win}}$ is of the order of the injected waveform duration, leastwise. Here classes are assigned: if a window of data contains an injection, then class $2$ (C2) is assigned; on the other hand, if that window does not contain an injection, class $1$ (C1) is assigned. This class assignment is applied to all windows of data. Next, from the set of all tagged windows we select four consecutive ones of C1, and four consecutive ones of C2. This procedure is applied to each segment of clean data $s_{\text{clean}}{}^l(t)$ and, for avoid confusion with notation, we depict a $k \mhyphen th$ windowed strain sample as:
\begin{eqnarray}
 s_{\text{win}}{}^k (t) &=& \left[s^k(t_0), s^k(t_1), ..., s^k(t_{N_{win}-1})\right]^T~, \label{eq:strain_sample}
\end{eqnarray}
where $N_{win}=f_s~T_{\text{win}}$ according to the time length of the samples, with $f_s$ the sampling frequency. In practice, as we initially have $501$ segments from H1 and $402$ segments from L1, then $501 \times 8 = 4,008$ and $402 \times 8 = 3,360$ strain samples, respectively, are generated. Consequently, index $k$ appearing in Eq.~\ref{eq:strain_sample} take values from $0$ to $4007$ for H1 detector data, and values from $0$ to $3359$ for L1 detector data. Besides, $T_{\text{win}}$ is a resolution measure, and as it will seen later, we run our code with several values of $T_{\text{win}}$ in order to identify which of them are optimal with respect to performances of CNNs.

\subsubsection{Wavelet transform}
\label{subsec:wavelet_transf}

Some works in GW data analysis have used raw strain time series directly as input to deep convolutional neural networks (e.g.~\cite{dGeH18a} and~\cite{tGnKiHbS19}), however we will not follow this approach. We rather decided to apply CNNs for what was designed in its origins~\cite{yL89}, and for what have dramatically improved last decades~\cite{yLyBgH15}, namely image recognition. Then, we need a method to transform our strain vectors to image matrices, i.e. grid of pixels. For this research, we decided to use the wavelet transform (WT), which in signal processing is a known approach for working in the time-frequency representation~\cite{Mallat-Book}. One of the great advantages of the WT is that, by using a localized wavelet as kernel (also called ``mother wavelet''), it allows to visualize tiny changes of the frequency structure in time, and therefore, improve the search of GW candidates which arise as non-stationary short-duration transients in addition to the detector noise\footnote{WT suits better to our purposes than standard procedures as, for instance, Fourier transform. The latter draw on kernels describing simple harmonic oscillations, i.e. without temporal localization; making it the default option for searches of steady-state signals instead of transient signals.}.

In general, there are several wavelets that can be used as kernel. Here we decide to work with the complex Morlet wavelet~\cite{rKjMaG87}, that in its discrete version has the following form:
\begin{equation}
  \psi(t_n,f_j) = \frac{1}{\sqrt{\sigma^t{}_j \sqrt{\pi}}}
              \exp\left[\frac{-t_n{}^2}{2 \left(\sigma^t{}_j\right){}^2}\right]
              \exp\left(2 i \pi f_j t_n \right)~, \label{eq:morlet_wavelet}
\end{equation}
having a Gaussian form along time and along frequency, with standard deviations $\sigma^t$ and $\sigma^f$, respectively. Moreover, these standard deviations are not independent of each other, because they are related by $\sigma^t{}_j = 1 / \left(2 \pi \sigma^f{}_j \right)$ and $\sigma^f{}_j=f_j/\delta w$, where $\delta w$ is the width of the wavelet and $f_j$ its center in the frequency domain.

It is important to clarify that Morlet wavelet has a Gaussian shape, then it does not have a compact support. For this reason, the mesh in which we defined Eq.~\ref{eq:morlet_wavelet} (i.e. the set of discrete values for time $t_n$ and frequency $f_j$) is infinite by definition. Further, resolutions $\Delta t = t_n - t_{n-1}$ and $\Delta f = f_j - f_{j-1}$ are solely constrained by our system resources and/or to what extent we want to economize these resources.

Then, to perform the WT of the strain sample $s_{\text{win}}{}^k(t)$ (defined by Eq.~\ref{eq:strain_sample}) with respect the kernel wavelet $\psi(t_n,f_j)$ (defined by Eq.~\ref{eq:morlet_wavelet}), we just need to compute the following convolution operation:
\begin{eqnarray}
  Ws^k\left[t_n,f_j\right] &=& \sum_{m=0}^{N_{win}-1} s^k(t_m) \psi^*(t_{m-n},f_j)~, \label{eq:wavelet_transform}
\end{eqnarray}
where $s^k(t_m)$ is the $m \mhyphen th$ element of the column vector $s_{\text{win}}{}^k(t)$. Besides, $n=0,1,...,N_{\text{time}}$ and $j=0,1,...,N_{\text{freq}}$, where $N_{\text{time}}$ and $N_{\text{freq}}$ define the size of each $k \mhyphen th$ image generated by the WT transform, being $Ws^k\left[t_n,f_j\right]$ just the $(n,j)$ element or pixel of the each generated image. In practice, we set our WT such that it outputs images with dimensions $N_{\text{time}}=4096$ and $N_{\text{freq}}=47$ pixels.

In general, the grid of pixels defined by all values $t_n$ and $f_j$ depend on the formulation of the problem. Here we chose frequencies varying from $f_0=40$Hz to $f_{N_{\text{freq}}}=500$Hz, with a resolution of $\Delta f=10$Hz, given that is consistent with the GW signals that we want to detect. In addition, as we apply the WT to each cleaned strain data sample, we have discrete time values varying from $t_0=0$s to $t_{N_{\text{time}}}=T_{\text{win}}-1/f_s$, with a resolution of $\Delta t=t_s=1/f_s$, where $f_s=4096$Hz is the sampling frequency of the initial segments of strain data.

Although the size of output images is not too large, we decided to apply a resizing to reduce them from $4096 \times 47$ pixels to $32 \times 16$ pixels. Keep in mind that as we need to analyse several thousand images, using their original size would be unnecessarily expensive for system resources. The resizing was performed by a bicubic interpolation, in which each pixel of a reduced image is a weighted average of pixels in the nearest 4-by-4 neighborhood.

\begin{figure*}[htp]
\begin{center}
  \includegraphics[width=8.6cm]{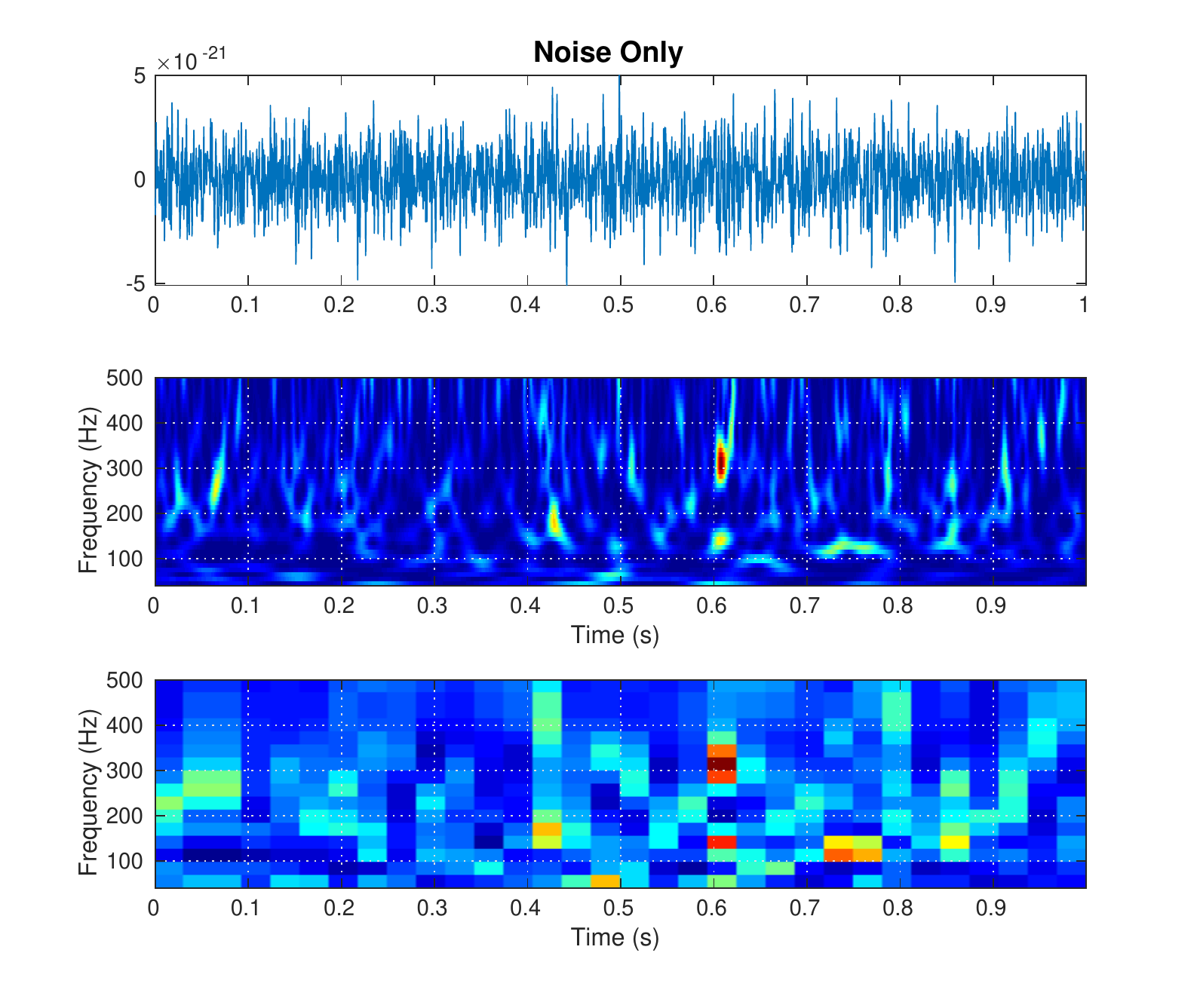}
  ~
  \includegraphics[width=8.6cm]{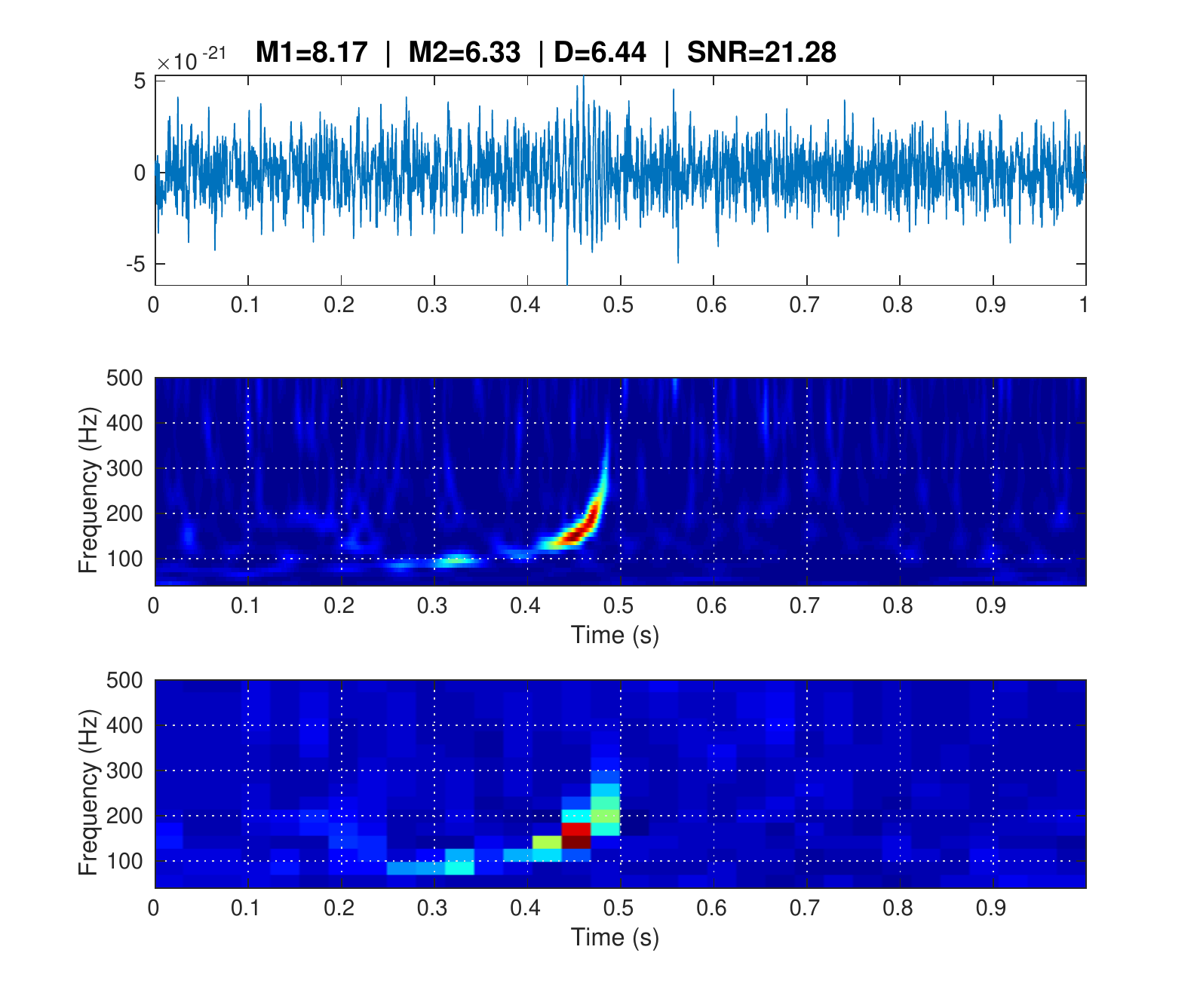}
   \caption{\label{fig:strainsamples} Visualization of two image samples of $T_{\text{win}}=1.0$s, generated from a strain data segment of length $4096$s and GPS initial time $932335616$, recorded at the L1 detector. In the left panel a sample of noise alone (class 1) is shown, and in the right panel a sample of noise plus GW (class 2). In the class 2 sample, masses $M1$ and $M2$ are in solar masses units, distance $D$ in Mpc, and SNR is the expected signal-to-noise ratio. Upper panel show strain data in the time domain, middle panels show time-frequency representations after apply a WT, and bottom panels show resized images by a bicubic interpolation.}
\end{center}
\end{figure*}

In summary, after applying the WT and the resizing to each strain data sample, we generated an image dataset $\left\{ {\bf X}_i, y_i \right\}_{i=1}^{N}$, where ${\bf X}_i \in \mathbb{R}^{N_{\text{time}}=32 \times N_{\text{freq}}=16}$ depicts each image as a matrix of pixels\footnote{In image processing, it is usual to specify the size of a image as $height \times width$. However, as we are defining our images by a standard $2D$ discrete plot ($x$ axis vs. $y$ axis), we prefered to invert the convention by putting sizes as $width \times height$.}, $y_i \in \left\{1,2\right\}$ the classes that we are working with, and $N_{\text{set}} \in \left\{4008,3360\right\}$ depending if we are using data from H1 or L1, respectively.

Fig.~\ref{fig:strainsamples} show two representative strain date samples (one belonging to C1 and other to C2) as a time domain signal, their time-frequency representation according to the WT with a Morlet wavelet as kernel, and its resized form. Both samples were generated from a strain data segment recorded by the L1 detector of $4094$s at GPS time $932335616$. Image sample at the right shows a GW transcient that has a variable frequency approximately between $100$Hz and $400$Hz. It is important to clarify that, before entering to CNNs, image samples are rearranged such that they the whole dataset has a size $N_{\text{set}} \times 32 \times 16 \times 1$, denoting $N_{\text{set}}$ images of size $32 \times 16$ pixels using $1$ channel for grayscale --sample input images shown in Fig.~\ref{fig:strainsamples} are in color just for illustrative purposes.

\subsection{CNN architectures}
\label{sec:CNN_architec}

Historically and before CNNs were developed, traditional feedforward Deep Neural Networks (DNNs, that are only ANNs with more than one hidden fully connected layer) were not able to produce good results as more data and/or layers were included~\cite{Goodfellow-Book}. For image recognition this was a big issue, and it was not overcomed until seminal works of Fukushima~\cite{kF80} and LeCun et al.~\cite{yLlByBpHF80}. Both works were bioinspired and developed new algorithms taking as reference experiments of Hubel and Wiesel~\cite{dhH59,dhHtnW59,dhHtnW68} about feature learning in visual cortex on cats and monkeys. Already mentioned work of LeCun et al. developed the first CNN architecture, namely the \texttt{LeNet-5} architecture.

What distingish CNNs from DNNs is that first ones use convolution operation in one or more of their layers, rather (or in adition to) fully connected layers~\cite{Goodfellow-Book}, leading to three advantages that cannot be found on second ones. Firstly, weight sharing that, unlike general matrix multiplication performed by fully connected layers, is a tremendous help for reducing system requeriments. Secondly, hierarchical learning that, by applying several convolutional kernels or filters, a CNN is able to extract features from the input image from low levels to higher levels. Finally, their significant improvements in performance with respect to classic ML algorithms provided that there are enough data. These advantages are just practical because, from a fundamental point of view, still there are not theories able to explain why CNNs have a small generalization error, or even, to establish analytical criteria to unequivocally choose a specific architecture for performing a particular task~\cite{cZ17}. Even so, there are several standards for CNNs design and certain mathematical guidelines that we can adopt to understand how a CNN learns from data inside the black box.

Our CNN algorithms consists in two main stages in terms of functionality, namely feature extraction and classification. First stage begins inputting images from a training dataset, and second stage ends ouputing predicted classes for each image sample --to be compare with actual classes. Obviously, classification is our ultimate goal, consisting in a perceptron stage plus an activation function as usual in ANNs. However, feature extraction is actually the core of CNNs, providing the ability of image recognition itself. Here there are three substages that are very common in many CNNs, including ours: convolution, which applies an affine transform; detection, for recognizing nonlinearities; and pooling, that performs a downsampling. These substages are implemented through layers that, as a whole, define a stack. Then, this stack can be connected to a second stack, and so on, until last stack is connected to the classifier. As it will be seen later, for our hyperparameter adjustments, we tested CNN architectures with $1$, $2$, and $3$ stacks.

\begin{figure*}[htp]
\begin{center}
  \includegraphics[width=16cm]{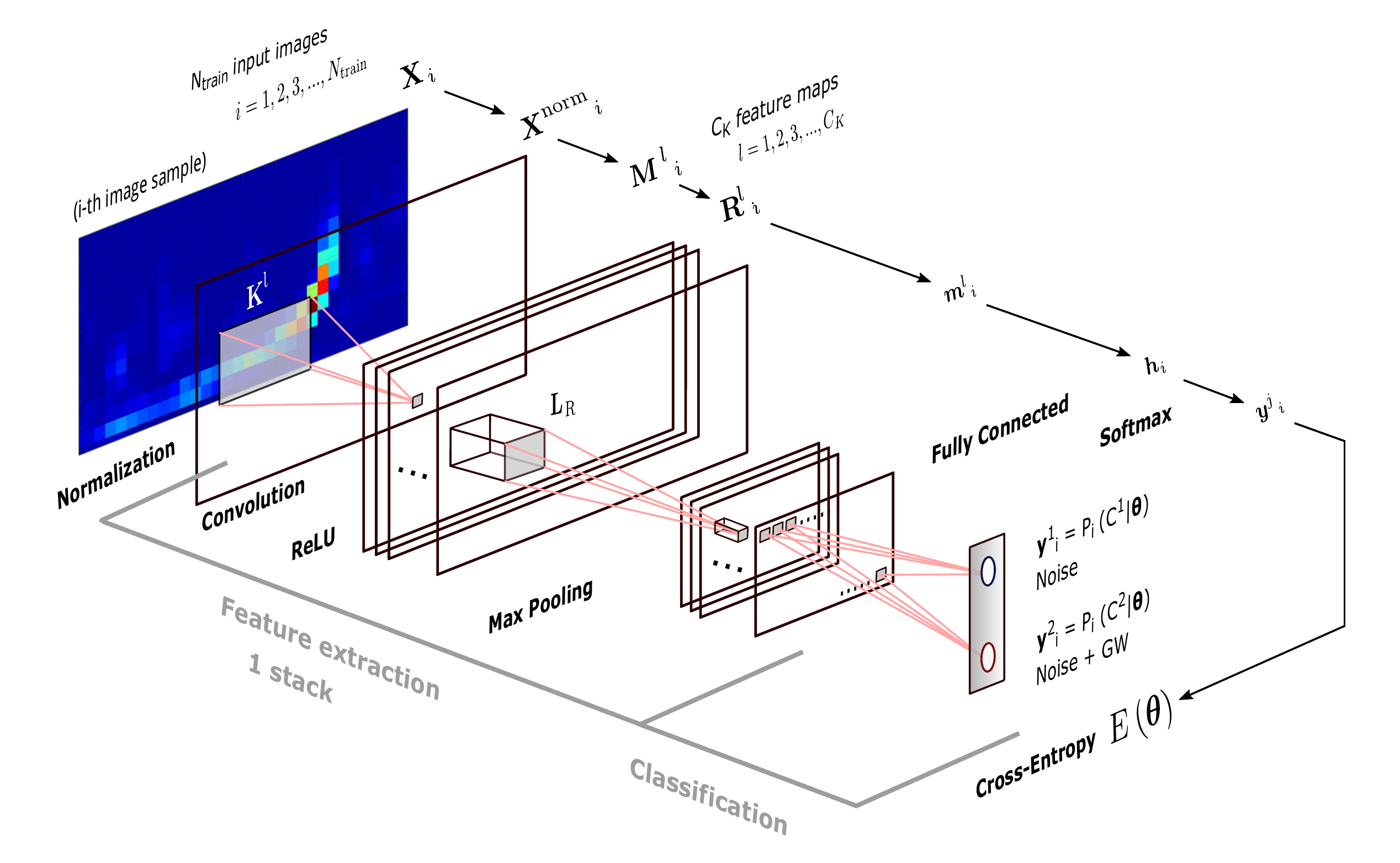}
  \caption{\label{fig:cnn_architecture} Single stack CNN architecture used as basis for this research. $N_{\text{train}}$ same size image samples are simultaneously inputted but, for simplicity, we detail the procedure for a single image --besides, although this image is shown colorized, training images dataset occupies just one channel. A $\boldsymbol{X}_i$ image feed the CNN, and a two-dimensional vector is outputted, giving us posterior probabilities of class 1 (noise alone) and class 2 (noise + GW), both conditioned by model parameters included in vector $\boldsymbol{\theta}$. After that, cross-entropy $E\left(\boldsymbol{\theta}\right)$ is computed. Notation for input(output) matrices and vectors is the same introduced in section~\ref{sec:CNN_architec} to mathematically describe the several kind of layers used in the CNNs.}
\end{center}
\end{figure*}

Knowing general functionality of the CNNs is a must but, for contributing with a white-box approach, we need to understand what each layer does. For this reason, we proceed to mathematically describe each kind of layer that were used --not only those involved in the already mentioned (sub)stages of the CNN, but also those that were required in our hands-on implementation with the \texttt{MATLAB Deep Learning Toolbox}~\cite{matlab-dl}. Take in mind that output of a layer is input of the next layer, as detailed in the Fig.~\ref{fig:cnn_architecture} for a single-stack CNN.\\

{\bf Image Input Layer}. Inputs images and applies a zero-center normalization. Denoting an $i \mhyphen th$ input sample as the matrix of pixels $\boldsymbol{X}_i \in \mathbb{R}^{N_{\text{time}} \times N_{\text{freq}}}$ belonging to a dataset of $N_{\text{train}}$ same size training images, this layer outputs the normalized image:
\begin{equation}
 \boldsymbol{X}^{\text{norm}}{}_i(j,k) = \boldsymbol{X}_i(j,k) - \frac{1}{N_{\text{train}}}\sum_{i=1}^{N_{\text{train}}}\boldsymbol{X}_i(j,k)~,
\end{equation}
where the second term is the average image of the whole dataset. Normalization, also known as histogram stretching, is useful for dimension scaling, making changes in each attribute, i.e. each pixel $(j,k)$ along all images, of a common scale. As normalization does not distort relative intensities too seriously and helps to enhance contrast of images, we can apply it to the entire training dataset, independently what class each image belong for.\\

{\bf Convolution Layer}. Convolves each image $\boldsymbol{X}^{\text{norm}}{}_i$ with $C_{K}$ sliding kernels of dimension $\mathcal{K}_{\text{time}} \times \mathcal{K}_{\text{freq}}$. This layer just apply an affine transform to input images. Denoting each $l \mhyphen th$ kernel by $\boldsymbol{K}^l$ with $l \in \left\{1,2,...,C_{K}\right\}$, this layer outputs $C_{K}$ feature maps, and each of them is an image composed by the elements or pixels:
\begin{widetext}
\begin{equation}
 \boldsymbol{M}^l{}_i (p,q) = \sum_{m} \sum_{n} \boldsymbol{X}^{\text{norm}}\left(m,n\right) \boldsymbol{K}^l\left(p-m+1,q-n+1\right) + b~, \label{eq:ConLayer}
\end{equation}
\end{widetext}
where $b$ is a bias term, and indices $p$ and $q$ run over all values that lead to legal subscripts of $\boldsymbol{X}^{\text{norm}}\left(m,n\right)$ and $\boldsymbol{K}^l\left(p-m+1,q-n+1\right)$. Depending on parametrization of subscripts $m$ and $n$, dimension of images $\boldsymbol{M}^l{}_i$ can vary. If we include width $M_{\text{time}}$ and height $M_{\text{freq}}$ of output maps (in pixels) in of a two-dimensional vector just for notation, these spatial sizes are computed by:
\begin{multline}
 ( M_{\text{time}}, M_{\text{freq}} ) \\
 = \frac{1}{\left(\text{str}_\text{time}, \text{str}_\text{freq}\right)}
   \left[( N_{\text{time}}, N_{\text{freq}}) - (\mathcal{K}_\text{time}, \mathcal{K}_{\text{freq}}) \right. \\
   + \left. 2(\text{pad}_\text{time}, \text{pad}_\text{freq}) \right]
      \left(1,1\right)~, \label{eq:size_output_conv}
\end{multline}
where str (i.e. stride) is the step size in pixels with which a kernel move above $\boldsymbol{X}^{\text{norm}}{}_i$, and padd (i.e. padding) denotes time rows and/or frequency columns of pixels added to $\boldsymbol{X}^{\text{norm}}{}_i$ for moving the kernel beyond the borders of the former. During the training, components of kernel and bias terms are iteratively learned from certain initial values; then, once the CNN has captured and fixed optimal values for these parameters, convolution is applied to all testing images.\\

{\bf ReLU Layer}. Applies the Rectified Linear Unit (ReLU) activation function to each neuron (pixel) of each feature map $\boldsymbol{M}^l{}_i$ obtained from the previous convolutional layer, outputting the following:
\begin{equation}
 \boldsymbol{R}^l{}_i(p,q) = \max \left\{ 0 , \boldsymbol{M}^l{}_i (p,q) \right\}~. \label{eq:ReLUlayer}
\end{equation}
In practice, this layer works as a detector of nonlinearities in input sample images; and its neurons can output true zero values, generating sparse interactions and reducing system requeriments. Besides, this layer does not lead to saturation in hidden units during the learning, because its form, given by Eq.~\ref{eq:ReLUlayer}, does not converge to finite asymptotic values\footnote{Saturation is an effect when an activation function locate in a hidden layer of a CNN converge rapidly to its finite extreme values, becoming the CNN insensitive to small variations of input data in most of its domain. In feedforward networks, activation functions as $\text{sigmoid}$ or $\text{tanh}$ are prone to saturation, hence they use are discouraged except when output layer of the CNN has a cost function able to compensate their saturation~\cite{Goodfellow-Book}, as for example the cross-entropy function.}.\\

{\bf Max Pooling Layer}. Downsamples each feature map $\boldsymbol{R}^l{}_i$ with the maximum on local sliding regions $\boldsymbol{L}_R$ of dimension $\mathcal{P}_{\text{time}} \times \mathcal{P}_{\text{freq}}$. Each pixel of a resulting reduced featured map $\boldsymbol{m}^{\ell}{}_i$ is given by the following:
\begin{equation}
 \boldsymbol{m}^l{}_i(r,s) = \max_{r,s} \left\{\boldsymbol{R}^l{}_i(p,q)\right\} ~,~~ \forall (p,q) \in \boldsymbol{L}_R ~,
\end{equation}
where ranges for indices $r$ and $s$ depend on spatial sizes of outputs maps; and these sizes, i.e. width $m_{\text{time}}$ and height $m_{\text{freq}}$, being included in a two-dimensional vector just for notation, are computed by:
\begin{multline}
 (m_{\text{time}},m_{\text{freq}}) \\
  = \frac{1}{{(\text{str}_\text{time},\text{str}_\text{freq})}}
    \left[(M_{\text{time}},M_{\text{freq}}) - (\mathcal{P}_{\text{time}},\mathcal{P}_{\text{freq}}) \right.\\
    \left. + 2 (\text{pad}_\text{time},\text{pad}_\text{freq}) \right] + (1,1) ~,\label{eq:size_output_maxpool}
\end{multline}
where padding and stride values have same meanings as in the convolutional layer. Interestly, max pooling layer leaves invariant output values under small translations in the input images, which could be useful for working with a network of detectors --case in which a GW signal from the same source has several versions depending on the number of detectors that are being considered, with its respective time delays. And more generally, this layer contribute to reduce overfiting, to reduce the number of neurons for each map, and hence, to reduce system requeriments.\\

{\bf Fully Connected Layer}. This is the classic perceptron layer used in classical ANNs and performs the binary classification itself. It maps all images $\boldsymbol{m}^{\ell}{}_i$ to the two-dimensional vector $\boldsymbol{h}_i$ by the affine transformation:
\begin{equation}
 \boldsymbol{h}_i = \boldsymbol{W}^{T} \boldsymbol{m}^{\text{flat}}{}_i + \boldsymbol{b} ~,
\end{equation}
where $\boldsymbol{m}^{\text{flat}}{}_i$ is a vector of $N_{\text{fc}}$ dimensions, with $N_{\text{fc}}$ the total number of neurons considering all input feature maps, $\boldsymbol{b}$ a two-dimensional bias vector, and $\boldsymbol{W}$ a weight matrix of dimension $2 \times N_{\text{fc}}$. Similarly to convolutional layer, elements of $\boldsymbol{W}$ and $\boldsymbol{b}$ are the model parameters to be learn in the training, beginning from certain initial values appropriately chosen. Notice that matrix $\boldsymbol{m}^{\text{flat}}{}_i$ become flatter, in a single column vector, all feature maps $\boldsymbol{m}^l{}_i$ (with $l=1,2,...,C_{K}$); therefore information about topology of sample images (i.e. their edges) is lost.\\

{\bf Softmax Layer}. Applies the softmax activation function to each component $j$ of vector $\boldsymbol{h}_i$
\begin{equation}
 \boldsymbol{y}^j{}_i = \text{softmax}(\boldsymbol{h}^j{}_i)
                      = e^{\boldsymbol{h}^j{}_i} / \sum_j e^{\boldsymbol{h}^j{}_i} ~, \label{eq:softmax_layer}
\end{equation}
where $j=1,2$ depending on the class. Softmax layer is the multiclass generalization of sigmoid function, and we include it in the CNN because by definition, transform real output values of fully connected layer in probabilities. In fact, according to~\cite{jB89}, output vales $\boldsymbol{y}^j{}_i$ can be interpreted as posterior distributions of class $c^j$ conditioned by model parameters. That is to say $y^j{}_i(\boldsymbol{\theta}) = P_i(c^j|\boldsymbol{\theta})$, where $\boldsymbol{\theta}$ is a multidimensional vector containing all model parameters. This is the reason why it is common to refer to the quantity $y^j{}_i(\boldsymbol{\theta}) \in \left[0,1\right]$ as the output \textit{score} of the CNN.\\

{\bf Classification Layer}. Stochastically takes $\tilde{N}<N_{\text{train}}$ samples and computes the cross-entropy function:
\begin{eqnarray}
 E\left(\boldsymbol{\theta}\right)
 &=& -\ln \mathcal{L} \left(\boldsymbol{\theta} | y^1{}_i, y^2{}_i\right)
 = -\ln \prod_{i=1}^{\tilde{N}} \underbrace{P_i\left(c^1|\boldsymbol{\theta}\right)}_{y^1{}_i}
                            \underbrace{P_i\left(c^2|\boldsymbol{\theta}\right)}_{y^2{}_i} \nonumber\\
 &=& -\sum_{i=1}^{\tilde{N}} \left(\ln y^1{}_i + \ln y^2{}_i\right) ~, \label{eq:cross-entropy}
\end{eqnarray}
where $y^1{}_i$ and $y^2{}_i$ are the two posterior probabilites outputted by softmax layer and $\mathcal{L}$ a likelihood function. In general, cross-entropy is a measure of the \text{risk} of our classifier and, following a discriminative approach~\cite{Bishop-Book} to estimate model parameters included in $\boldsymbol{\theta}$, Eq.~\ref{eq:cross-entropy} actually defines the maximum likelihood estimation for these parameters. Then, at this point what we need is a learning algorithm for maximizing the likelihood function $\mathcal{L} \left(\boldsymbol{\theta} | y^1{}_i, y^2{}_i\right)$ or, equivalently, minimizing $E\left(\boldsymbol{\theta}\right)$, with respect to model parameters\footnote{This approach estimates model parameters through a feedforward learning algorithm from classification layer to previous layers of the CNN. Alternatively, considering that posterior probability outputted by softmax layer is $P_i(c^j|\boldsymbol{\theta}) = P_i(\boldsymbol{\theta}|c^j) P(c^j) / P(\boldsymbol{\theta})$ because of the Bayes theorem, other approach is to maximize likelihood funcion $P_i(\boldsymbol{\theta}|c^j)$ with respect to model parameters. This approach is called generative and will be not considered in our CNN model. Anyway, in subsection~\ref{subsec:ROC_analyses} we will present the simplest examples of generative models that we considered for purposes of comparison with our CNN algorithms, namely Naive Bayes classifiers.}. This algorithm will be introduced in subsection~\ref{subsec:training}.

Above mathematical descriptions layers is by no means exhaustive regarding to the kind of layers that can be included in a CNN. One could add a dropout layer for regularization purposes, and/or a dilated convolutional layer for capture more contextual information in which fine details of images lie, among others; but in our case, as a first approach, it was not necessary. Finally, details about the three architectures that we used in our experiments are shown in Table~\ref{tab:particular_CNNs}. These architectures differ in the mount of stacks, number kernels $C_{K}$, size of kernels, size of max pooling regions, number of activations, and number of weights and biases to be learn for a given initial values detailed in in subsection~\ref{subsec:training}.

\begin{table*}
    \centering
    	\begin{tabular}{lll}
     	\toprule
		{\bf \textit{Layer}}			& {\bf \textit{Activations}}	~~~			& {\bf \textit{Learnables}} \\
									& {\bf \textit{per image sample}} ~~~ 	& {\bf \textit{per image sample}} \\
		\midrule

		{\bf Image Input}			& $32 \times 16 \times 1$	& --	    \\
		\midrule

        \begin{tabular}{@{}l} {\bf Convolution} of size $5 \times 4$, $C_{K}$ kernels \vspace{.05cm} \\
		                      \small Strides: $1$, Paddings: $0$ \end{tabular}
							& $28 \times 13 \times C_{K}$ &
		\begin{tabular}{@{}l} Weights: $5 \times 4 \times 1 \times C_{K}$ \vspace{.05cm} \\
		                      Biases: $1 \times 1 \times C_{K}$ \end{tabular}
		\vspace{.1cm} \\
		{\bf ReLU}					& $28 \times 13 \times C_{K}$	& --		\vspace{.1cm} \\
		\begin{tabular}{@{}l} {\bf Max Pooling} of size $2 \times 2$ \vspace{.05cm} \\
							 \small Strides: $2$, Paddings: $0$ \end{tabular}
							& $14 \times 6 \times C_{K}$	& --		\vspace{.1cm} \\
		\midrule
		
		\begin{tabular}{@{}l} {\bf Convolution} of size $5 \times 4$, $C_{K}$ kernels \vspace{.05cm} \\
		                      \small Strides: $1$, Paddings: $0$ \end{tabular}
							& $10 \times 3  \times C_{K}$	&
		\begin{tabular}{@{}l} Weights: $5 \times 4 \times C_{K} \times C_{K}$ \vspace{.05cm} \\
		                      Biases: $1 \times 1 \times C_{K}$ \end{tabular}
		\vspace{.1cm} \\
		{\bf ReLU}					& $10 \times 2  \times C_{K}$	& --		\vspace{.1cm} \\
		\begin{tabular}{@{}l} {\bf Max Pooling} of size $2 \times 2$ \vspace{.05cm} \\
							 \small Strides: $2$, Paddings: $0$ \end{tabular}
							& $5 \times 1 \times C_{K}$	& --		\vspace{.1cm} \\
		\midrule

		\begin{tabular}{@{}l} {\bf Convolution} of size $4 \times 1 \times$, $C_{K}$ kernels \vspace{.05cm} \\
		                      \small Strides: $1$, Paddings: $0$ \end{tabular}
							& $2 \times 1 \times C_{K}$	&
		\begin{tabular}{@{}l} Weights: $4 \times 1 \times C_{K} \times C_{K}$ \vspace{.05cm} \\
		                      Biases: $1 \times 1 \times C_{K}$ \end{tabular}
		\vspace{.1cm} \\
		{\bf ReLU}					& $2 \times 1 \times C_{K}$	& --		\vspace{.1cm} \\
		\begin{tabular}{@{}l} {\bf Max Pooling} of size $2 \times 1$ \vspace{.05cm} \\
							 \small Strides: $1$, Paddings: $0$ \end{tabular}
							& $1 \times 1 \times C_{K}$	& --		\vspace{.2cm} \\

		\midrule

		{\bf Fully Connected}       & $1 \times 1 \times 2$		&
		\begin{tabular}{@{}l} Weights: $2 \times C_{K}$ \vspace{.05cm} \\
		                      Biases: $2 \times 1$ \end{tabular}
		\vspace{.1cm} \\
		{\bf Softmax}               & $1 \times 1 \times 2$		& --		\\
		\midrule

        {\bf Ouput Cross-Entropy}   & --							& --		\\
        \bottomrule		
	\end{tabular}
    \caption{\label{tab:particular_CNNs}CNN architecture of $3$ stacks used for this research. The number of kernels in convolutional layers is variable, and it took values $C_{K} \in \left\{8,12,16,20,24,28,32\right\}$. Part of this illustration is also valid for CNN architectures of $1$ or $2$ stacks (also implemented in this research), in which image input layer is followed by the next three or six layers as feature extractor, respectively, and then by the fully connected layer until the output cros-entropy layer in the classification stage. All these CNNs were implemented with the \texttt{MATLAB Deep Learning Toolbox}.}
\end{table*}

\subsection{Model training}
\label{subsec:training}
As detailed, model parameters are included in vector $\boldsymbol{\theta}$, that in turn appears in the cross-entropy function (\ref{eq:cross-entropy}). Starting from given initial values, these parameters have to be learned by an minimization of the cross-entropy, taking $\tilde{N} < N_{\text{train}}$ random image samples, i.e. a mini-batch. For model parameters updating, we drawn on the known gradient descent algorithm, including a momentum term to boost iterations. Denoting model parameters at the $r \mhyphen th$ iteration or epoch as $\boldsymbol{\theta}^{(r)}$, then its updating at the $(r+1) \mhyphen th$ iteration is given by the following optimization rule:
\begin{equation}
 \boldsymbol{\theta}^{(r+1)} = \boldsymbol{\theta}^{(r)} - \alpha \boldsymbol{\nabla}_{\boldsymbol{\theta}^{(r)}}
             E \left(\boldsymbol{\theta}^{(r)}\right)
             + \gamma \left(\boldsymbol{\theta}^{(r)} - \boldsymbol{\theta}^{(r-1)} \right) ~, \label{eq:sgd_mom}
\end{equation}
where we have that $\boldsymbol{\nabla}_{\boldsymbol{\theta}^{(r)}}$ is the gradient with respect to model parameters and $E$ the cross-entropy, in addition to two empirical quantities to be set by hand, namely the learning rate $\alpha$ and the momentum $\gamma$. Given that we are computing the cross-entropy with $\tilde{N}<N_{\text{train}}$ random samples, the above rule is called mini-batch stochastic gradient descent (SGD) algorithm. Besides, for all our experiments we set a learning rate $\alpha=1.00$, a momentum $\gamma=0.9$, and a mini-batch size of $\tilde{N}=128$ image samples.

With regard to initialization of weights, we draw on Glorot initializer~\cite{xGyB10}. This scheme independently samples values from a uniform distribution with a mean equal to zero and a variance given by $2/\left(n_{\text{in}}+n_{\text{out}}\right)$, where $n_{\text{in}}=\mathcal{K}_{\text{time}}\mathcal{K}_{\text{freq}}$ and $n_{\text{out}}=\mathcal{K}_{\text{time}}\mathcal{K}_{\text{freq}}C_{K}$ for convolutional layers, and $n_{\text{in}}=C_{K}~\text{size}(\boldsymbol{m}^{\text{flat}}_i)$ and $n_{\text{out}}=2C_{K}$ for the fully connected layer --remember that $C_{K}$ is the number of kernels of dimension $\mathcal{K}_{\text{time}} \times \mathcal{K}_{\text{freq}}$ and $\boldsymbol{m}^{\text{flat}}_i$ the vector inputted by the fully connected layer. All biases, on the other hand, are initialized with zeros.

Finally, whether we work with data from H1 or L1 detector, $N_{\text{train}}$ will be quite greater than $\tilde{N}=128$; nevertheless, its specific value depends on our global validation technique, to be explained in the subsection~\ref{subsec:validation}.

\subsection{Global and local validation}
\label{subsec:validation}

For this research, we used only real LIGO strain data with a given and limited number of CBC hardware injections. In practice, this approach reduces the instrumental ability to arbitrarily generate big datasets, unlike the classical approach in which, one can perform an unlimited number of software injections with numerical and/or analytical templates. Even, when using syntetic data and because of system resources limitations, the order of magnitude of generated templates is quite small compared to the \textit{big data} regime of petabytes and beyond. Therefore, a global validation technique to reach good statistical confidence and to perform a fair model evaluation, actually is required. For these reasons, we implemented the $k$-fold CV technique~\cite{fMjT68}, that consists in the following recipe: i) split the original dataset into $k$ nonoverlapping subsets to perfom $k$ trials, then ii) for each $i \mhyphen th$ trial use the $i \mhyphen th$ subset for testing leaving rest of data for training, and finally iii) compute the average of each performance metric across all trials.

It is known that the value of $k$ in $k$-fold CV defines a trade-off between bias and variance. When $k=N$ (i.e. leave-one-out cross-validation), the estimation is unbiased but variance can be high and, when $k$ is small, the estimation will have lower variance but bias could be a problem~\cite{Hastie-Book}. For several classic ML algorithms, previous works have suggested that $k=10$ represent the best trade-off option (\cite{Breiman-Book,sWnI94,rK95}), then we decided to take this value as a first approach. But even more, following works~\cite{aMrSrP05} and~\cite{jK09}, we decided to perform $10$ repetitions of the $10$-fold cross-validation process~\cite{Japkowicz-Book}, in order to reach a good stability of our estimates, to present fair values of the cross-entropy function given the stochastic nature of our resampling approach, and more important, obtain information about the distribution of accuracy (and other standard metrics) in which there is involved uncertainty.

Moreover, $k$-fold CV helps to aliviate the artificiality introduced by balanced dataset, because the initial splitting into k folds is totally random. In research~\cite{tGnKiHbS19} authors warns about the fact that CNNs not only capture GW templates alone, but also transfers to the test stage the exact same probability distribution given in the training set. This claim is true but, it is important to take in mind that working with balanced datasets, as a first approach, is simply motivated by the fact that many of standard performance metrics gives excessive optimistic results on classes of higher frequency in imbalanced dataset, and dealing with arbitrarily unbalanced datasets is not a trivial task. Anyway, including $k$-fold CV as a random resampling, starting from a balanced dataset, is desirable for statistical purposes.

With sizes of our datasets detailed at the end of subsection~\ref{subsec:wavelet_transf} and the $10$-fold CV, we have a training set of $N_{\text{test}}= \text{floor}[4008/10] = 400$ samples obtained from H1 detector, and of $N_{\text{test}}=\text{floor}[3360/10]=336$ samples from L1 detector. Consequently, $N_{\text{train}}=3607$ for H1 data, and $N_{\text{train}}=3607$ for L1 data.

Local validation, that is to say validation performed within a learning epoch, is also an crucial ingredient of our methodology. In particular, our algorithm splits the training set of $N_{\text{train}}$ data into two subsets, one for the training itself ($0.9N_{\text{train}})$ and other for validation ($0.1N_{\text{valid}}$). Validation works as a preparatory mini test which is useful for monitoring learning and generalization during the training process.

With regard to regularization techniques, local validation was performed once per $ \text{floor}[N_{\text{train}} / \tilde{N}]$ epochs and cross-entropy was monitored with a validation patience $p=5$ (value given by hand) which simply means that if $E^{(r+1)} \geq E^{(r)}$ occurs $p$ times during the validation, then training process is automatically stopped. Besides, to avoid overfitting, training samples were randomized before training and before validation, solely in the first learning epoch. This randomization is performed such that link between each training image $\boldsymbol{X}_i$ and its respective class $y_i$ is left intact, because we do not want to ``mislead'' our CNN when it learns from known data.

\subsection{Performance metrics}
\label{subsec:performance_metrics}

Once our CNN is trained, the goal is predicting classes of unseen data, i.e. data on which the model was not trained, and achieves a good performance. Hence performance metrics in the test are really crucial. Considering last layer of our CNNs, a metric that is natural to monitor during the training and validation process is the cross-entropy. Other metrics that we use come from counting predictions. As our task has to do with a binary classification, these metrics can be computed from the elements of a $2 \times 2$ confusion matrix, namely true positives (TP), false positives (FP) or type I errors, true negatives (TN), and false negatives (FN) or type II errors, as it is detailed in Table~\ref{tab:table_metrics}.

\begin{table*}
  \noindent
  \renewcommand\arraystretch{1.5}
  \setlength\tabcolsep{0pt}
  \begin{tabular}{c >{\bfseries}r @{\hspace{0.7em}}c @{\hspace{0.4em}}c @{\hspace{0.7em}}l}
    \multirow{10}{*}{\rotatebox{90}{\parbox{3.1cm}{\bfseries\centering\small Predicted values}}} & 
      & \multicolumn{2}{c}{\bfseries\centering\small Actual values} & \\
    & & ~ & ~ & \bfseries \small total \\
    & ~ & \MyBox{True}{Positives (TP)} & \MyBox{False}{Positives (FP)} & P$'$ \\[2.4em]
    & ~ & \MyBox{False}{Negative (FN)} & \MyBox{True}{Negatives (TN)} & N$'$ \\
    & \small total & P & N &
  \end{tabular}  
  ~~~~~~
  \centering
  \begin{tabular}{lll}
    \toprule
    Metric   & ~~~ Definition & ~~~~ What does it measure? \\
    \midrule
    Accuracy   &  ~~~ $\frac{TP+TN}{P+N}$ & ~~~~ How often a correct\vspace{-.1cm} \\
               &                          & ~~~~ classification is made\\
    \vspace{.2cm}    
    Precision  &  ~~~ $TP/P'$      & ~~~~ How many examples retrieved\vspace{-.3cm} \\
               &                  & ~~~~ as relevant are truly relevant\\
    \vspace{.2cm}
    Recall     &  ~~ $TP/P$      & ~~~~ How many truly relevant \vspace{-.3cm} \\
               &                  & ~~~~ examples are retrieved \\
    \vspace{.2cm}
    Fall-out   &  ~~ $FP/N$      & ~~~~ How many examples retrieved \vspace{-.3cm} \\
               &                  & ~~~~ as relevant are non-relevant \\
    \vspace{.2cm}
    F1 score   &  ~~ $2 \frac{\text{Precision} \times \text{Recall}}{\text{Precision} + \text{Recall}}$
               & ~~~~ Harmonic mean of precision\vspace{-.3cm} \\
               &                  & ~~~~ and recall.\\
    \vspace{.2cm}
    G mean1   &  ~~ $\sqrt{\text{Recall} \times \text{Fall-out}}$
               & ~~~~ Geometric mean of recall\vspace{-.3cm} \\
               &                  & ~~~~ and fall-out.\\
    \bottomrule
  \end{tabular}
  \caption{Confusion matrix for a binary classifier and its consequent standard performance metrics. In general, to have a complete view of the classifier, it is suitable to draw on, at least, accuracy, precision, recall, and fall-out. F1 score and G mean1 are useful metrics for imbalanced multilabel classifications, and they also give important moderation features which help in model evaluation. Each metric has its probabilistic interpretation.}
  \label{tab:table_metrics}
\end{table*}

Accuracy is the most used standard metric in binary classifications. Besides, all metrics shown in the left panel of Table~\ref{tab:table_metrics} depend on a choosen threshold as crucial part our decision criterion. Depending of this threshold, and the output score for an input image sample, our CNN algorithms assign class $1$ or class $2$. Although threshold is fixed for all these metrics, one can also vary it for generating the well known Receiving Operating Characteristic (ROC)~\cite{tF06} and Precision-Recall~\cite{jDmG06} curves, among others. The first curve describes the performance of the CNN in the fall-out (or false positive rate) vs. recall (or true positive rate) space, and the second one in the recall vs. precision space. Given that the whole curves are generated by varying the threshold, each point of those curves represents the performance of the CNN given a specific threshold. As we worked with balanced datasets, plotting ROC curves will be enough. Finally, F1 score and G mean1 are metrics that summarize in a single metric pairs of other metrics and being, in general, useful for imbalanced multiclass classifications. Anyway, we decided to compute these two last metrics, because they give useful moderation features for performance evaluation --more details in subsection~\ref{subsec:confu_matrices}.

In order to ensure that our results are statistically significant, we also want to perform a shuffling. As it was mentioned in subsection \ref{subsec:validation}, our algorithm already performs a randomization over the training set, before the training and before the validation in order to prevent overfitting. However, shuffling applied here is more radical because it broke the link between training images and their respective classes, and it is made by random permutation over indices $i$ solely for the matrices $\boldsymbol{X}_i$, belonging to training set $\left\{\boldsymbol{X}_i,y_i\right\}_{i=1}^{N_{\text{train}}}$, before each the training. Then, if this shuffling is present and our results are truly significant, we expect that accuracy in testing be lower than that computed when shuffling no is present, reaching values around $0.5$ --as this is the chance level for a binary classifier. This will be visually explored, looking at dispersion of mean accuracies in boxplots and, more formally, confirmed by a paired-sample t-test of statistical inference. For each distribution of mean accuracies, with and without shuffling, we can define sets $\mathcal{D}$ and $\mathcal{D}_{\text{shuff}}$, respectively. Then, with means of each of these sets, namely $\mu$ and $\mu_{\text{shuff}}$, the goal is testing the null hypothesis $H_0 : \mu - \mu_{\text{shuff}} = 0$, and then, we just would need to compute the p-value, i.e. the probability of resulting accuracies be possible assuming that the mentioned null hypothesis is correct, given a level of significance --more details about shuffling and consequent statistical tests are presented in subsection~\ref{subsec:shuffling}.

\section{Results and discussion}
\label{sec:results}

\subsection{Learning monitoring per fold}
\label{subsec:learning_monitor}

While the mini-batch SGD was running, we monitored the cross-entropy and accuracy evolution along epochs. This is the very first check to ensure our CNN were properly learning from data and our local validation criteria stopped the learning algorithm in the right moment. If our CNNs were correctly implemented, we expected that cross-entropy be minimized to reach values as close as possible to $0$, and accuracy reach values as close as possible to $1$. If this check gave wrong results, then there would be no point in computing subsequent metrics.

\begin{figure*}[ht]
\begin{center}
  \includegraphics[width=8.2cm]{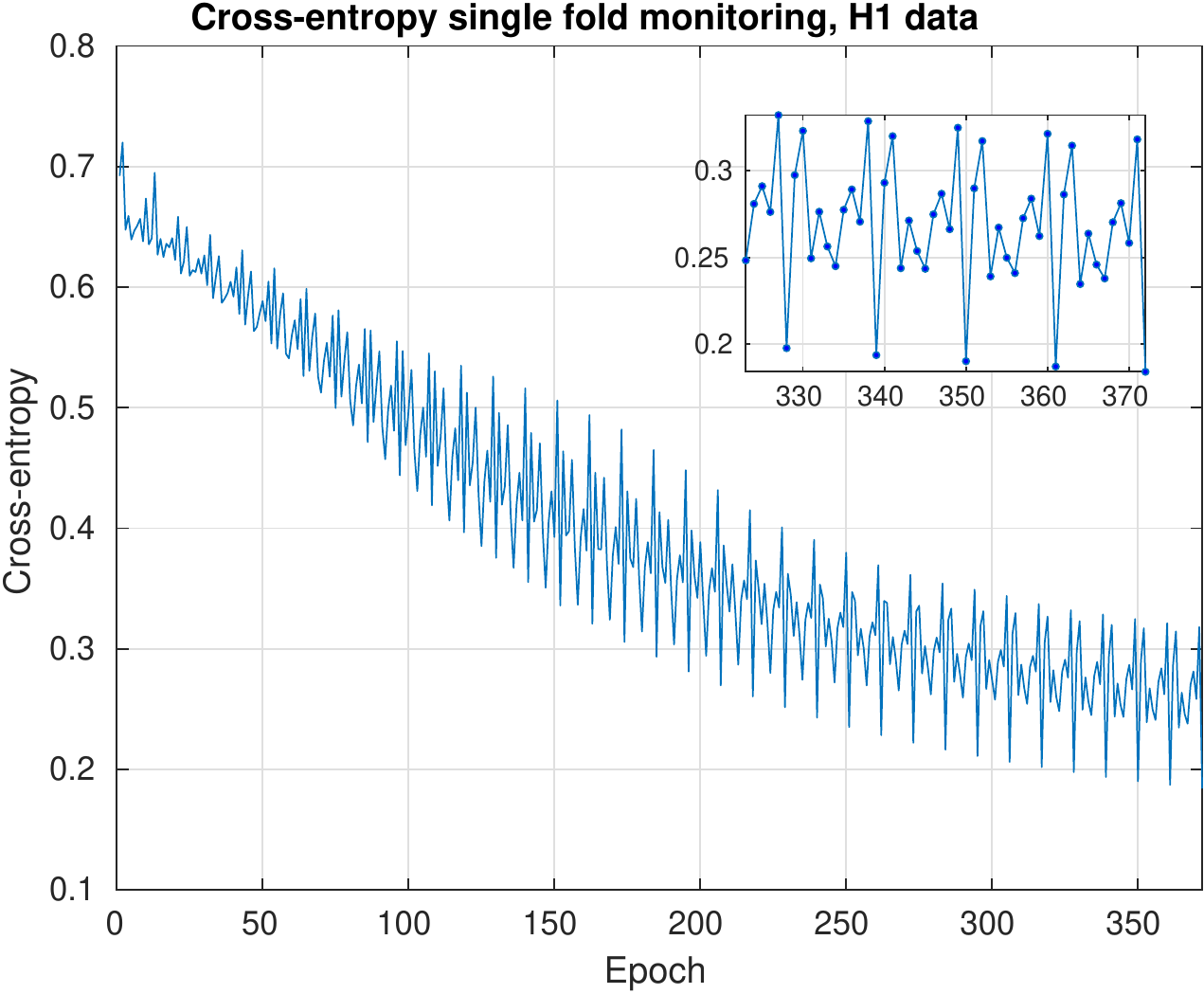} ~~~~
  \includegraphics[width=8.2cm]{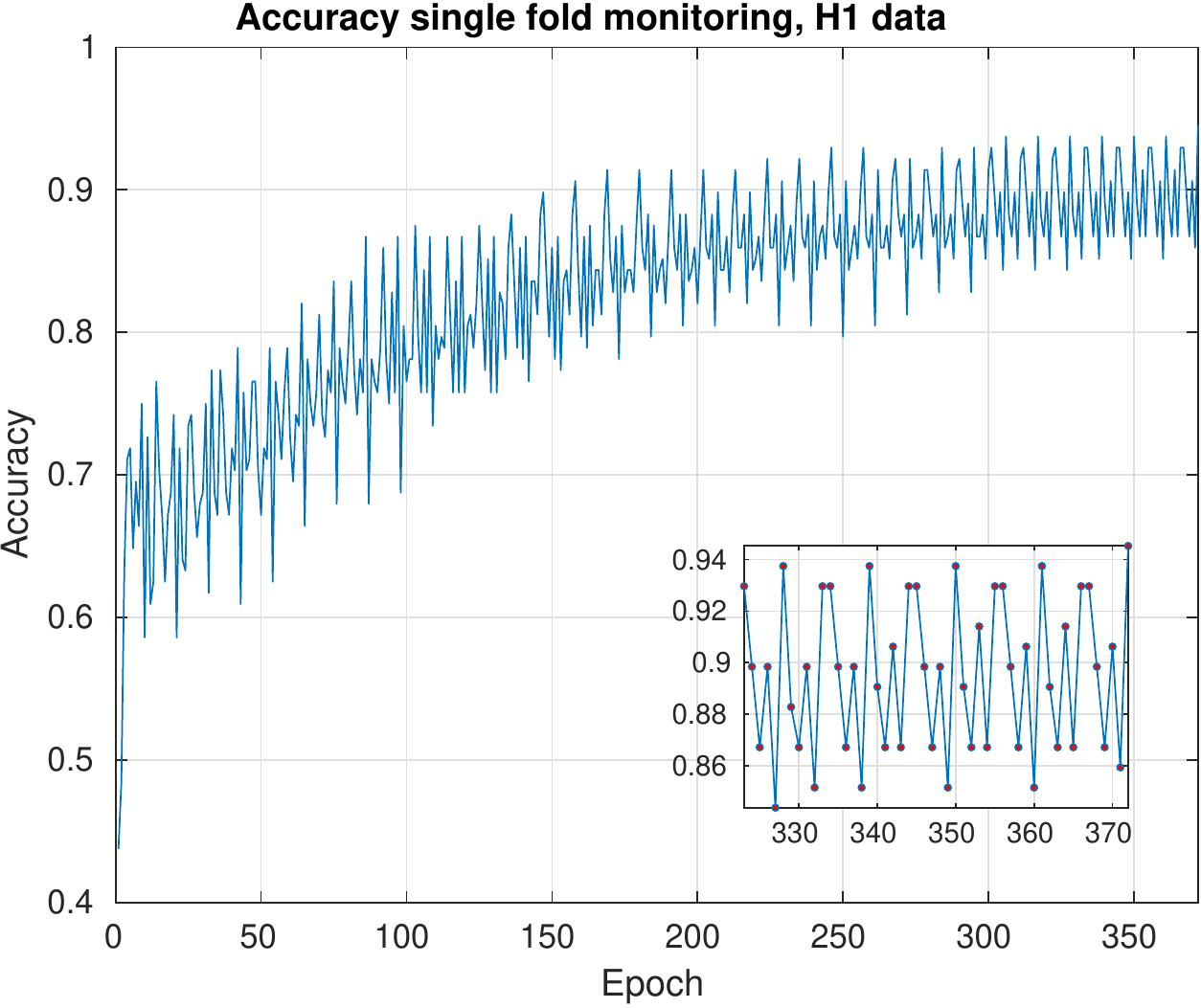} \\
  \vspace{.5cm}
  \includegraphics[width=8.2cm]{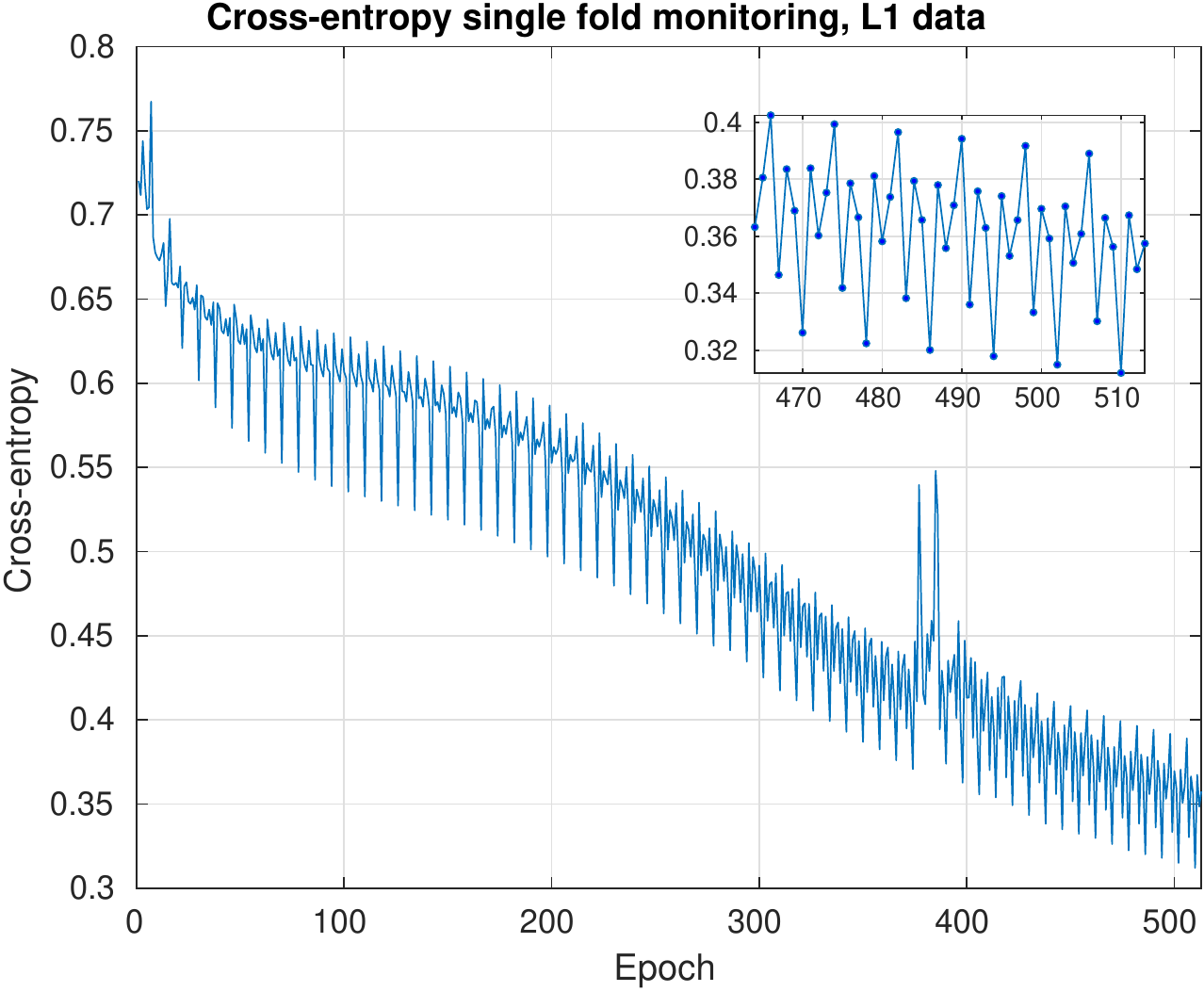} ~~~~
  \includegraphics[width=8.2cm]{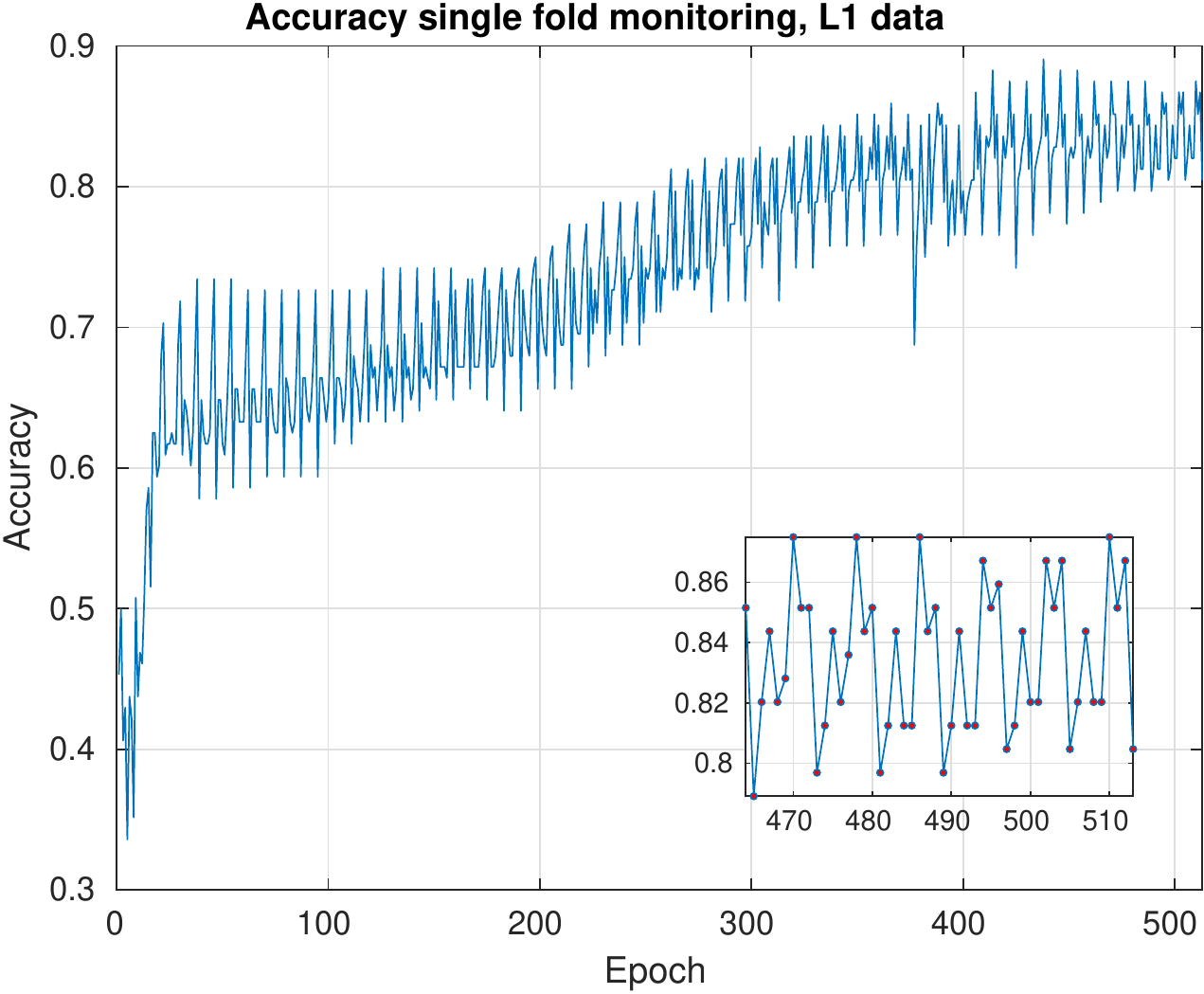}
  \caption{\label{fig:Training_evolution} Evolution of cross-entropy (left panels) and accuracy (right panels), in function of epochs of learning process, using data from H1 detector (upper panels) and L1 detector (lower panels). On trending, cross-entropy decreases and accuracy increases, even if there is a clear stochastic component due to the mini-batch SGD learning algorithm. Here we set length of sliding windows in $T_{\text{win}}=0.50$s, and the CNN architecture with $2$ stacks and $20$ kernels. Some anomalous peaks appear when data from L1 detector is used, but this does not affect general trends of mentioned metrics.}
\end{center}
\end{figure*}

Two representative examples of this check, using data from H1 and L1 detectors, are shown in Fig.~\ref{fig:Training_evolution}. Both were performed during a single fold of a $10$-fold CV experiment, from the first to the last mini-batch SGD epoch. Besides, here we used a time resolution of $T_{\text{win}}=0.50$s with $2$ stacks, and $20$ kernels in convolutional layers. Notice that cross-entropy shows decrease trends (left panels) and accuracy increase trends (right panels). Total number of epochs for H1 data was $372$, and for L1 data was $513$, which means that the CNN has more difficulties to learn parameters with L1 than with H1 data. When the CNN finish its learning process, cross-entropy and accuracy reach values $0.184$ and $0.945$, respectively, using H1 data; and $0.357$ and $0.805$, respectively, using L1 data.

Notice, from all plots in Fig.~\ref{fig:Training_evolution}, that fluctuations appear. This is actually expected, since in the mini-batch SGD algorithm a randomly number of samples $\tilde{N}<N_{\text{train}}$ are taken, then stochastic noise is introduced. Besides, when using data from L1 detector, some anomalous peaks appear between epoch $350$ and $400$ but this is not a problem because CNN normally continues its learning process and trendings in both metrics are not affected. At the end, we can observe this resilience effect because of our validation patience criterion, that is implemented to prevent our CNN algorithm prematurely stops and/or to dispense with manually adjust the total number of epochs for each learning fold.

Still focusing on the SGD fluctuations, zoomed plots in Fig.~\ref{fig:Training_evolution} show their orden of magnitude --the highest peak minus the lowest peak. When we work with H1 data, cross-entropy fluctuations are about $0.130$ and accuracy fluctuacions are about $0.090$. On the other hand, when we learn from L1 data, both cross-entropy and accuracy fluctuations are about $0.080$. Here it should be stressed that although mini-batch SGD perturbations contribute with its own uncertainty, when we compute mean accuracies among all folds in the next subsection~\ref{subsec:hyp_setting} we will see that magnitude of these perturbations do not totally influence the magnitude of data dispersion present in distribution of mean accuracies.

\subsection{Hyperparameter adjustments}
\label{subsec:hyp_setting}

Our CNN models introduce several hyperparameters, namely number of stacks, size and mount of kernels, size of pooling regions, number of layers for each stack, stride, and padding, among many others. To present a systematic study for all hyperparameters is beyond the scope of our research. However, given CNN architectures shown in Table~\ref{tab:particular_CNNs}, we decided to study and adjust two of them, namely the number of stacks and the number of kernels in convolutional layers. In addition, although it is not an hyperparameter of the CNN, the time length $T_{\text{win}}$ of the samples is a resolution that also required to be set to reach an optimal performance, then we included it in the following analyses.

\begin{figure*}[ht]
\begin{center}
  \includegraphics[height=6.6cm]{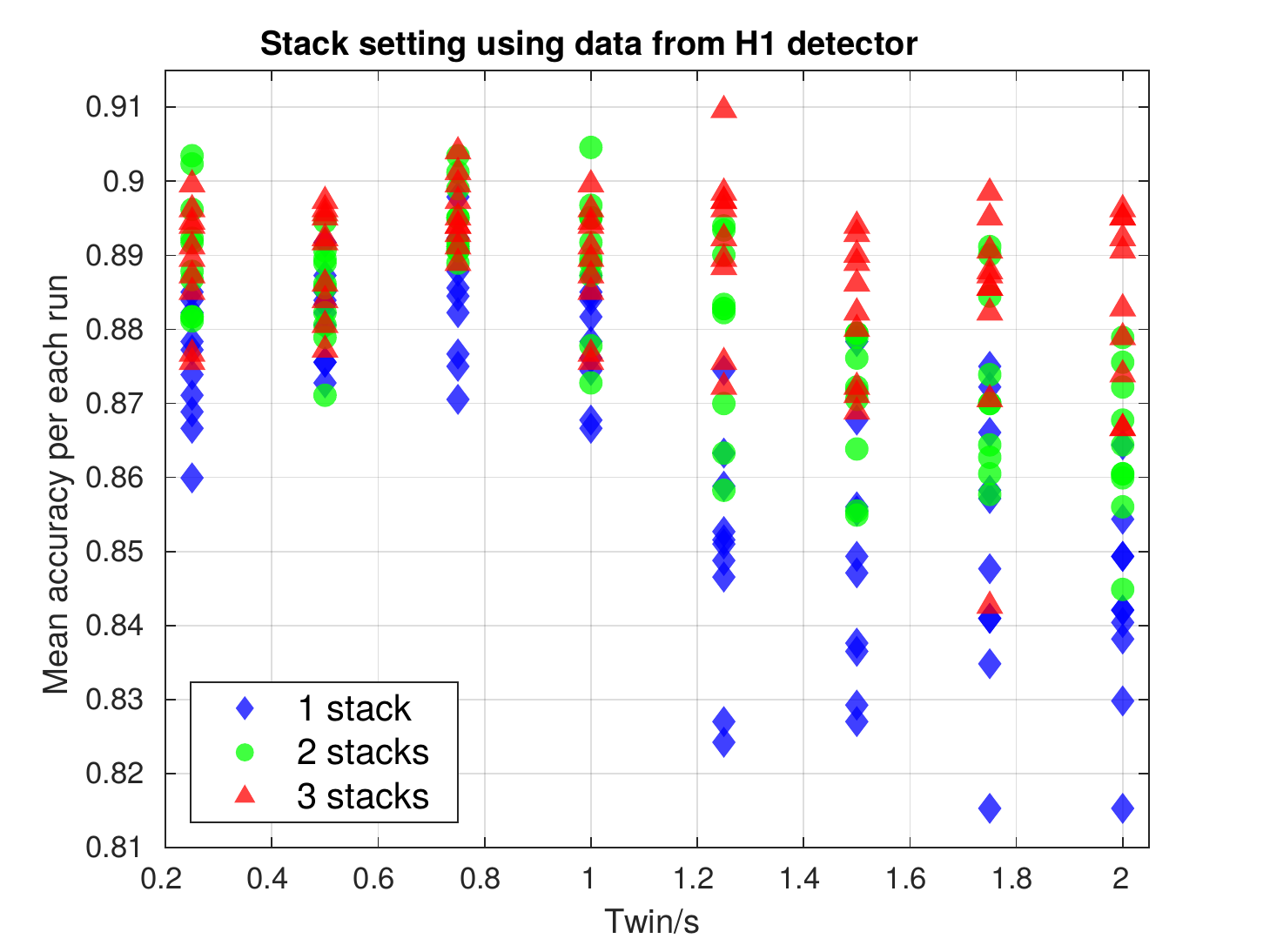} 
  \includegraphics[height=6.4cm]{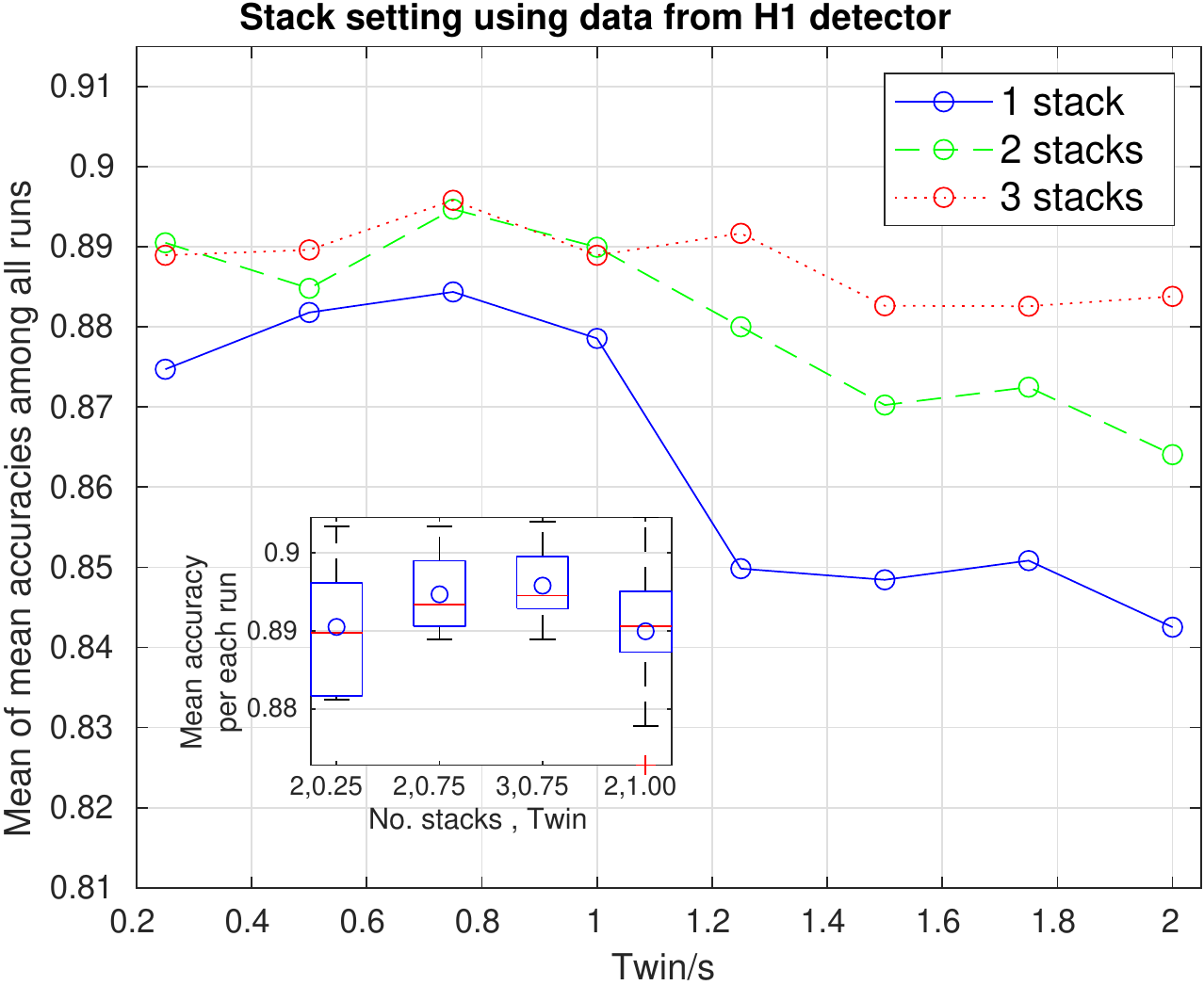} \\
  \vspace{.5cm}
  \includegraphics[height=6.6cm]{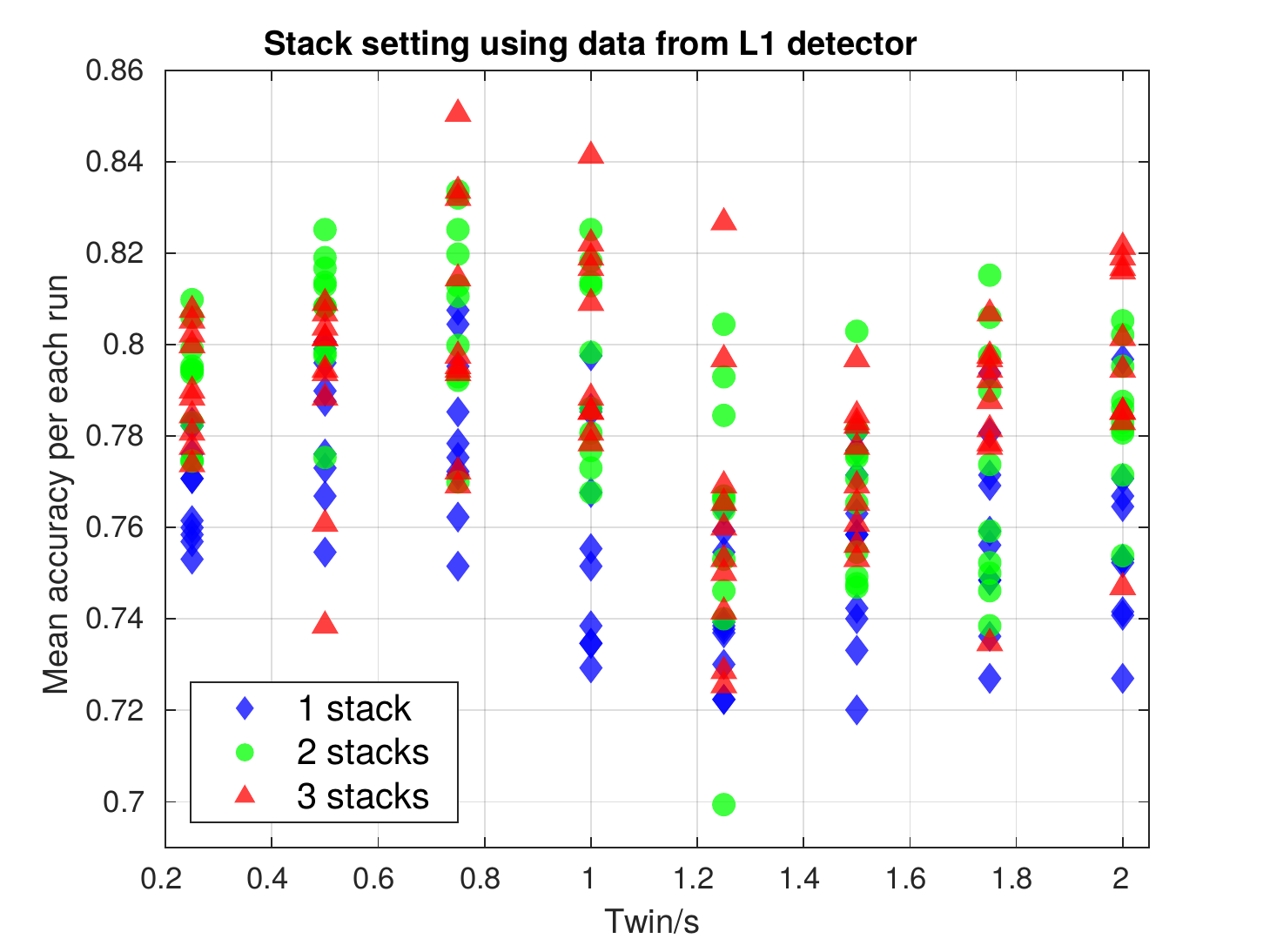}
  \includegraphics[height=6.4cm]{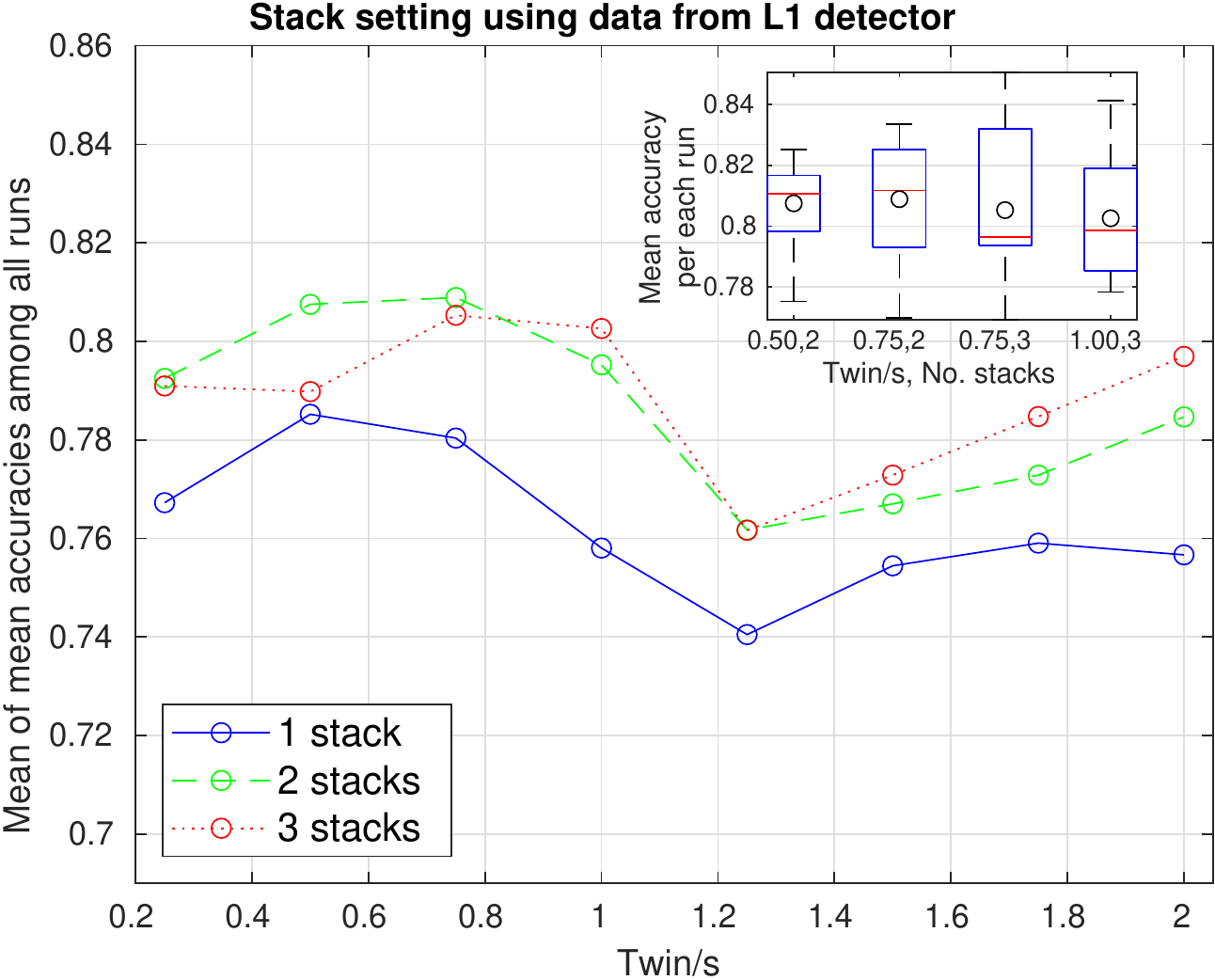}
  \caption{\label{fig:Acc_twins_stacks} Hyperparameter adjustment to find best CNN architectures, i.e. number of stacks, and time resolution $T_{\text{win}}$. Left panels show all samples of mean accuracies, and right panels the mean of mean accuracies among all runs; all in function of $T_{\text{win}}$ for $1$, $2$, and $3$ stacks in the CNN. Besides, some small boxplots are included inside right plots to have more clear information about dispersion and skewness of distribution of mean accuracies. Based on this adjustment, we conclude that $T_{\text{win}}=0.75$s, with $3$ stacks (H1 data) and with $2$ stacks (L1 data), are optimal settings. This adjustment were performed with $20$ kernels in convolutional layers, repeating a $10$-fold CV experiment $10$ times.}
\end{center}
\end{figure*}

First hiperparameter adjustment is shown in Fig.~\ref{fig:Acc_twins_stacks} and it was implemented to find optimal number of stacks and time length resolution, according to resulting mean accuracies. Left panels show distribution of mean accuracy for all $10$ repetitions or runs of the entire $10$-fold CV experiment, in function of $T_{\text{win}}$. Each of these mean accuracies, that we can denote as $Acc_i{}^{CV}$, is the average among all fold-accuracies of a $i \mhyphen th$ run of the $10$-fold CV. In addition, right panels show mean of mean accuracies among all $10$ runs, i.e. $\overline{Acc}=\left(1/10\right)\sum_{i=1}^{10}Acc_i{}^{CV}$, in function of $T_{\text{win}}$. Inside right plots we have included small boxplots that, as it will be seen next, are useful to study dispersion and skewness of mean accuracy distributions --circles inside boxplots are distribution mean values. Line plots show contributions of our three CNN architectures, i.e. with $1$, $2$, and $3$ stacks.

Consider top panels of Fig.~\ref{fig:Acc_twins_stacks} for H1 data. Notice from the right plot that, for all CNN architectures, mean of mean accuracies shows a trend to decrease when $0.75\text{s} \leq T_{\text{win}} \leq 2.00\text{s}$ and this decrease occurs more pronouncedly when we work with less stacks. Besides, when $0.25\text{s} \leq T_{\text{win}} \leq 0.75\text{s}$, a slight increase appears, even if local differences are of the order of SGD fluctuations. Anyway, given our mean accuracy sample dataset, the highest mean of mean accuracies, about $0.895$, occurs when $T_{\text{win}}=0.75$ with a CNN of $3$ stacks. Then, to decide if this setting is optimal, we need to explore in Fig.~\ref{fig:Acc_twins_stacks} left upper plot together with boxplots inside right upper plot. From the left plot, we have that not only the mentioned setting give high mean accuracy values, but also $T_{\text{win}}=0.25$s and $T_{\text{win}}=1.00$s, both with $2$ stacks, $T_{\text{win}}=0.75$s with $2$ and $3$ stacks, and even $T_{\text{win}}=1.25$s with $3$ stacks; all these settings reach a mean accuracy greater than $0.9$. Setting of $T_{\text{win}}=1.25$s with $3$ stacks can be discarded because its maximum mean accuracy clearly is an outlier and, to elucidate what of remaining settings is optimal, we need to explore boxplots.

Here it is crucial to assimilate that the optimal setting to choose actually depends on what specifically we have. If the dispersion does not worry us too much and we want to have a high probability of occurence for many high values of mean accuracy, setting of $T_{\text{win}}=0.25$s with $2$ stacks is the best, because its distribution has a slighly negative skewness concentrating most of mass probability to upper mean accuracy values. On the other hand, if we prefer to have more stable estimates working with less dispersion at the expense of having a clear positive skewness (in fact, having a high mass concentration in a region that does not reach as high mean accuracy values as the range from median to third quartile in the previous setting), setting of $T_{\text{win}}=0.75$s with $3$ stacks is the natural choice. In practice, we would like to work with greater dispersions if they help to reach highest mean accuracy values but, as all our boxplots have similar maximum values, we decide to maintain our initial choice of $T_{\text{win}}=0.75$s with $3$ stacks for H1 data.

From left upper scatter plot of Fig.~\ref{fig:Acc_twins_stacks} can be seen that, regardless number of stacks, data dispersion in $1.25\text{s} \leq T_{\text{win}} \leq 2.00\text{s}$ is greater than in $0.25\text{s} \leq T_{\text{win}} \leq 1.00\text{s}$, even if in the former region dispersion slightly tends to decrease as we increase the number of stacks. This actually is a clear visual hint that, together with the evident trend to decrease mean accuracy as $T_{\text{win}}$ increase, motivate to discard all settings for $T_{\text{win}} \geq 1.25$s. However, this hint is not present in bottom panels, in which a trend of decrease and then increase appears (which is clearer in the right plot), and data dispersion of mean accuracy distributions are similar almost for all time resolutions. For this reason, and although the procedure for hyperparameter adjustment is the same as upper panels, one should be cautious, in the sense that decisions here are more tentative, specially if we have prospect to increase the mount of data.

Anyway, given our current L1 data and based on scatter distribution plot and line mean of mean accuracies plot, we have that best performance(s) should be among settings of $T_{\text{win}}=0.5$s with $2$ stacks, $T_{\text{win}}=0.75$s with $2$ and $3$ stacks, and $T_{\text{win}}=1.00$s with $3$ stacks. Now, exploring boxplots inside right panel we notice that, even if settings of $T_{\text{win}}=0.75\text{s}, 1.00\text{s}$ with $3$ stacks reach the highest mean accuracy values, their positive skewness toward lower values of mean accuracies is not great. Then, the two remaining settings, which in fact have negative skewness toward higher mean accuracy values, are the optimal options, and again choosing one or other will depend of what extent we tolerate data dispersion. Unlike upper panels, here a larger dispersion increase probality to reach higher mean accuracy values, therefore we finally decide to work with the setting of $T_{\text{win}}=0.75$s with $2$ stacks for L1 data.

\begin{figure*}[ht]
\begin{center}
  \includegraphics[height=6.5cm]{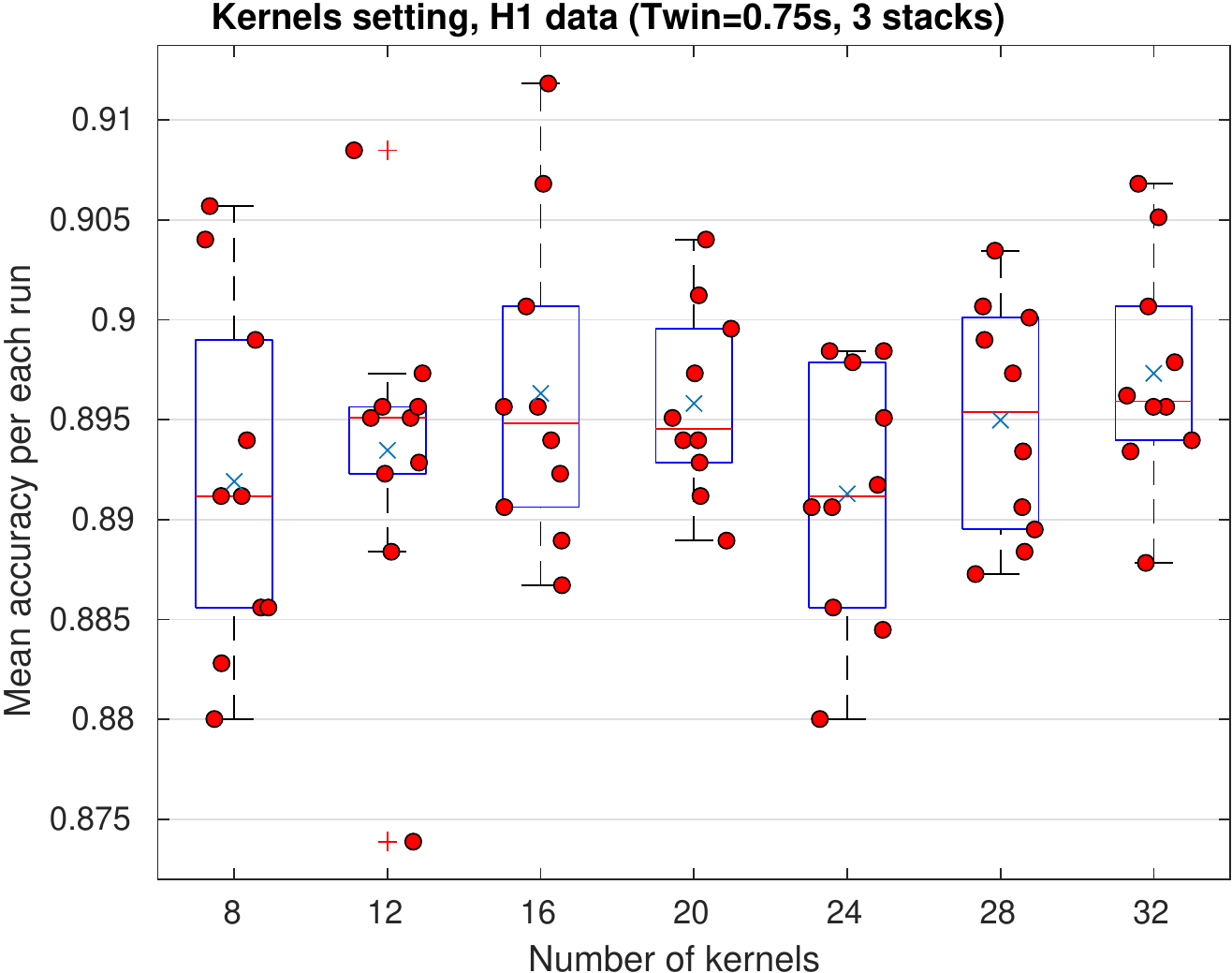}
  ~~~~~
  \includegraphics[height=6.5cm]{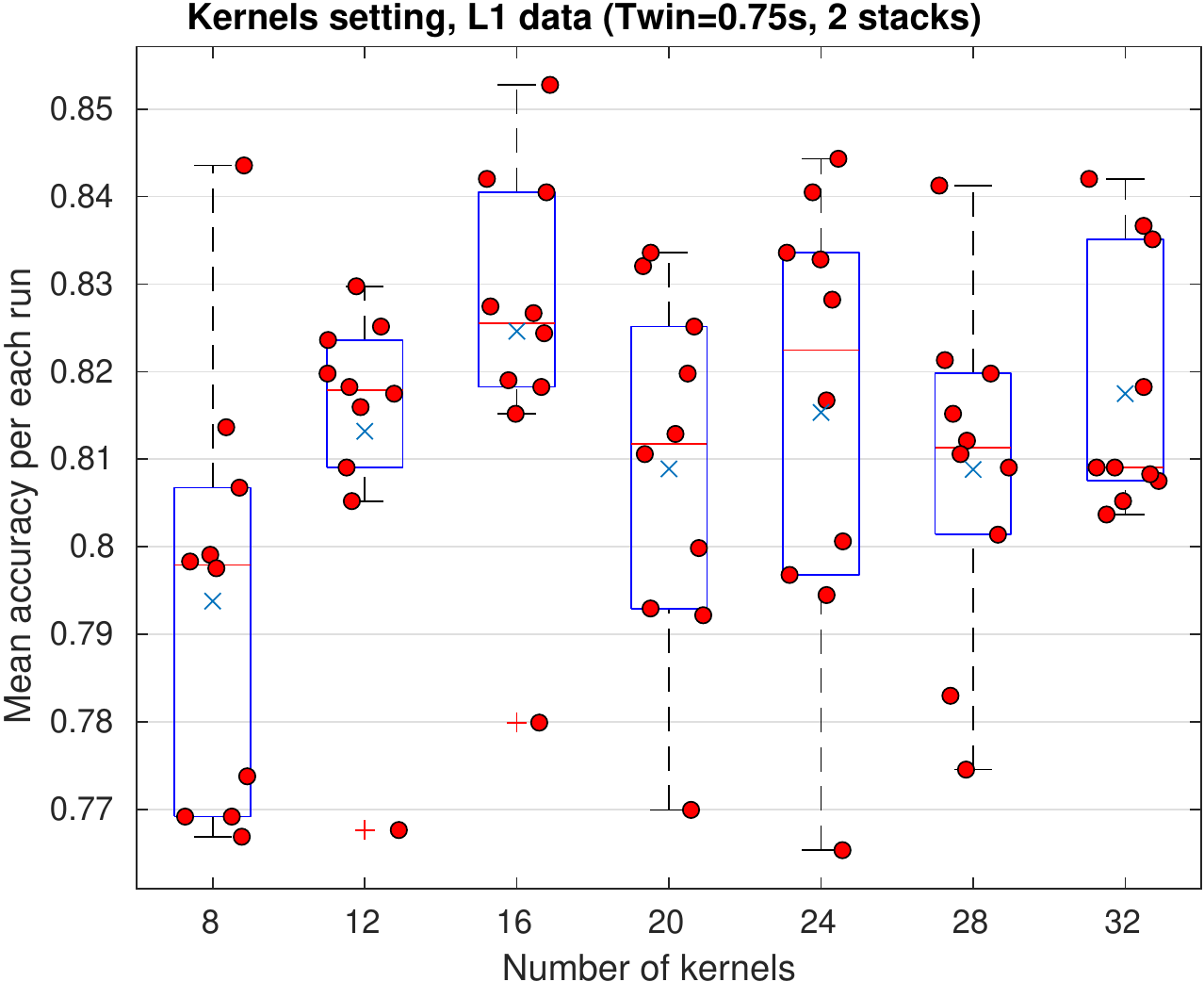}
  \caption{\label{fig:Acc_twins_kernels} Adjustment to find the optimal number of kernels in the CNN architecture. Both panels show boxplots for the distributions of mean accuracies in function of the number of kernels, with values of $T_{\text{win}}$ and mount of stacks found in previous hyperparameter adjustments. Based on location of mean accuracy samples, dispersion, presence of outliers, and skewness of distributions, CNN architectures with $32$ kernels and $16$ kernels are the optimal choices to reach the highest mean accuracy values, when working with data from H1 and L1 detectors, respectively.}
\end{center}
\end{figure*}

To find the optimal mount of kernels in convolution layers, we perfomed the adjustment shown in Fig.~\ref{fig:Acc_twins_kernels}, again separately for data from each LIGO detector. Considering the information provided by previous adjustment, we set  $T_{\text{win}}=0.75$s, and the number of stacks in $3$ for H1 data and $2$ for L1 data. Then, once the $10$-fold CV was run $10$ times as usual, we generated boxplots for CNN configurations with several mount of kernels as it was advanced in Table~\ref{tab:particular_CNNs}, including all mean accuracies for each run marked (red circles). Besides, average for each boxplot is included (blue crosses). Random data horizontal spreading inside each boxplot was made to avoid visual overlap of markers, and it does not mean that samples was obtained with number of kernels different from already speficied in horizontal axis.

Let us concentrate on kernels adjustment for H1 data in the left panel of Fig.~\ref{fig:Acc_twins_kernels}. From these results we have that a CNN with $12$ kernels give us more stable results by far, because most of its mean accuracies lie in the smallest dispersion region --discarding outliers, half of mean accuracies are concentrated in a tiny interquartile region located near to $0.895$. On the other hand, CNN configuration with $24$ kernels is the least suitable setting among all, not because its mean accuracy values are low \textit{per se} (values from $0.800$ to $0.898$ are actually good), but rather because, unlike other cases, the nearly zero skewness of its distribution is not prone to boost sample values beyond third quartile as it is appeared. Configuration with $8$ kernels has a distribution mean very close to the setting with $24$ kernels, and even, reaches two mean accuracy values about $0.905$. Nonetheless, given that settings with $16$, $20$, $28$, and $32$ have mean of mean accuracies greater or equal to $0.895$ (and hence, boxplots located towards relative higher mean accuracy values), these last four configurations offer the best options. At the end, we decided to work with $32$ kernels, because this setting group a whole set of desirable features: the highest mean of mean accuracies, namely $0.893$, a relatively low dispersion, and a positive skewness defined by a pretty small range from the first quartile to the median.

Kernels adjustment for L1 data is shown in right panel of Fig.~\ref{fig:Acc_twins_kernels}. Here the situation is easier to analize, because performance differences appears visually clearer than those for data from H1. Settings with $8$, $20$, and $28$ kernels lead to mediocre performances, specially the first one which has a high dispersion and $70\%$ of its samples are below $0.8$ of mean accuracy. Notice that, like adjustment for H1 data, setting with $12$ kernels shows the smallest dispersion (discarding a outlier below $0.77$), where we have mean accuracies from $0.805$ to $0.830$, and againly, this option will be suitable if we would very interested in to reach stable estimates. We decided to pick up setting with $16$ kernels, that has the highest distribution mean, $0.825$, $50\%$ of mean accuracy samples above the distribution mean, an aceptable data dispersion (without counting the clear outlier), and a relatively small region from the minimum  to the median.

In summary, based on all above adjustments, the best time resolution is $T_{\text{win}}=0.75$s, with a CNN architecture of $3$ stacks and $32$ kernels when working with data from H1 detector, and $2$ stacks and $16$ kernels when working with data from L1 detector. We use only these hyparameter settings hereinafter.

Now, let us finish this subsection reporting an interesting additional result. As it was mentioned at the end of subsection~\ref{subsec:learning_monitor}, we can ask to what extent magnitude of perturbations from the mini-batch SGD algorithm influence dispersion of mean accuracy distributions. Here we can compare the order of magnitude of SGD perturbations and dispersion present in boxplots. In previous subsection, we had that when $T_{\text{win}}=0.50$s and we work with a CNN architecture of $2$ stacks and $20$ kernels in convolution layers, the order of magnitude of SGD fluctuations in accuracy is about $0.090$. Curiously, this value is much greater than dispersion of data distribution shown in left panel of Fig.~\ref{fig:Acc_twins_stacks}, which in turn reach a value of $0.025$, that is to say, $3.6$ times smaller. These results are good news, because apart of showing that stochasticity of mini-batch SGD perturbations do not totally define dispersion of mean accuracy distributions, it seems that our resampling approach, actually contributes to smooth stochastic effects of mini-batch SGD perturbations, and hence to decrease uncertainty in mentioned distributions. This is a very important result that could serve as motivation and standard guide to future works --surprisingly, most of previous works in DL and GW data analysis that use stochastic optimization algorithms, no standard resampling ML techniques are applied.

\subsection{Confusion matrices and standard metrics}
\label{subsec:confu_matrices}

\begin{figure*}
\begin{center}
  \includegraphics[height=5.5cm]{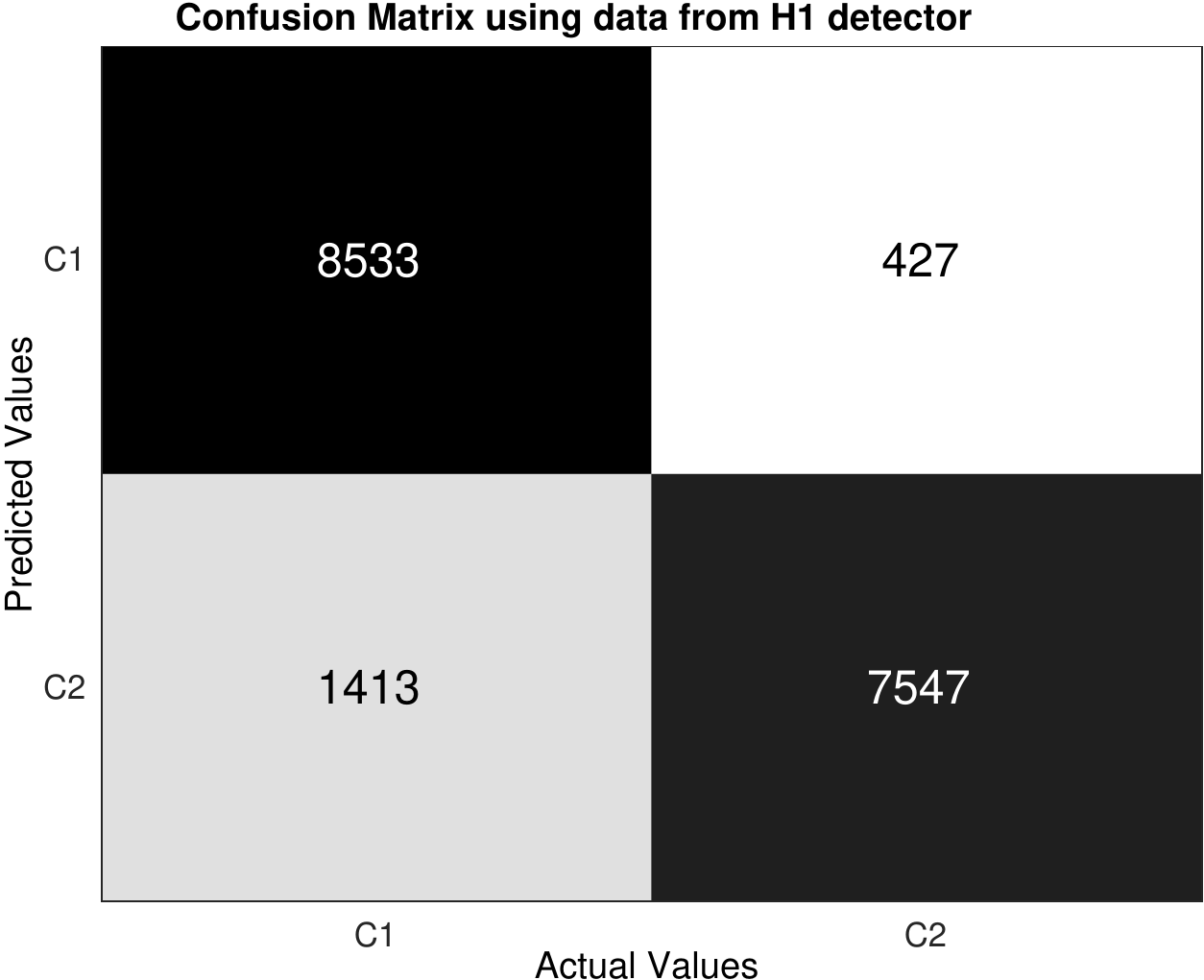}
  ~~~~~~~~
  \includegraphics[height=5.5cm]{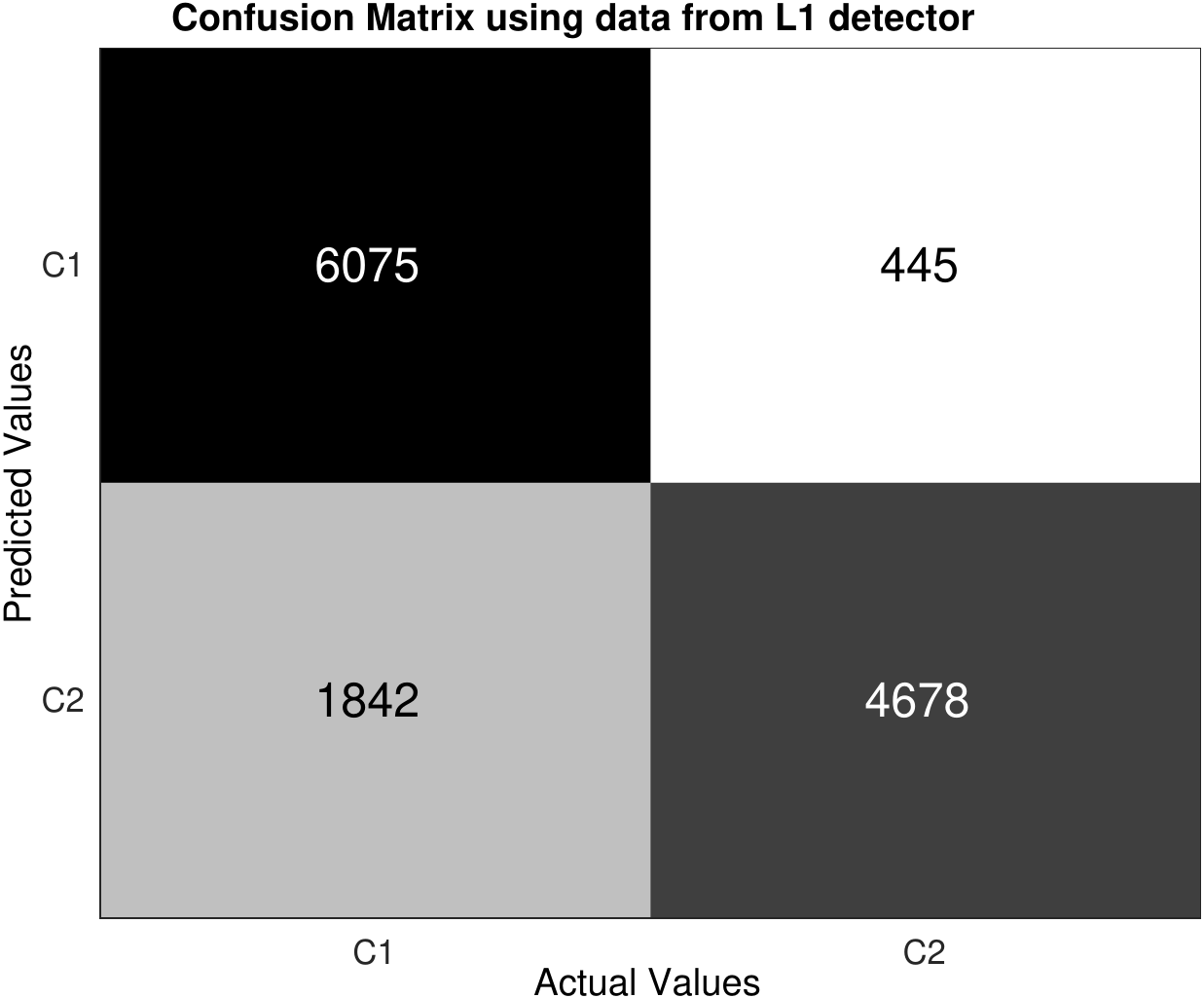}
  \caption{\label{fig:confu_results} Confusion matrices computed from the testing. C1 is the label for noise only and C2 the label for noise plus GW injection. Our CNN has $32$ kernels and $3$ stacks when is run with H1 data, and $16$ kernels and $2$ stacks when is run with L1 data. Time resolution is $T_{\text{win}}=0.75$s. Working with H1 data, $8533/(8533+427) \approx 0.952$ predicitions for C1 are correct and $1413/(8533+1413) \approx 0.142$ predictions for C2 are incorrect and, working L1 data, $6075/(6075+445) \approx 0.932$ predicitions for C1 are correct and $1842/(6075+1842) \approx 0.233$ predictions for C2 are incorrect. These results show that our CNN classifies very precisely noise samples at the cost of reaching a not less number of false GW predictions}
\end{center}
\end{figure*}

In general, accuracy gives information about the probability of a successful classification, either if we are classifying a noise alone sample (C1) or noise plus GW sample (C2); that is to say, it is a performance metric with multi-label focus. However, we would like to elucidate to what extent our CNNs are profient to separately detect samples of each class, then it is useful to introduce peformance metrics with single-label focus. A standard tool is the confusion matrix, that is shown in Fig.~\ref{fig:confu_results} depending on data from each detector. As we are under a resampling regime, each element of confusion matrices is computed considering the entire mount of $100~N_{\text{test}}$ detections, which in turn are resulting from concatenating all prediction vectors of dimension $N_{\text{test}}$ outputted by the $10$ runs of the $10$-k fold CV.

A first glance to confusion matrices shown in Fig.~\ref{fig:confu_results} reveals that our CNNs have a better performance detecting noise alone samples than detecting noise plus GW samples, because (C1,C1)\footnote{We are using the notation $\left(\text{row}, \text{column}\right)$ to represent each element of a confusion matrix.} element is greater than (C2,C2) for both matrices. Yet amount of successful predictions of noise plus GW are reasonable good because they considerably surpass a totally random performance --described by successful detections or the order of $50\%$ of total negative samples.

Moreover, from Fig.~\ref{fig:confu_results} we have that based on wrong predictions, CNNs are more likely to make a type II error than type I error, because (C2,C1)$>$(C1,C2) for both confusion matrices. If we think more carefully, this result leads to an advantage and a disadvantage. The advantage is that our CNN performs a ``conservative'' detection of noise alone samples in the sense that a sample will be not classified as beloging class 1 unless the CNN is very sure, that is to say the CNN is quite precise to detect noise samples. Using H1 data, $8533/(8533+427) \approx 0.952$ of samples predicted as C1 belong this class; and using L1 data, $6075/(6075+445) \approx 0.932$ of samples predicted as C1 belong this class. This is an important benefit if, for instance, we wanted to apply our CNNs to remove noise samples from a segment of data with a narrow marging of error in addition to other detection algorithm and/or analysis focused on generating triggers. Nonetheless, the disadvantage is that a not less number of noise samples are lost by wrongly classifying them as GW event samples. In terms of false negative rates, we have that $1413/(8533+1413) \approx 0.142$ of actual noise samples are misclassified with H1 data, and $1842/(6075+1842)\approx 0.233$ of actual noise samples are misclassified with L1 data. This would be a serious problem if our CNN were implemented to decide if an individual trigger is actually a GW signal and not a noise sample --either Gaussian or non-Gaussian noise.

Take in mind that, according to statistical decision theory, there will always be a trade-off between type I and type II errors~\cite{Kay-Book}. Hence, given our CNN architecture and datasets, it is not possible to reduce value of (C2,C1) element without increasing value of (C1,C2) element. In principle, keeping the total number of training samples, we could generalize the CNN architecture for a multi-label classification to further specify the noise including several kind of glitches as was implemented in works as~\cite{mReC18} and~\cite{aIeCfMjP19}. Indeed, starting from our current problem, such multiclass generalization could be motivated to redistribute current false negative counts (C2,C1) in new elements of a bigger confusion matrix, where several false positive predictions will be converted to new sucessful detections located along a new longer diagonal. Nonetheless, it is not clear how to keep constant the bottom edge of the diagonal of the original binary confusion matrix when the number of noise classes is increased; not to mention that this approach can be seen as a totally different problem instead of a generalization, actually.

\begin{figure*}[htp]
\begin{center}
  \includegraphics[width=13cm]{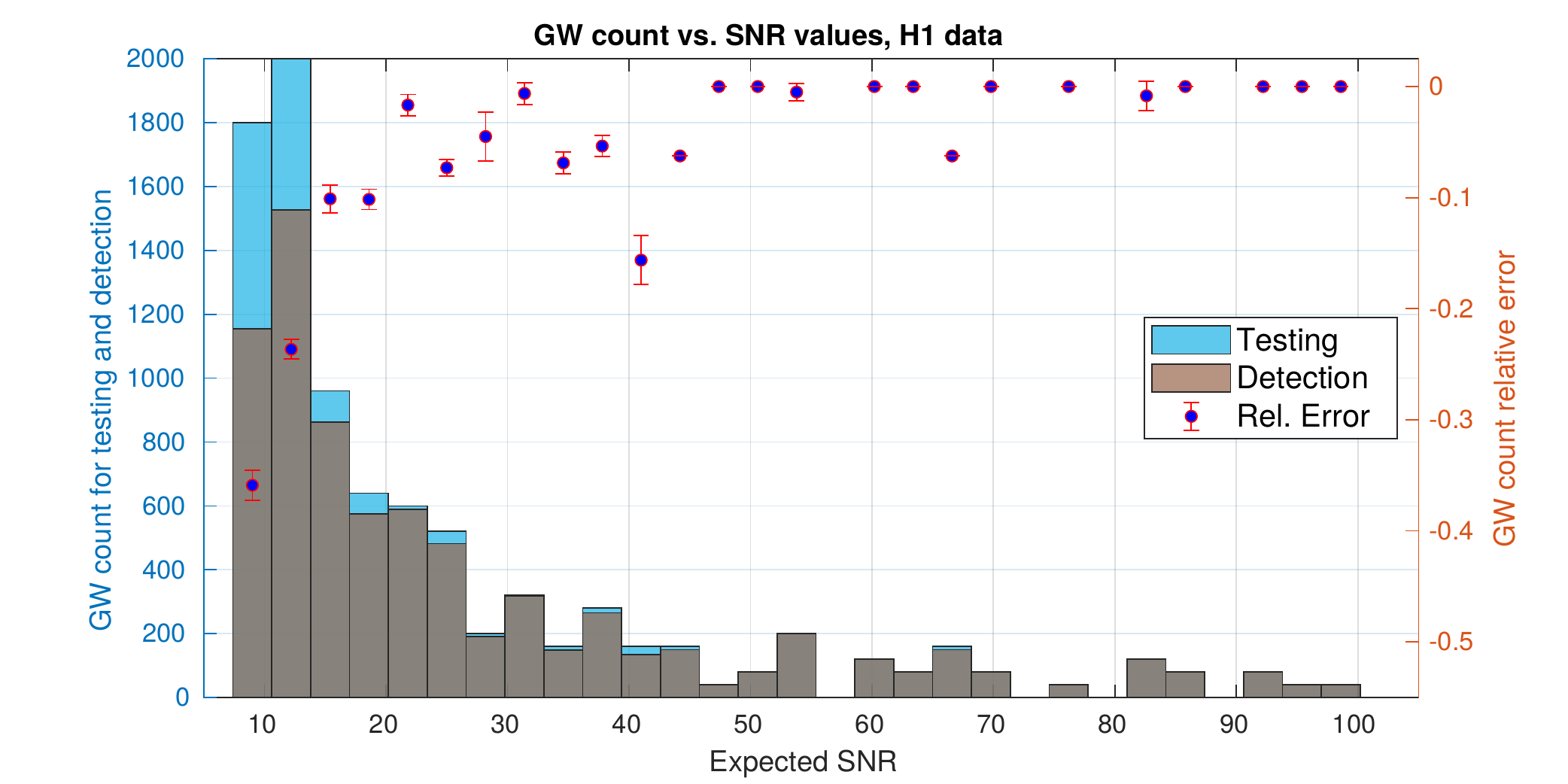}\\
  \vspace{.5cm}
  \includegraphics[width=13cm]{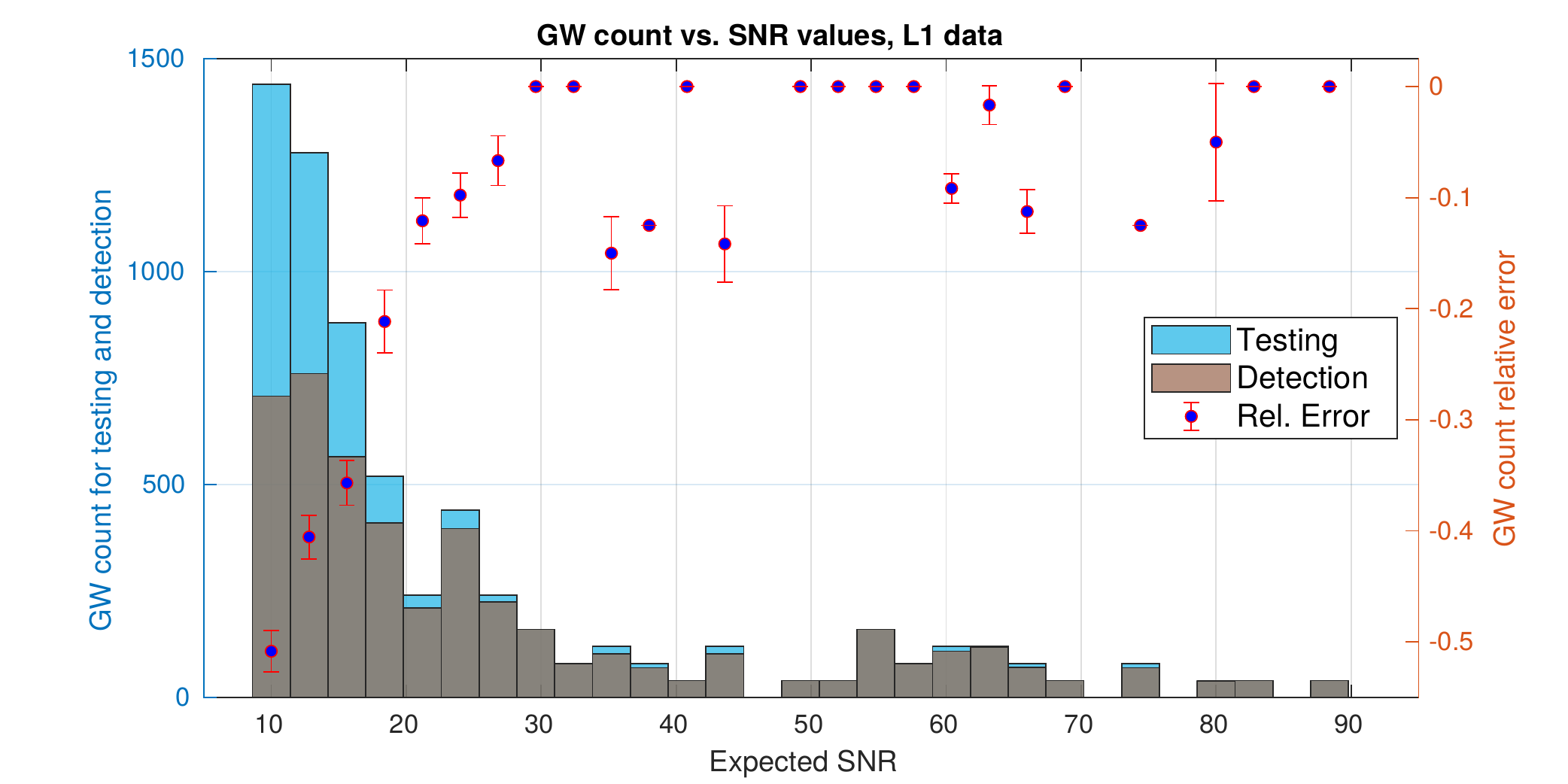}
  \caption{\label{fig:CountGW_SNR} Histograms for counting GW samples present in test set and GW samples detected by the CNN. This count was made from all $100~N_{\text{test}}$ predictions because we have $10 \times 10 = 100$ SGD learning-testing runs. Besides, relative error between both histograms is shown as scatter points, with their respective standard deviations computed from the $10$ repetitions of the $10$-fold CV routine. From plots we have that our CNN detects more CBC GW events insofar they have a SNR $\geq 21.80$ for H1 data (upper panel), and SNR $\geq 26.80$ for L1 data (bottom panel).}
\end{center}
\end{figure*}

With regard to misclassified GW event samples, despite they are quite less than misclassified noise samples, we would like to understand more about them. Then, we decided to study the ability of the CNN to detect GW events depending on the values of their expected SNR --values that are provided with LIGO hardware injections. Results are shown in Fig.~\ref{fig:CountGW_SNR}; upper panel with data from H1 detector, and lower panel with data from L1 detector. Both panels include a blue histogram for actual (injected) GW events that come from the testing set, a gray histogram with GW events detected for the CNN, and the bin-by-bin discrepancy between both histograms as scatter points. As a first approach, we defined this bin-by-bin discrepancy as the relative error:
\begin{equation}
 Rel~Err(i) = [N_{det}(i)-N_{test}(i)]/N_{det}(i) ~,
\end{equation}
where $N_{det}$ and $N_{test}$ are the detected GW count and injected GW count respectively, and index $i$ represent a bin. Here we set $29$ same-length bins for both histograms, starting from a lower edge SNR$=9.0$ to a upper edge SNR$=101.8$ for H1 data, and from SNR$=10.0$ to SNR$=91.2$ for L1 data, respectively. For testing histograms, count of events come from our $100~N_{\text{test}}$ predictions given our resampling regime.

By comparing most bins that appear on both panels of Fig.~\ref{fig:CountGW_SNR}, we have that detected GW count is greater as more actual injections in the testing set there are. Besides, most GW events are concentrated in a region of smaller SNR values. For H1 data, most events are in the first six bins, namely from SNR$=9.0$ to SNR$=28.9$; with $6520$ actual GW events and $5191$ detected GW, representing aprox. the $72.77\%$ and $68.78\%$ of the total number of actual GW events and detected GW events, respectively. For L1 data, on the other hand, most events are in the first seven bins, from SNR$=10.0$ to SNR$=29.6$; with $5040$ actual GW events and $3277$ detected GW events, representing aprox. the $77.30\%$ and $70.05\%$ of the total number of actual GW events and detected GW events, respectively.

Above information about counts is relevant but, the most important results come from relative errors. From these we have that, in both panels, a clear trend of detecting a greater percentage of actual GW events as long as those events has greater SNR values. Besides, if we focus in upper panel of Fig.~\ref{fig:CountGW_SNR}, corresponding to H1 data, we have that in first four bins, GW count relative errors are the greatest; beginning with $-0.3589$ and ending with $-0.1016$. Then, from fifth bin at SNR$=21.80$, relative errors stochastically approaches to zero --indeed, relative error exactly equal to zero is reached in $10$ of $29$ bins. For L1 data, shown in lower panel, we observe a similar behaviour of relative error. In first six bins the greatest relative errors appears; from $-0.5083$ to $-0.0977$. Next, from seventh bin at SNR$=26.80$, relative errors stochastically approaches to zero. Indeed, here we have that relative error value exactly equal to zero is reached for the first time at smaller SNR value than with H1 data, although once zero values begin to appear, relative errors further away from zero than with H1 data also appear. This last result is statistically consistent with the fact that, according to Fig.~\ref{fig:confu_results}, negative predictive value ($\text{NPV}=\left(\text{C2},\text{C2}\right)/\left[\left(\text{C2},\text{C1}\right)+\left(\text{C2},\text{C2}\right)\right]$) is smaller in the confusion matrix for H1 data than for L1 data, with $\text{NPV} \approx 0.842$ and $\text{NPV} \approx 0.717$, respectively.

For bin-by-bin discrepancies shown in Fig.~\ref{fig:CountGW_SNR}, we include error bars. These are standard deviations and each of these was computed from distributions of 10 relative errors because of the $10$-fold CV experiment is repeated $10$ times. From the plots we observe that standard deviations do not approach to zero as their SNR increase, meaning that stochasticity introduced in such standard deviations by our resampling cannot seem to be smoothed by selecting certain SNR values.

It is important to reiterate that here we applied the CNN algorithm under a realistic approach in the sense that GW events are given by the hardware injections provided by LIGO, and therefore, all SNR values are given in the strain data with no possibility to be directly handled in the numerical relativity templates before software injections\footnote{In addition to SNR values, frequency of occurrence of GW events also represents an important challenge to generate a more realistic dataset emulating record of astrophysical data, even though this leads us to work with highly imbalanced datasets. Even, for a more realistic situation, we could internally described each bin of histograms in Fig.~\ref{fig:CountGW_SNR} as a random sampling in which the counts themselves take random values, following a distribution --indeed, this hypothesis is usually assumed to perform systematic statistical comparisons between two histograms.}. And this is desirable, because in real experimental conditions, SNR values of GW signals depend solely on the nature of the astrophysical sources and the conditions of the detectors --aspects that obviously are not modifiable during an observation run. If the CNN is able to deal with this limiting scenario beforehand, it does not learn more than what is strictly necessary, avoiding overoptimistic results or even, underperformance. Indeed, because of this aspect is that we can transparently concluded that our CNN \textit{per se} is more sensitive to stocastically detect GW signals when SNR $\geq 21.8$ for H1 data, and when SNR $\geq 26.80$ for L1 data.

\begin{table*}[ht]
    \begin{minipage}{.45\linewidth}
      \centering
        \small{\bf Standard metrics with H1 data}\\
        \vspace{.3cm}
        \begin{tabular}{lllll}
			\toprule
			Metric       & Mean      & Min       & Max       & SD   \\
			\midrule
			Accuracy     & $0.897$  & $0.888$  & $0.907$  & $0.00564$ \\
			Precision    & $0.952$  & $0.941$  & $0.968$  & $0.00852$ \\
			Recall       & $0.858$  & $0.844$  & $0.873$  & $0.00894$ \\
			Fall-out     & $0.0534$ & $0.0374$ & $0.0647$ & $0.00880$ \\
			F1 score     & $0.903$  & $0.894$  & $0.912$  & $0.00509$ \\
			G mean1      & $0.213$  & $0.179$  & $0.237$  & $0.0185$ \\
			\bottomrule
		\end{tabular}
    \end{minipage}%
    \begin{minipage}{.45\linewidth}
      \centering
        \small{\bf Standard metrics with L1 data}\\
        \vspace{.3cm}
        \begin{tabular}{lllll}
			\toprule
			Metric       & Mean      & Min       & Max       & SD   \\
			\midrule
			Accuracy     & $0.825$  & $0.780$  & $0.853$ & $0.0198$ \\
			Precision    & $0.932$  & $0.905$  & $0.956$ & $0.0161$ \\
			Recall       & $0.768$  & $0.724$  & $0.793$ & $0.0197$ \\
			Fall-out     & $0.0869$ & $0.0560$ & $0.127$ & $0.0208$ \\
			F1 score     & $0.842$  & $0.804$  & $0.866$ & $0.0168$ \\
			G mean1      & $0.256$  & $0.211$  & $0.303$ & $0.0282$ \\
			\bottomrule
		\end{tabular}
    \end{minipage}
    \caption{\label{tab:tab_metrics_results}Standard metrics computed from values of confusion matrices shown in Fig.~\ref{fig:confu_results}. We set our CNN with $32$ kernels and $3$ stacks with H1 data, and with $16$ kernels and $2$ stacks with L1 data. Time resolution is $T_{\text{win}}=0.75$s. Except of G mean1 reaching moderate results that could help to avoid over pessimistic (or optimistic) results, accuracy, precision, recall, fall-out and F1 score shown that the CNN has a better performance working with H1 data.}
\end{table*}

Continuing with our analysis, Table~\ref{tab:tab_metrics_results} shows a summary of several metrics that we previously defined on Table~\ref{tab:table_metrics} --againly, these metrics were computed by counting the entire mount of $100~N_{\text{test}}$ predictions given that we repeated $10$ times a $10$-fold CV experiment. From the table, we have that working with L1 data, we observe that recall has a mean value telling us that $76.8\%$ of noise alone samples are retrieved. Given these results, if we want to have chances of recovering most noise alone samples of a segment of data on our side in order to, for instance, increase in the short-term our catalogues of glitches or to full analyse strain data in real-time observation to filter them, this CNN could be not the best option because its sensitivity is not great. Mean recall is slightly better with H1 data, $85.8\%$, but not as great as to considerably improve sensitivity. Notice, on the other hand, that mean precision and mean fall-out show that our CNN is quite precise classifying noise alone samples, because once it label a set of samples as that, for L1 data we have that $93.2\%$ of them are actually noise alone, and just $8.69\%$ are GW signals. Even, for H1 data results are better, because mean precision is $95.2\%$ and a mean fall-out is $5.34\%$. At the end, this disparity between recall and precision is summarized in F1 score. For H1 data, F1 score is $0.903$, and for L1 data is $0.842$. In both cases, mean F1 score reaches a moderate performance with numerical values lying between values of mean recall and mean precision. Besides, although fall-out plus precision theoretically is exactly $1$, here we are considering means among several stochastic realization of theses metrics; then summation slightly differs in $0.0189$ for L1 data, and $0.00540$ for H1 data.

As F1 score has the limitation of leaving out true negatives samples, it is recomendable to report it together with G mean1. Values for the mean of this metric is also shown in Table~\ref{tab:tab_metrics_results}, namely $0.213$ for H1 data and $0.256$ for L1 data. These two values are low because, by definition, G mean1 is mainly susceptible to the sensitivity of the CNN. In fact, these results elucidate a useful feature of G mean1, namely that it is works as a warning for avoiding overoptimistic performance reports based solely on accuracy. Notice, on the other hand, that mean of G mean1 shows a slightly better performance for L1 than for H1 data; showing that G mean1 also contributes to avoid excesive pessimistic interpretations when accuracy, or other metrics, reach lower relative results\footnote{For a $N$-labels classification, imbalanced datasets, and $N>2$, accuracy has a serious risk to becomes a pessimistic metric, and working with single-label focus metrics would be impractical when $N$ is quite larger, because we would need $N$ metrics to detail the model performance. Hence the needed of draw on metrics as F1 score and G mean1.}.

Dispersion of metrics is also shown in Table~\ref{tab:tab_metrics_results}. For data from a given detector (H1 or L1), we observe that standard deviations of accuracy, precision, recall, and fall-out are of the same order of magnitude. This is expected because these metric were computed directly from the same resampling of data predictions. Besides, for H1 or L1 data, we have that standard deviation of F1 score is also of the same order of magnitude as other metrics. However, with G mean1 we observe a slightly smaller dispersion with L1 data than with H1 data, which is in consistency with the little improvement reported in the mean of the G mean1. Anyway, this reported improvement is actually marginal, because all other metrics reports a better performance of our CNN working with data from H1 detector. In the next subsection we give more reasons to reach this conclusion.

\subsection{ROC comparative analyses}
\label{subsec:ROC_analyses}

As it was mentioned in subsection~\ref{subsec:training}, all performance metrics shown in Table~\ref{tab:table_metrics} depend on a choosen fixed threshold for assigning a class per image sample. Until now, previous analyses used a threshold of $0.5$ by default; but for generating ROC curves, it is necessary to vary this thereshold from $0$ to $1$. In general, ROC curves visually shows to what extent our binary CNN classifier, depending on thresholds, define the trade-off between recovered noise alone samples and GW events samples wrongly classified as noise alone samples. Moreover, ROC curves are used to contrast performances of a model learning from different datasets, or more widely, to compare performances of different models. Here we will present two ROC comparative analyses, one for H1 data and other for L1 data, where each one will contrast performances of our CNN with other two classic ML models, namely Naive Bayes (NB) and Support Vector Machines (SVM). Comprehensive details about NB and SVM classifiers are not necessary for our purposes. Even so, in Appendices~\ref{sec:App_NB} and~\ref{sec:App_SVM} we introduce brief definitions of NB and SVM models just for clarification.

\begin{figure*}[htp]
\begin{center}
  \includegraphics[width=8.2cm]{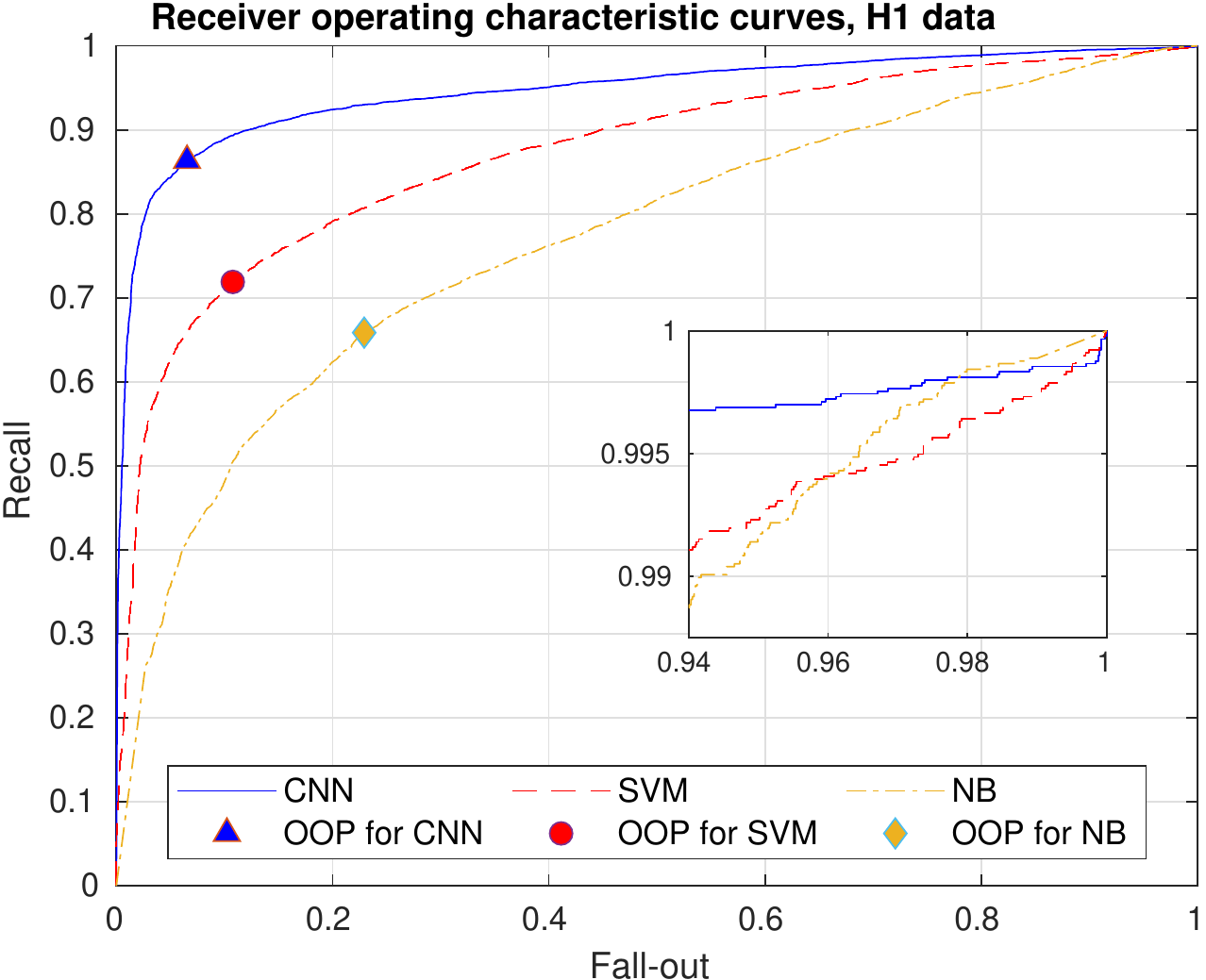}
  ~~~~~
  \includegraphics[width=8.2cm]{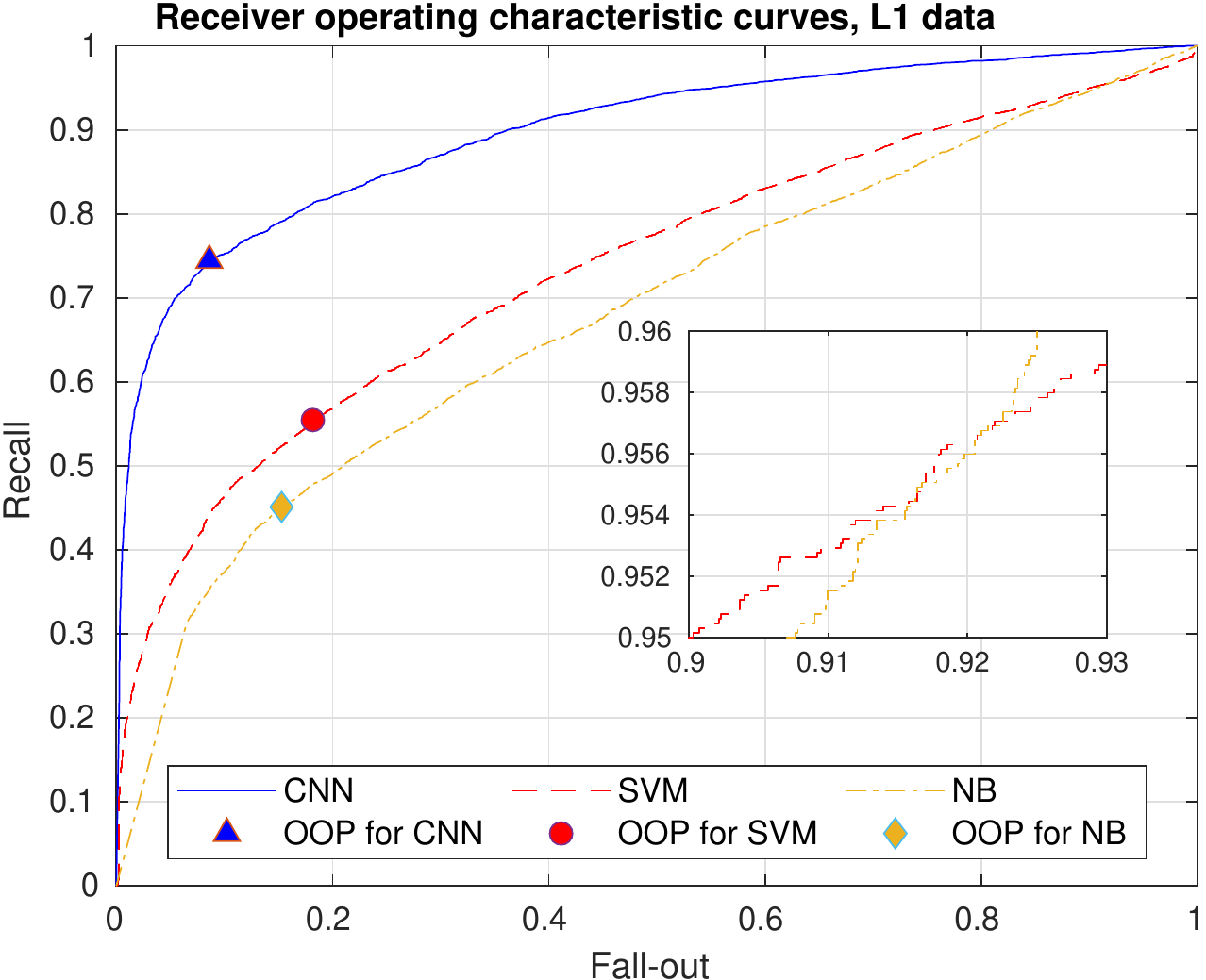}
  \caption{\label{fig:ROC_curves} ROC curves for the CNNs, a SVM classifier with a linear kernel, and a Gaussian NB classifier. Data from H1 detector (left panel) and L1 detector (righ panel) were used. For each ROC curve, its optimal operating point (OOP) is also shown. We set $T_{\text{win}}=0.75$s and same CNN hyperparameter adjustments used in previous studies. The general trend, for almost all thresholds, is that our CNN has the best performance, followed by the SVM classifier, and finally by the NB classifier. Zoomed plots shown small changes in performances near to point $(1,1)$, even though these do not affect the general trend.}
\end{center}
\end{figure*}

NB and SVM models need vectors as input, then we apply a reshaping operation: each image sample $\boldsymbol{X}_i \in \mathbb{R}^{N_{\text{time}} \times N_{\text{freq}}}$ is flattened in a vector $\boldsymbol{x}_i \in \mathbb{R}^{N_{\text{time}}N_{\text{freq}} \times 1}$ such that all columns of $\boldsymbol{X}_i$, from the first to the last, are concatenated as a one big single column. For NB model, we assume that our train set follows a Gaussian distribution, with mean and variance obtained from to the maximum likelihood estimation (MLE). For SVM model, on the other hand, we applied a normalization for each component of $\boldsymbol{x}_i$ along all training samples, and we used a linear kernel.

Take in mind that there is not a definitive criteria to generate ROC curves under the resampling regime. Then, following the same approach taken in subsection~\ref{subsec:confu_matrices} for computing confusion matrices, we considered the whole set of $100~N_{\text{test}}$ predictions made by our $10 \times 10=100$ learning-testing process. In practice, this approach avoids to average point-by-point and helps to smooth ROC curves through by increasing its number of discrete generative steps. For all ROC curves we set $T_{\text{win}}=0.75$ for the strain samples, and for ROC curves describing performance ouf our CNN, we used same hyperparameters adjustments for stacks and kernels that, in subsection~\ref{subsec:hyp_setting}, we found are the best.

Results of our comparative analyses are shown in Fig.~\ref{fig:ROC_curves}. Notice that, in both panels, we have that all ROC curves, in general, are quite distance from the 45-degree diagonal of totally random performance, which is fairly good. Even so, depending on the used dataset, their have different performances. When the models learns (and test with) H1 data, their performance are better than with L1 data. Now, if we focus separately on each panel, we have that, almost for all thresholds, CNN model has the best performance, NB model the worst performance, and SVM model is in the middle. However, as it is shown in zoomed plots, we have that ROC curves in both panels has some peculiarities. In the left zoomed plot, we observe that our CNN has the best performance only until its ROC curve reaches $\left(\text{fall-out},\text{recall}\right)=\left(0.979,0.998\right)$, because NB classifier becomes the best and remains that way until the end --in fact, performance of SVM model already had been surpassed by performance of NB model from the point $(0.960,0.994)$. From what happens next, very near to the north-east edge $(1,1)$, we should not make any strong conclusion, because we are very close to the totally random performance and, therefore, results are mainly perturbations. In the right zoomed plot we observe that only after the point $(0.923,0.957)$ in the ROC space, NB model becomes better that SVM model, and the CNN classifier always has the best performance.

Notice that, on all ROC curves, a specific point have been highlighted. This is called ``optimal operating point'' (OOP) and corresponds to the particular optimal threshold (OT) in which a classifier has the best trade-off between the costs of missing noise aline samples, cost$\left(FN\right)$, against the costs of raising false noise aline detections, cost$\left(FP\right)$. In the ROC space, this trade-off is defined by isocost lines of constant expected cost:
\begin{multline}
 \text{cost}_{\text{exp}} =
 P \left[ 1 - \text{Recall}\right]\text{cost}\left(FN\right) \\
 + N \left[\text{Fall-out}\right]\text{cost}\left(FP\right) ~,~~~~~~
 \label{eq:expec_cost}
\end{multline}
where $P=TP+FN$ and $N=FP+TN$. Assuming, as a first approach, that cost$\left(FN\right)=$ cost$\left(FP\right)=0.5$, then OOP is just the point lying on the ROC curve that intersects the 45 degree isocost line closest to the north-west corner $(0,1)$ of the ROC plot\footnote{If cost$\left(FN\right)$ and cost$\left(FP\right)$ were different and/or the dataset were imbalanced with respect classes C1 and C2, OOP and OT would be near one of its extremes. Then, in that situation, ROC analysis would be more sensitive to statistical fluctuations, making difficult to take statisfically significant decisions with respect to the class with much less detections and/or samples. This situation would require more data for dealing with the imbalance or alternative analysis as precision-recall curves or cost curves.}. For each ROC curve, their OOP, OT, and expected costs, are included in the table~\ref{tab:ROC_comparison}. Notice from this table that, for the CNN classifier, OT with H1 data is not closer to the exact fifty-fifty chance value of $0.5$ than OT with L1 data, which shows that default threshold of $0.5$ actually is choosen by convention and not because it is a limit as performance of our classifier improves. Relative difference beetwen OT and $0.5$ has nothing to do with performance, but rather with skewness of classes and/or cost of misclassifications. We also include the optimal expected cost which is computed with Eq.~\ref{eq:expec_cost} and defines the isocost curve in which the OOP lies. Notice that smaller values of $\text{cost}_{\text{cost}}$ define isocost curves closer to $(0,1)$ in the ROC space.

\begin{table*}
    \centering
    	\begin{tabular}{llllll}
     	\toprule
		Data & Model & Optimal operating point & Optimal threshold & Optimal $\text{cost}_{\text{exp}}$ & AUC\\
		\midrule
		\multirow{3}{*}{H1} & CNN  & $\left(0.0654,0.863\right)$ & $0.430$ & $0.0505$ & $0.946$  \\
		     & SVM  & $\left(0.108,0.719\right)$ & $0.228$ & $0.0972$ & $0.872$  \\
		     & NB   & $\left(0.229,0.659\right)$ & $0.379$ & $0.143$  & $0.768$  \\
        \multirow{3}{*}{L1} & CNN  & $\left(0.0860,0.744\right)$ & $0.472$ & $0.0854$ & $0.897$  \\		
		     & SVM  & $\left(0.182,0.555\right)$ & $0.418$ & $0.157$ & $0.736$  \\
		     & NB   & $\left(0.153,0.451\right)$ & $0.920$ & $0.175$ & $0.681$  \\
		\bottomrule
		\\
	\end{tabular}
	\caption{\label{tab:ROC_comparison} Additional metrics computed from ROC curves in Fig~\ref{fig:ROC_curves}. From the area under the ROC curve (AUC), we have that, for both datasets, the CNN has the best performance (the highest AUC value), followed by the SVM with a middle performance, and finally by the NB classifier with the worst performance. Notice that, for the CNN, better performance do not mean an optimal threshold closer to the defaul $0.5$ value.}
\end{table*}

In general, relative performance between models can change depending if their ROC curves intersects. Because of this we would like to have a metric for summarizing, regardless thresholds, performance of a model in a single scalar. Here we used the total area under the ROC curve (AUC)~\cite{aB96}; this is a standard metric that gives the performance of the CNN averaged over all possible trade-offs between TP predictions and FP predictions. Moreover, we can use this metric to made a final choice among all models; the best model corrrespond to the highest AUC value. In practice, we computed this metric by a trapezoidal approximation and its results are also included in the table~\ref{tab:ROC_comparison}. We have that, for both datasets, $\text{AUC}_{\text{NB}}<\text{AUC}_{\text{SVM}}<\text{AUC}_{\text{CNN}}$, allowing us to conclude that, among the three models, the CNN definitely has the best performance, followed by the SVM classifier, and finally by the NB classifier.

\subsection{Shuffling and output scoring}
\label{subsec:shuffling}

\begin{figure}[htp]
\begin{center}
  \includegraphics[width=8.0cm]{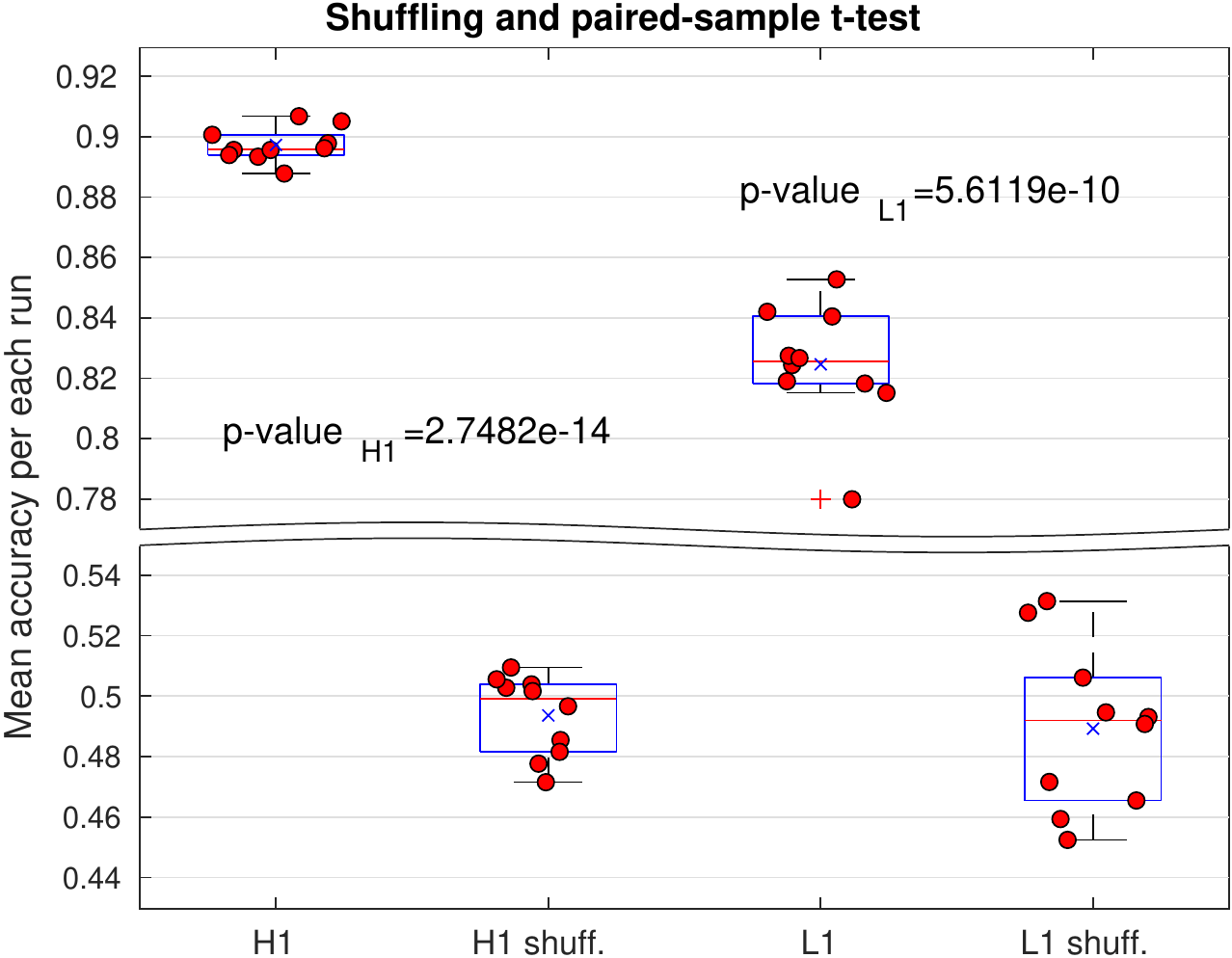}
  \caption{\label{fig:shuff_boxplots} Shuffling and paired-sample t-test for elucidating if predictions of our CNN are statistically significant. Boxplots for mean accuracies resulting of each experiment of the $10$-fold CV, are shown. By shuffling, we mean that our training set is shuffled before each training, such that links between each sample and its known class is broken. Besides, p-values of the mentioned paired-sample t-test are including, showing that predictions of the CNN are significantly different from a totally random process --with a significance level of $5\%$.}
\end{center}
\end{figure}

As it was mentioned in subsection~\ref{subsec:performance_metrics}, two related analysis to ensure that results are statistically significant were performed. The first one was run our CNN algorithm including a shuffling of training samples before each training, with the peculiarity of removing links between each sample and its known label. A comparison of distribution of the mean accuracies along all runs of the $10$-fold CV experiment, with and without shuffling, is shown in Fig.~\ref{fig:shuff_boxplots} --remember that each point of the boxplots, i.e. a $i \mhyphen th$ mean accuracy or $Acc_i{}^{CV}$ as was defined subsection~\ref{subsec:hyp_setting}, come from the $i \mhyphen th$ run of the whole $10$-fold CV. From this plot we have that shuffling radically affects results. Whether we work with data from H1 detector or L1 detector, and if shuffling is present, distribution of mean accuracy moves towards lower values and increase its dispersion. With H1 data, mean of mean accuracies decreases from $0.897$ to $0.494$ and standard deviation increases from $0.564$ o $1.34$; and with L1 data, mean of mean accuracies falls from $0.825$ to $0.489$ and standard deviation grows from $1.98$ to $2.73$.

Moreover, if shuffling is not or is present, boxplot of mean accuracies has positive or negative skewness, respectively. This makes sense because, without shuffling, the higher mean accuracies the greater is the effort of the CNN for reaching performances with those accuracies; there is not free lunch and we expect to have a higher concentration of samples below the median than above the median and, therefore, a positive skewness. On the other hand, if we have shuffling, we known from the basics of probability that adding new points, one by one, to the mean accuracy distribution, is actually a stochastically symmetric process around $0.5$ --theoretical limit if we have an infinite number of points in the distribution, i.e. an infinite number of runs of the $k$-fold CV experiment. Then, given that here we obtained medians slightly below of $0.5$ ($0.499$ with H1 data and $0.492$ with L1 data), it is expected there is a higher concentration of points above medians, and then, boxplots with negative skewness; because this works as a balance to maintain the symmetry of stochastic occurrences (i.e. boxplot points) around the $0.5$ mean accuracy value.

Descriptive statistics is a reasonable analysis but, to make a formal conclusion about the significance of our results, we performed a sample-paired t-test. Therefore, we firstly define the mean accuracy datasets
\begin{equation}
\mathcal{D} = \left\{Acc_i{}^{CV}\right\}_{i=1}^{10} \text{~,~}
\mathcal{D}_{\text{shuff}} = \left\{Acc_i{}^{CV,\text{shuff}}_i\right\}_{i=1}^{10} ~,
\end{equation}
without and with shuffling, respectively. Then, with means of each dataset at hand, $\mu$ and $\mu_{\text{shuff}}$, the task is test the null hypothesis $H_0: \mu - \mu_{\text{shuff}} = 0$ by computing the $p \mhyphen$value that is defined as:
\begin{equation}
 p = \frac{\mu-\mu_{\text{shuff}}}{\sigma/\sqrt{N_D}} ~. 
\end{equation}

Then, assuming a significance level $\alpha=0.05$ (a standard similarity threshold between $\mathcal{D}$ and $\mathcal{D}_{\text{shuff}}$), we have that: i) if $p>\alpha=0.05$, then we accept $H_0$, or ii) if $p<\alpha=0.05$, then we reject $H_0$. Results for $p \mhyphen$values are shown in Fig.~\ref{fig:shuff_boxplots}, namely $2.7482 \times 10^{-14}$ with H1 data and $5.6119 \times 10^{-10}$ with L1 data. These values are much less than $\alpha=0.05$, hence we reject null hypothesis and conclude that, for a significance level of $5\%$ (or confidence level of $95\%$), distribution $\mathcal{D}$ is significantly different from $\mathcal{D}_{\text{shuff}}$. This is actually a quite good result.

\begin{figure*}[htp]
\begin{center}
  \includegraphics[width=8.8cm]{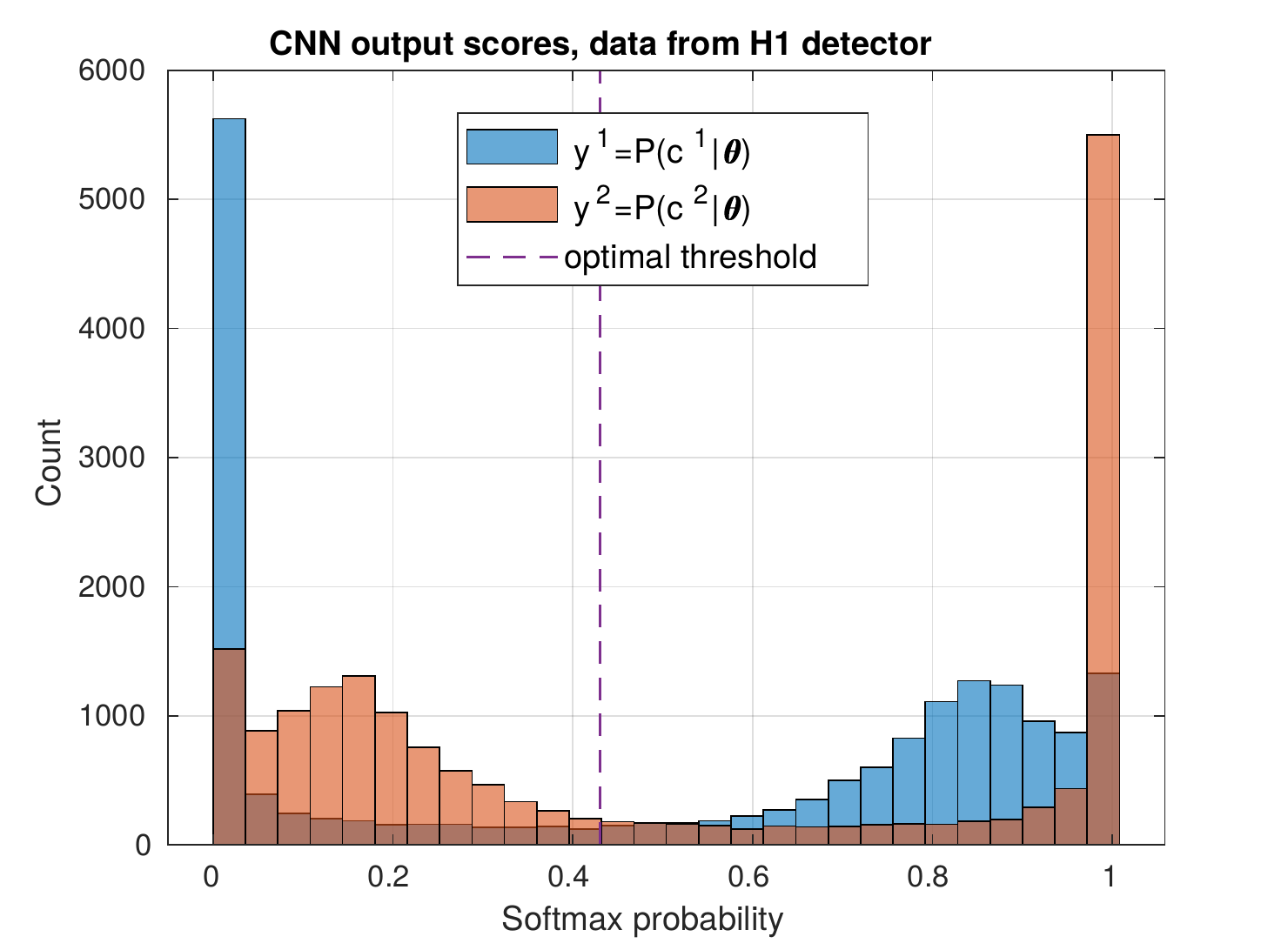}
  ~
  \includegraphics[width=8.8cm]{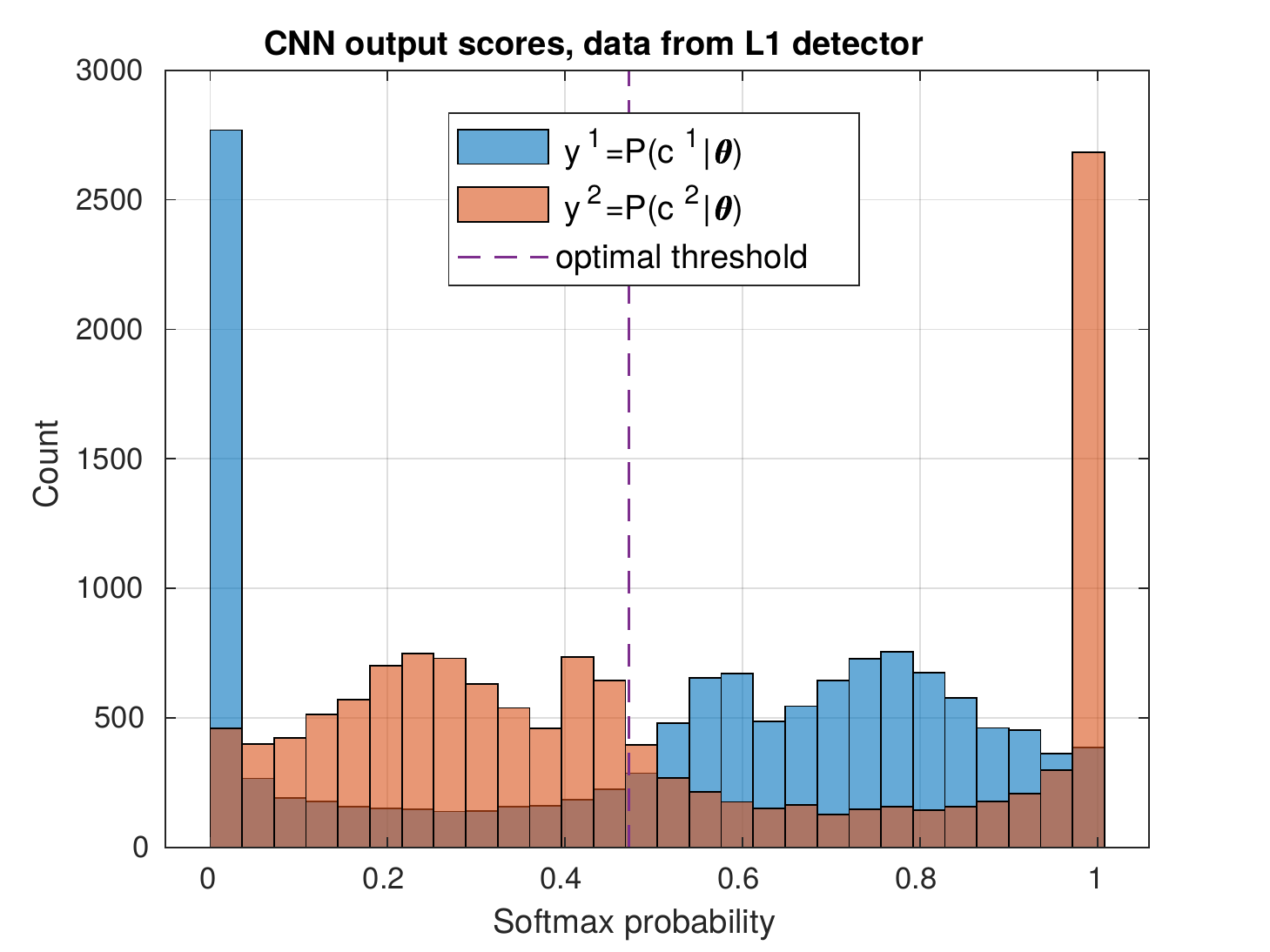}
  \caption{\label{fig:softmax_probs} Histogram of CNN output probabilistic scores, $y^1$ and $y^2$, working with data from H1 detector (left panel) and from L1 detector (right panel). As it can be observed from the skewness towards edge bins, the CNNs are more optimistic predicting GW samples than predicting noise alone samples. Besides, with H1 data there is less uncertainty than with L1 data, because with L1 data there is more population of occurrences in middle bins. Here we counted all $100~N_{\text{test}}$ predictions.}
\end{center}
\end{figure*}

As final analysis, we focus on output scoring of the CNNs. As it was explained in subsection~\ref{sec:CNN_architec}, our CNNs output scores that are probabilities generated by the softmax layer. After the training, these probabilities are defined by our classes, $c^1$ (noise alone) and $c^2$ (noise plus GW), conditioned by model parameters within vector $\boldsymbol{\theta}$ once they have already be learned; namely $y^j{}(\boldsymbol{\theta}) = P(c^j|\boldsymbol{\theta})$ (with $j=1,2$) for each input image sample. Histograms describing distribution of these probabilities, considering all our $100~N_{\text{test}}$ predictions, are included in Fig.~\ref{fig:softmax_probs} --all histograms were made using $28$ same-length bins. Here we have important results.

Firstly, in both panels of Fig.~\ref{fig:softmax_probs} we have that distribution for $y^1$ and distribution for $y^2$ are multimodal, and each one has three different modes or maximums. In addition, we observe that both distributions are asymmetric. Given a multimodal distribution, there are not a univocal definition of its center; it can be its mean, its median, its center of probability mass, among others. Here we decided to define the center of distribution for as the optimal threshold (OT), because this metric is directly related to our decision criteria for assigning a class to the output score. The closer to the OT a probabilistic occurence is located, the greater uncertainty for taking a decision about what class a CNN actually is predicting with that probability. OT values were already computed and presented in Table~\ref{tab:ROC_comparison}, and we included them in panels of Fig.~\ref{fig:softmax_probs} as dashed lines.

For $y^1$ probability we have that in the left-hand side of OT there is a low concentration of occurrences until before left edge bin, $\left[0,0.0360\right]$, having the greatest mode of the whole distribution; this edge bin has $0.31$ of all occurrences counted along all bins if we work with H1 data, and $0.21$ of all occurrences with L1 data. In contrast, in the right-hand side of OT we have more dispersed occurrences around the two remaining modes --with H1 data, one of these reimaining modes is located at edge right bin $\left]0.9720,1.0080\right]$, and with L1 data, no mode is located at edge right bin. Distribution of $y^2$ is similar to that of $y^1$ except now it is inverted along horizontal axis, then the highest mode is at the right edge bin, two remaining modes at the left-hand side of OT, among others. For $y^2$, fraction of counted occurrences in right edge bin is also the same, $0.31$ with H1 data and $0.21$ with L1 data.

Above results actually mean that, given our datasets, our CNNs are more optimistic predicting GW samples than predicting noise alone samples, or equivantly, more pessimistic predicting noise alone samples than predicting GW events. Hence, even though our input datasets are exactly class-balanced, predictive behavior of our CNNs is highly class dependent. Under a frequentist approach, asymmetric shape of distributions for $y^1$ and $y^2$ is the statistical reason why the CNNs have a high precision and a not negligible false negative rate or, said more simply, why the CNNs are more ``conservative'' classifying samples as noise alone, than classifying as GW events --and this is coherent with that we interpreted from confusion matrices shown in Fig.~\ref{fig:confu_results}.

It is also important to notice how, depending of data, distribution of occurrences, either for $y^1$ or $y^2$, change. Remember that, from previous ROC comparative analyses, we found that working with H1 data reach a better performance than working with L1 data, and here we also observed this improvement from another point of view. Because, the more uncertainty our CNNs have for predicting a specific class, occurrences are more concentrated (i.e. skewed) towards OT. Even, if our network was not learn anything, e.g. because of a shuffling as we did previously applied, then we would have that all probabilities are distributed as a Gaussian centered a the default threshold, $0.5$, i.e. a totally random performance --although it is not explicitly included here, we checked this random values with our code and visualizations.

\section{Conclusions}
\label{sec:conclusions}

In this work, we applied CNN algorithms for detection of GW signals from CBCs, through a binary classication with samples of non-Gaussian noise only and samples of non-Gaussian noise plus GW injections. Being a crucial part of the data pre-processing, we applied a Morlet wavelet transform, to convert our time series vectors (i.e. strain data) in time-frequency matrices (i.e. image data). And the resulting images were automatically decoded in the convolutional stacks of our CNNs. Besides, images in time-frequency representation were reduced in such a way that all our CNNs were easily run in a single local CPU, reaching good performance results.

The significant novel contribution of our work is adopting a resampling white-box approach, motivated by the need to advance towards a statistically informed understanding of uncertainties involved in CNNs. Moreover, as a manner to reproduce a more realistic experimental setting for testing our CNN algorithms, we draw on single-interferometer data from LIGO S6 run, considering raw strain data with noise and hardware injections of GW signals solely; that is to say, we removed the instrumental freedom of generating distributions of simulated GW signals as intense as one want.

Because of introducing stochasticity by repeated $10$-fold CV experiments, almost all tasks were more complex than in a simple deterministic fashion but, it also forced us to acquire useful tools to overcome too much optimistic or too much pessimistic evaluations of CNN algorithms. Hyperparameter adjustments required careful interpretations of mean accuracy distributions. We tested several CNN architectures and found two that achieve optimal peformances, one with $3$ stacks and $32$ kernels for data from H1 detector, and other with $2$ stacks and $16$ kernels for data from L1 detector, and in both cases with a resolution $T_{\text{win}}=0.75$s. Besides, we found that stochasticity introduced by mini-batch SGD in accuracies is smoothed by the resampling, which is achieved by reducing perturbations in a factor of $3.6$. This result serves as recommendation for future works, to run CNNs in the resampling regime.

From analyses of confusion matrics and standard metrics, we found that whether working with H1 data or L1 data, CNN algorithms are quite precise to detect noise samples but not sensitive enough to recover GW signals. This results mean that, in general, CNN algorithms are better suitable for noise reduction than generation of GW triggers. However, these conclusions are not totally definitive, because if we considering only GW events of SNR $\geq 21.8$ working with H1 data, and SNR $\geq 26.8$ working with L1 data, CNN algorithms actually could be considered as a tentative algorithm for GW detection. Here we stress the label \textit{tentative}, because mentioned conditions for SNR values actually depend on the nature of initial datasets inputted by the CNNs that, in our case, still are class-balanced. At this point, it is evident that more research is necessary to begin dealing with arbitrarily class-imbalanced data.

With ROC curves, we also compared the CNNs with other two classic ML models, NB and SVM, reaching that CNNs have much better performances than mentioned classic ML models --result that is consistent with what have been reported these recent years in most works of DL applied to GW detection. From ROC analyses we also found optimal thresholds, that are very useful parameters to establish a statistical decision criterion for the classification, namely $0.430$ for H1 data, and $0.472$ for L1 data.

In order to elucidate if our predictions are statistically significant, we performed a paired-sample t-test, obtaining that performace of CNN algorithms are significant different to that of a totally random classifier, with a confidence level of $95\%$. Finally, thank to our white-box approach, we found that probabilistic scores are asymmetrically distributed, implying that predictive behaviour of the CNNs is highly class dependent. Here we concluded that discrepancy between precision and sensitiviy of the CNNs can be statistically explained by the nature of the score distributions, which in turn come from mathematical formulation of the softmax activation function that we included as penultimate layer in the classification stage of the CNNs. This conclusion implies that softmax layer works, not only as tool to generate probalistic scores useful for the binary classification, but also to measure uncertainties of the CNNs given the datasets.

We presented a detailed cross-disciplinary exploration to seriously deal with uncertainties of CNN algorithms. In particular, we show that for achieving this goal is really fundamental to have clear statistical information about advantages and disadvantages of CNN algorithms. In general, we highly recommend that future works focused on testing DL algorithms in GW analysis, should focus on the problem of how to establish more realistic experimental settings in their metodologies, and how to deal with difficulties that arise from these settings, paying more attention to uncertainties. At the end, this is one of the main pathways to claim that DL techniques are real alternatives to standard GW data analysis pipelines.

\section{Acknowledgments}
\label{sec:acknowl}

M.D.M and A.I.N. acknowledge the support of PRODEP, $511$-$6/18$-$16280$. C.M. thank to PROSNI-UDG. This research has made use of data, software and/or web tools obtained from the Gravitational Wave Open Science Center (\url{https://www.gw-openscience.org}), a service of LIGO Laboratory, the LIGO Scientific Collaboration and the Virgo Collaboration. LIGO is funded by the U.S. National Science Foundation. Virgo is funded by the French Centre National de Recherche Scientifique (CNRS), the Italian Istituto Nazionale della Fisica Nucleare (INFN) and the Dutch Nikhef, with contributions by Polish and Hungarian institutes. 

\appendix

\section{Naive Bayes classifiers}
\label{sec:App_NB}

Naive Bayes (NB) is a family of classifiers based on the known Bayes theorem and the naive assumption that all observed data are independent of each other and identically distributed (i.i.d.). Considering a binary classification, we have that given an $i \mhyphen th$ input feature vector ${\boldsymbol x}_i = \left(x_{i1},x_{i2},...,x_{id}\right)$\footnote{Given the dimension of images in the time-frequency representation, and keeping consistency with our notation, $d=N_{\text{time}}N_{\text{freq}}$.}, a NB model outputs two posterior categorical probabilities, namely:
\begin{equation}
 y^j{}_i({\boldsymbol x}_i) = P_i(c^j|{\boldsymbol x}_i) ~~,~~ \text{with~~} j={1,2} ~.
\end{equation}
Then, class of the highest output probability is chosen as the prediction for the input vector ${\boldsymbol x}_i$ based on the maximum a posteriori (MAP) estimation as decision criteria. Under the hood, NB classifiers output posterior probabilities that, according to Bayes theorem, are:
\begin{equation}
 P_i(c^j|{\boldsymbol x}_i) \propto P_i({\boldsymbol x}_i|c^j) P(c^j) ~, \label{eq:NB_prop}
\end{equation}
where we can ignore the marginal normalization constant, i.e. the numerator of the right-hand side that would convert the proportionality in a equality, because of class predictions are made regardless of it. Now, to compute probabilities appearing at the right-hand side of Eq.~\ref{eq:NB_prop}, we need three ingredients. Firstly, the likelihood function that, assuming that our features are independent and i.i.d., is computed by:
\begin{equation}
 P_i({\boldsymbol x}_i|c^j)=\sum_{k=1}^{d} P_i(x_{ik}|c^j) ~.
\end{equation}
Secondly, prior marginal probability $P(c^j)$, that are computed as relative frequency of class $c^j$ in the train set. Finally, we have that values of conditional probabilities $P_i(x_{ik}|c^j)$ will depend on our assumption about how features are distributed. Indeed, this last assumption is an important aspect, because it will define what kind of NB classifier we will working with.

For our study, we chose $P_i(x_{ik}|c^j)$ as Gaussian distribution, in which its mean and variance (i.e. our model parameters) are computed from the train set through the maximum likelihood estimation (MLE). Besides, under a frequentist apporach, MLE is the standard procedure of maximizing likelihood function --or for minimizing log-likelihood function as the cost function. In practice, we implemented our Gaussian NB classifier with the \texttt{MATLAB Statistics and Machine Learning Toolbox}~\cite{matlab-ml}.

\section{Support Vector Machine classifiers}
\label{sec:App_SVM}

Support Vector Machines (SVM) are a family of regressors and binary classifiers that are based on a optimization and geometrical framework. For a SVM classifier, we start with an $i \mhyphen th$ input feature vector ${\boldsymbol x}_i = \left(x_{i1},x_{i2},...,x_{id}\right)$. Next, assuming that data are not linearly separable, we apply a transformation $\phi$ in order to map our data in a feature space of $\tilde{d}>d$ dimensions, leading to the new feature vector $\phi \left( {\boldsymbol x}_i \right) = \left(\phi_{i1},\phi_{i2},...,\phi_{i\tilde{d}}\right)$. If our dataset is linearly separable in the higher dimensional space, then the binary SVM classifier outputs a (non-probabilistic) score given by the function:
\begin{equation}
 f({\boldsymbol x}) = {\boldsymbol w}^T \phi \left({\boldsymbol x}_i\right) + b ~,
\end{equation}
where ${\boldsymbol w} = \left(w_1,w_2,...,w_{\tilde{d}}\right)$ is a weight vector and $b$ a bias term, both to be learn. As a decision criteria, if $f({\boldsymbol x})>0$ we have that predicted class is 1 (positive) and, if $f({\boldsymbol x})<0$, predicted class is 2 (negative).

\begin{figure*}[!ht]
\begin{center}
  \includegraphics[width=13cm]{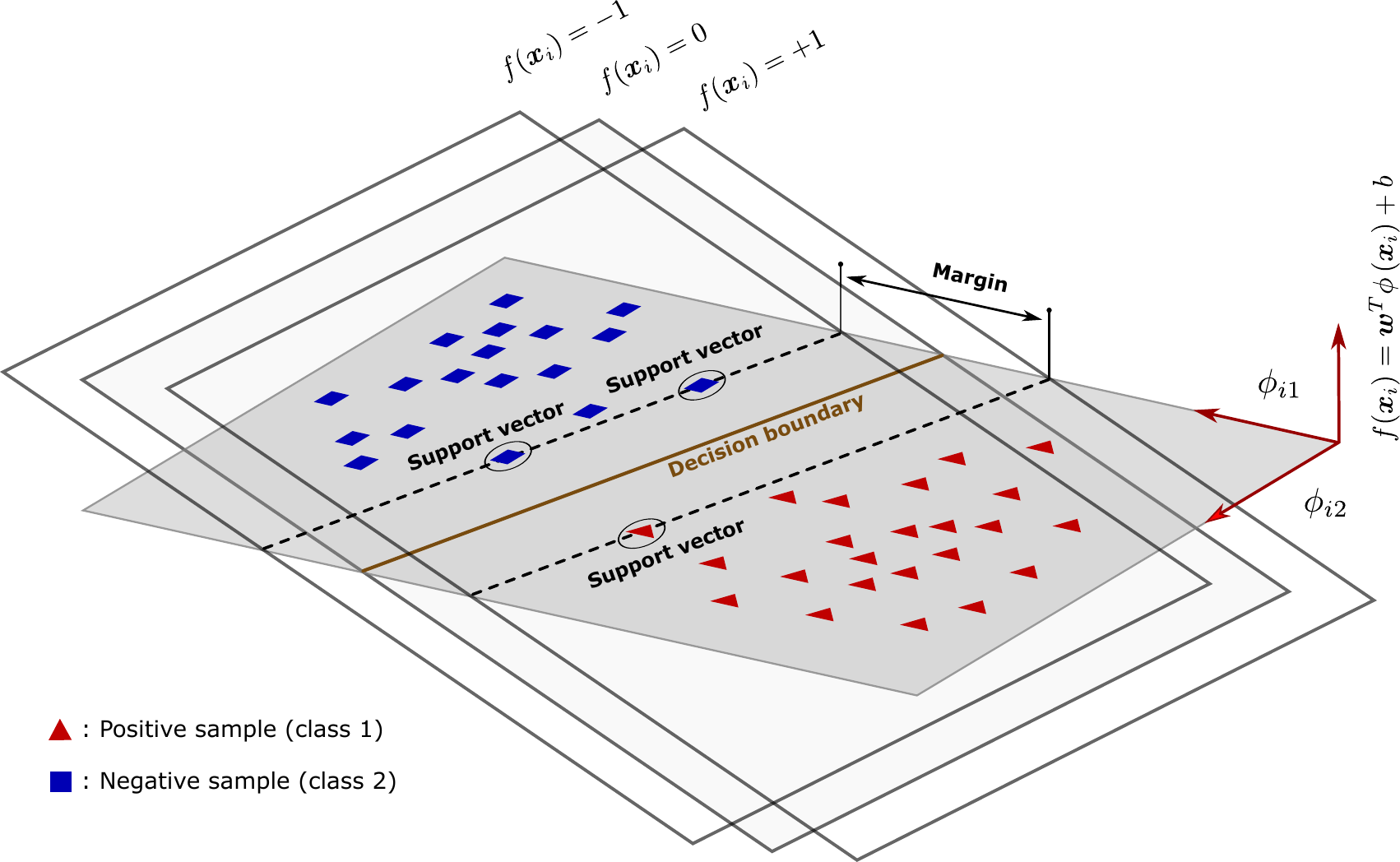}
  \caption{\label{fig:SVM_classifier} Geometric illustration of a SVM classifier. After apply a transformation $\phi$ to input feature vectors ${\boldsymbol x}_i$, dispersion of data samples ocurrs in a higher dimensional hypersurface defined by $\phi_{i1}$ and $\phi_{i2}$. The decision boundary is given by the intersection of hyperplane $f({\boldsymbol x})=0$ with the mentioned hypersurface. Besides, this intersection have a margin that is defined by the support vectors and the intersection lines of hyperplanes $f({\boldsymbol x})=\pm 1$ with the higher feature hypersurface.}
\end{center}
\end{figure*}

Beyond scoring, function $f({\boldsymbol x})$ has a geometrical intepretation. As it is illustrated in Fig.~\ref{fig:SVM_classifier}, we have that $f({\boldsymbol x})=0$ is an hyperplane that separates samples of one class from those of the other class, with a decision boundary located along the intersection line of hyperplane $f({\boldsymbol x})=0$ with the higher feature hypersurface (not necessarily a hyperplane) defined by the transformation $\phi$. Decision boundary is specifically formulated such that it has a margin defined by support vectors (i.e. the closest samples to the decision boundary), and intersections of hyperplanes $f({\boldsymbol x})=\pm 1$ with the mentioned higher feature hypersurface.

SVM illustrated in Fig.~\ref{fig:SVM_classifier} is called hard-margin SVM classifier, because it do not have any sample within the margin or margin violations. However, this approach generally is not very useful, because it is too sensitivity to outliers and could make unnecessarily difficult the search of a decision boundary for non linear datasets. The standard solution for this is to allow some margin violations, which leads to a soft-margin SVM classifier.

Then, assuming a soft-margin classification, learning of ${\boldsymbol w}$ and $b$ can be expressed as the objective:
\begin{eqnarray}
 && \text{minimize}_{~{\boldsymbol w},b,{\boldsymbol \zeta}}
 ~~ \left( \frac{1}{2} {\boldsymbol w}^T {\boldsymbol w} + C \sum_{i=1}^{d} \zeta_i \right)
 ~, \label{eq:SVM_objective} \\ 
 && \text{such that} ~~ t_i f({\boldsymbol x}) \geq 1 - \zeta_i ~~\text{with}~~
 i = 1,2,..,d, \label{eq:SVM_constraint}
\end{eqnarray}
where $\zeta^i \geq 0$ is a slack variable that measures how much the $i \mhyphen th$ instance is allowed to violate the margin, $C>0$ a hyperparameter defining the trade-off between minimization of $\frac{1}{2} {\boldsymbol w}^T {\boldsymbol w}$ (to increase margin as much as possible) and minimization of $\zeta^i$ (to reduce as much as possible margin violations), and $t_i = \pm 1$ depending if the $i \mhyphen th$ instance is class 1 or class 2, respectively. Eqs.~\ref{eq:SVM_objective} (objective) and~\ref{eq:SVM_constraint} both together denote a convex and linearly constrained optimization problem, wich is a particular case of a Quadratic Programming (QP) problem.

Notice that, because of constrain given by Eq.~\ref{eq:SVM_constraint}, computing $\phi({\boldsymbol x}_i)$ could be resource intensive, and even computationally prohibitive, if we have many features. However, this can be dodged with the kernel trick. Because of the Represent Theorem~\cite{Wahba-Book}, ${\boldsymbol w}$ can always be rewritten as a linear combination ${\boldsymbol w} = \sum_{i=1}^{N} \alpha_i \zeta_i {\boldsymbol x}_i$. Then, Eqs.~\ref{eq:SVM_objective} and~\ref{eq:SVM_constraint} can be expressed as:
\begin{eqnarray}
 && \text{minimize}_{~\alpha}
 \left[ \frac{1}{2} \sum_{i=1}^{d} \sum_{j=1}^{d} \alpha_i \alpha_j t_i t_j \phi \left({\boldsymbol x}_i\right)^T
    \phi\left({\boldsymbol x}_j\right) \right. \nonumber\\
 && ~~~~~~~~~~~~~~~~~~~~~~~~~~~ \left. - \sum_{i=1}^{d} \alpha_i \right] ~, \\ \label{eq:SVM_dual_objective}
 && \text{such that} ~~ \alpha_i \geq 0 ~~,~~ \text{with } i = 1,2,..,d, \label{eq:SVM_dual_constraint}
\end{eqnarray}
which is the dual for of our linearly constrain objetive, and also a QP problem. Moreover, we notice that the non-linear transformation appearing as a dot product in Eq.~\ref{eq:SVM_dual_objective} always can be rewritten as:
\begin{equation}
 \phi \left({\boldsymbol x}_i\right)^T \phi\left({\boldsymbol x}_j\right) = \text{K}\left({\boldsymbol x}_i,{\boldsymbol x}_j\right) ~,
\end{equation}
where function $\text{K}$ is called kernel. This kernel depends on original features and can take several forms depending on the problem we are dealing with. For this research, we implemented a linear kernel, $\text{K}\left({\boldsymbol x}_i,{\boldsymbol x}_j\right) = {\boldsymbol x}^T_i {\boldsymbol x}_j$, being the default option for a binary SVM classifier.

In general, there are several techniques to solve QP problems, both in its primal and dual forms. A detailed exploration of these are beyond the scope of this paper. However, it is worthy to mention that in our SVM classifier we implemented a Sequential Minimal Optimization (SMO) routine with the \texttt{MATLAB Statistics and Machine Learning Toolbox}. For more details about SMO, see reference~\cite{rFpCcLB96}.

\bibliographystyle{unsrt}
\bibliography{refs} 

\end{document}